\documentclass[%
reprint,
superscriptaddress,
 amsmath,amssymb,
 aps,prx,floatfix
]{revtex4-2}

\usepackage{newtxtext}
\usepackage{newtxmath}
\usepackage{subfigure}
\usepackage{float}
\usepackage{graphicx}
\usepackage{dcolumn}
\usepackage{bm}
\usepackage{cancel}
\usepackage[normalem]{ulem}

\newcommand{\rb}{{\mathbf r} }
\newcommand{\qv}{{\mathbf q} }
\newcommand{\jv}{{\mathbf j} }

\renewcommand{\Re}{\operatorname{\mathbb{R}e}}
\renewcommand{\Im}{\operatorname{\mathbb{I}m}}

\usepackage[colorlinks,citecolor=blue]{hyperref}

\begin{document}

\preprint{APS/123-QED}

\title{Physical limits to membrane curvature sensing by a single protein}

\author{Indrajit Badvaram}
\affiliation{%
 Department of Biophysics, Johns Hopkins University, Baltimore, MD
}

\author{Brian A. Camley}
\affiliation{%
    William H. Miller III Department of Physics \& Astronomy, Johns Hopkins University, Baltimore, MD
}%
\affiliation{%
 Department of Biophysics, Johns Hopkins University, Baltimore, MD
}


\begin{abstract}
Membrane curvature sensing is essential for a diverse range of biological processes. Recent experiments have revealed that a single nanometer-sized septin protein can distinguish between membrane-coated glass beads of one micron and three micron diameters, even though the septin is orders of magnitude smaller than the beads. This sensing ability is especially surprising since curvature-sensing proteins must deal with persistent thermal fluctuations of the membrane, leading to discrepancies between the bead's curvature and the local membrane curvature sensed instantaneously by a protein. Using continuum models of fluctuating membranes, we investigate whether it is feasible for a protein acting as a perfect observer of the membrane to sense micron-scale curvature either by measuring local membrane curvature or by using bilayer lipid densities as a proxy. To do this, we develop algorithms to simulate lipid density and membrane shape fluctuations. We derive physical limits to the sensing efficacy of a protein in terms of  protein size, membrane thickness, membrane bending modulus, membrane-substrate adhesion strength, and bead size. To explain the experimental protein-bead association rates, we develop two classes of predictive models: i) for proteins that maximally associate to a preferred curvature, and ii) for proteins with enhanced association rates above a threshold curvature. We find that the experimentally observed sensing efficacy is close to the theoretical sensing limits imposed on a septin-sized protein, and that the variance in membrane curvature fluctuations sensed by a protein determines how sharply the association rate depends on the curvature of the bead.
\end{abstract}

\maketitle

\section{\label{sec:level1}Introduction\protect}
Membrane curvature is ubiquitous  throughout cell biology \cite{mcmahon2005membrane,mcmahon2015membrane,stachowiak2013cost}: proteins that sense membrane curvatures can help locate the axis of cell division, determine cell polarity, facilitate membrane remodeling, and serve as a cue for intracellular signaling \cite{zimmerberg2006proteins,ramamurthi2010protein, singh2022sensing, bassereau20182018,lou2018role}.
These proteins often act in tandem by binding with each other to sense curvature cooperatively. However, in the case of septin proteins, recent experiments have shown that in addition to sensing curvatures via cooperative filament formation \cite{bridges2016micron,beber2019membrane}, even a {\it single} septin protein can distinguish between micron-scale membrane curvatures and preferentially bind to membranes adhered to glass beads of different curvatures with different association rates \cite{cannon2017unsolved,cannon2019amphipathic, shi2022kinetic}. How do proteins only a few nanometers in size effectively sense membrane curvatures that are hundreds of times larger than themselves, on the order of micrometers? This sensing ability is even more remarkable considering that biological membranes undergo persistent thermally-driven undulations \cite{safran2018statistical}. Even if a protein could perfectly measure the instantaneous shape of the membrane at the nanometer scale, these undulations drive the membrane away from its average shape, confounding the protein's attempts to measure the membrane's curvature. How can a protein reliably make a measurement of micron-scale curvature in this noisy environment?

\begin{figure}[ht]
\centering
    \includegraphics[width=0.47\textwidth]{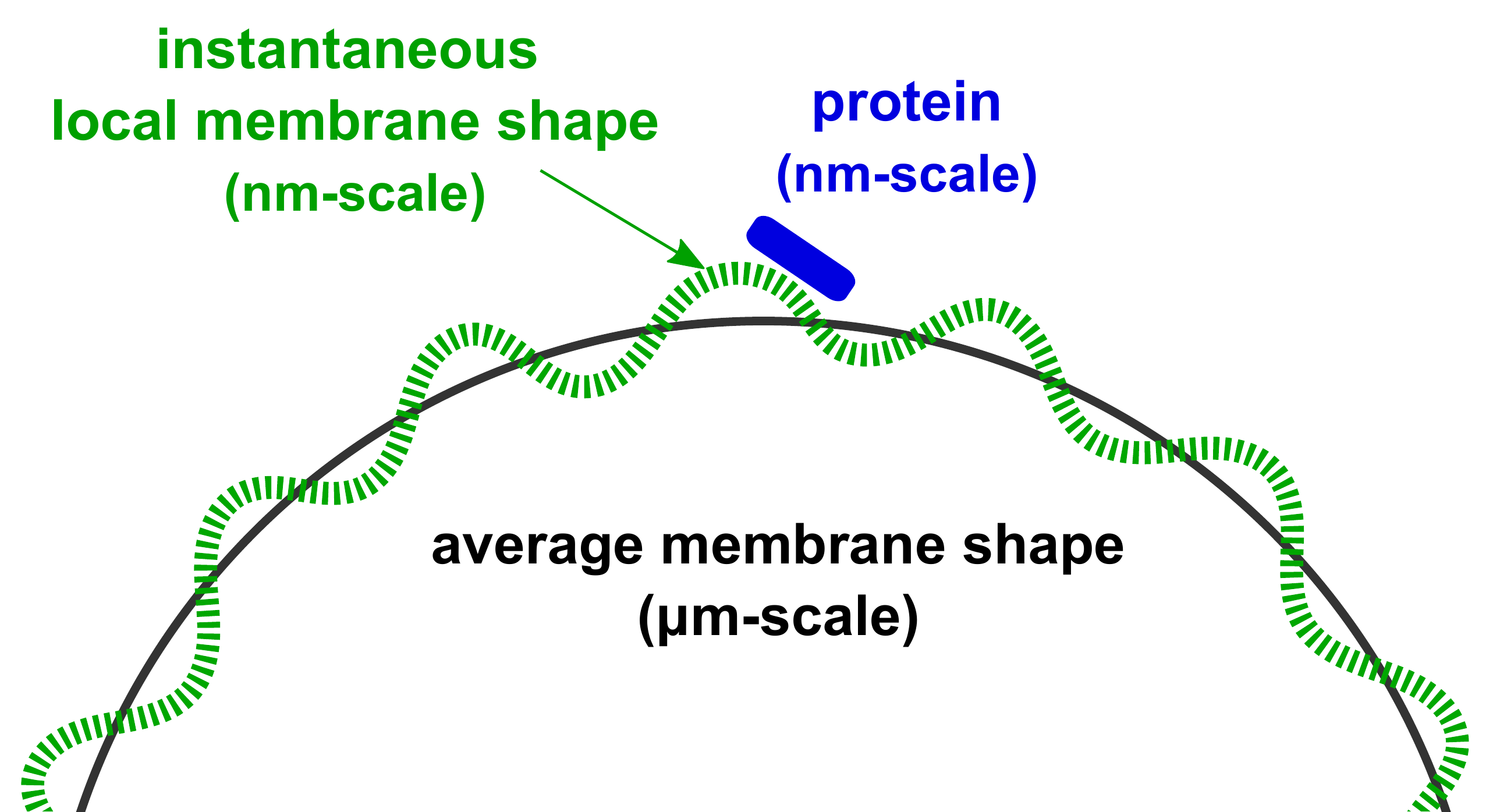}
    \caption{Thermal fluctuations of the membrane lead to discrepancies between the instantaneous local membrane shape present at the protein's location and the average membrane shape. Even if the protein were a perfect detector of curvatures, observing the local curvature without error, the local shape may not be indicative of the overall membrane shape. A nanometer-sized protein is challenged by relative differences in scale: steep local undulations have radii of curvature on the scale of nanometers, while the membrane on average may be much flatter, with a radius of curvature on the scale of micrometers (not to scale in this illustration).}
    \label{fig:SmallProtLargeProt}
\end{figure} 

Curvatures also induce deviations in the packing of lipids in the membrane bilayer, and proteins with amphipathic helices insert themselves into bilayers \cite{drin2010amphipathic,drin2007general}. Proteins that sense micron-scale curvature may then be, instead of measuring the shape directly, using a proxy for local curvature related to lipid packing to sense membrane shape \cite{cui2011mechanism}. We conjecture that one such proxy might be the deviations in lipid densities between the upper and lower monolayers of the membrane bilayer.

 Although there are descriptions of molecular mechanisms employed by proteins when sensing nanometer-scale curvatures \cite{antonny2011mechanisms}, curvature sensing at the micron-scale is less well-understood \cite{assoian2019cellular}. Previous theoretical studies have modeled the thermodynamics of curvature sensing \cite{baumgart2011thermodynamics} and the effects of helix insertion on the membrane's energy \cite{fu2021continuum,campelo2014sensing}. Here, we take a different approach: we ask how precisely a protein could measure the micron-scale curvature of a membrane if it made a perfect measurement of local membrane shape or local lipid density, subject to the inevitable thermal fluctuations of the membrane. This gives us the fundamental physical limits to curvature sensing for an idealized protein, akin to  Berg and Purcell's classic work on the limits of ligand concentration sensing for a perfect detector of a finite size \cite{berg1977physics} and later follow-ups \cite{bialek2005physical,endres2009maximum,kaizu2014berg,ten2016fundamental}. Our result builds on the larger literature of sensing limits in different contexts, including gradient sensing \cite{hu2010physical,camley2018collective,ipina2022collective,mugler2016limits,nwogbaga2022physical}, flow sensing \cite{fancher2020precision}, and sensing the mechanical properties of heterogeneous materials \cite{beroz2017physical,beroz2020physical}. To quantify curvature sensing in this way, we define a signal-to-noise ratio (SNR) to indicate how well a protein is able to extract useful information about the membrane's shape despite stochasticity. To support our analytical models, we develop algorithms to simulate membranes whose fluctuations in height are coupled to fluctuations in bilayer lipid densities. 

Motivated by the experiments in \cite{cannon2019amphipathic,shi2022kinetic}, we explore the theoretical limits to a protein's ability to distinguish between membrane-adhered beads of different sizes. For a bead of radius $R$, the mean membrane curvature on average is $C = \frac{1}{R}$. However, if we probe the membrane over a region of a protein size $a$, the curvature will differ from this value (Fig. \ref{fig:SmallProtLargeProt}). We derive that in the absence of membrane-substrate adhesion, the curvature fluctuations sensed by a protein of size $a$ and membrane bending modulus $\kappa$ are $\sqrt{\langle C_a ^2 \rangle} = \frac{1}{a}\sqrt{\frac{k_B T}{16 \pi \kappa}}$. Since $a \ll R$ and $\frac{k_B T}{\kappa} \approx 1/20$, the magnitude of these curvature fluctuations is very large relative to the average curvature $1/R$---so even with a perfect measurement, a protein could observe values of curvature far from $1/R$, indicating the difficulty in curvature sensing on a freely fluctuating membrane. Including membrane-bead adhesion suppresses these fluctuations, amplifying the protein's ability to distinguish between micron-sized beads. We also study the signal-to-noise ratio associated with measuring changes in lipid density, and find that measuring lipid densities compares well to an ideal curvature measurement when the protein is small, or when the membrane is thick and resistant to in-plane compression. Lastly, to explain the experimentally observed protein-bead association rates, we show how fluctuation variances like $\langle C_a^2 \rangle$ and sensing SNR can determine the association rate of a protein to membrane-adhered beads of different curvatures. To do so, we develop two classes of models: the preferred curvature model, for proteins that associate maximally to a specific membrane curvature, and the curvature threshold model, for proteins that exhibit enhanced binding to all curvatures above a threshold.
Using the preferred curvature model for our estimates of membrane-substrate adhesion, we find that measurements of curvature by
a single septin protein are close to the fundamental limit of sensing accuracy. With the curvature threshold model, we show that the sharpness in the transition to enhanced association above a threshold is determined by the variance in membrane curvature fluctuations sensed by the protein.

\section{\label{sec:level2}Models and simulation Methods\protect}

\begin{figure}[htb]
\centering
    \includegraphics[width=0.48\textwidth]{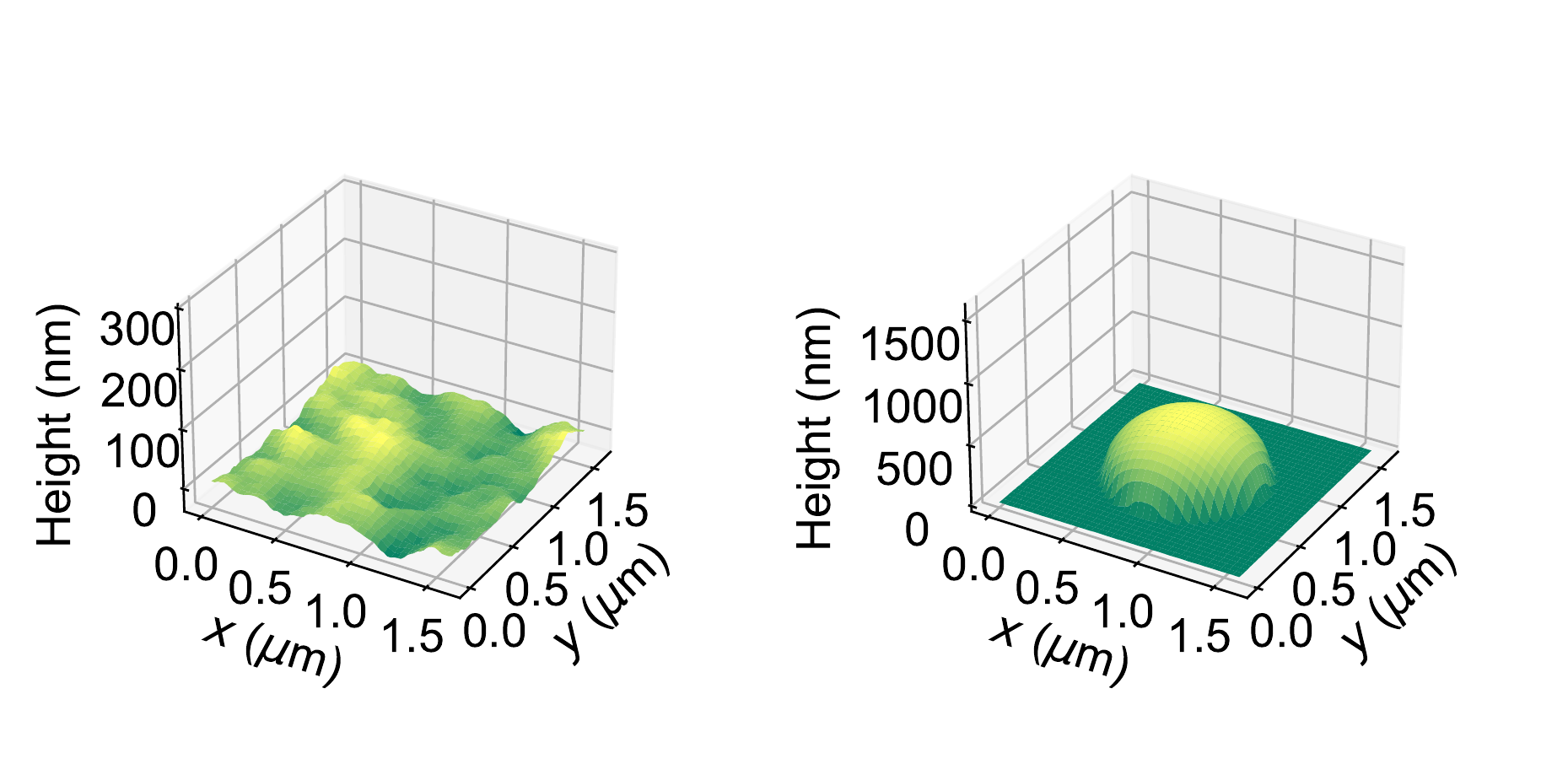}
    \caption{Snapshots of thermally-fluctuating simulated membranes: i) (left) a freely fluctuating flat membrane with no membrane-substrate adhesion, and ii) (right) a membrane adhered to a bead of radius $R = 500$ nm with adhesion strength $\gamma = 10^{13}$ J/m$^4$. System size is $L$ = 1.6 $\mu$m. Note that due to the large difference in size scales and strong membrane-substrate adhesion, fluctuations are not apparent in the plot on the right.  Simulation parameters: Table \ref{tab:ModelParameters}. }
    \label{fig:FlatAndBead_Surface}
\end{figure}

\subsection{\label{ModelingAdhesion}Modeling membrane, bead, and membrane-bead adhesion\protect}

We represent the shape of the membrane in terms of its height $h(\mathbf{r})$ above a two-dimensional plane as a function of position $\mathbf{r} = (x, y)$, i.e. using Monge gauge \cite{brown2008elastic}. To induce a curvature similar to the bead of \cite{cannon2019amphipathic,shi2022kinetic}, we model a substrate with a spherical bump on a flat surface (Fig. \ref{fig:FlatAndBead_Surface}). We assume that the adhesion energy between the bead and membrane arises from a harmonic potential,
\begin{equation}
\label{AdhesionEnergyReal}
    E_\text{adh} = \frac{\gamma}{2} \int \mathrm{d}\mathbf{r} (h(\mathbf{r}) - h_\text{bead}(\mathbf{r}))^2,
\end{equation}
where $\gamma$ is the strength of membrane-substrate adhesion in units of J/m$^4$, $h_\text{bead}(\mathbf{r})$ traces the height of the bead at each position in the $xy$ plane and serves as the equilibrium height, and the integral $\int \mathrm{d}\mathbf{r}$ is over the $xy$ plane. This harmonic potential approximates more detailed potentials, e.g. the Mie potential of \cite{schmidt2014signature}  or van der Waals interactions \cite{swain1999influence}; see Appendix \ref{app:justifyAdhesionStrength}. The height field $h_\text{bead}(\mathbf{r})$ corresponds to the $z$-axis height of a hemisphere of radius $R$, centered in the plane, i.e.  $h_\text{bead}(\mathbf{r}) = \sqrt{R^2 - 2s^2 + 2s(x + y) - x^2 - y^2}$ with $s = L/2$ for $\mathbf{r}$ within $R$ of the bead center $(s,s)$. We set $h_\textrm{bead}(\mathbf{r}) = 0$ outside this region: the membrane is adherent to a flat substrate outside of the bead region.

\subsection{\label{MembraneCoupling}Energy of membrane height and density changes\protect}

\begin{figure}[ht]
\centering
    \includegraphics[width=0.45\textwidth]{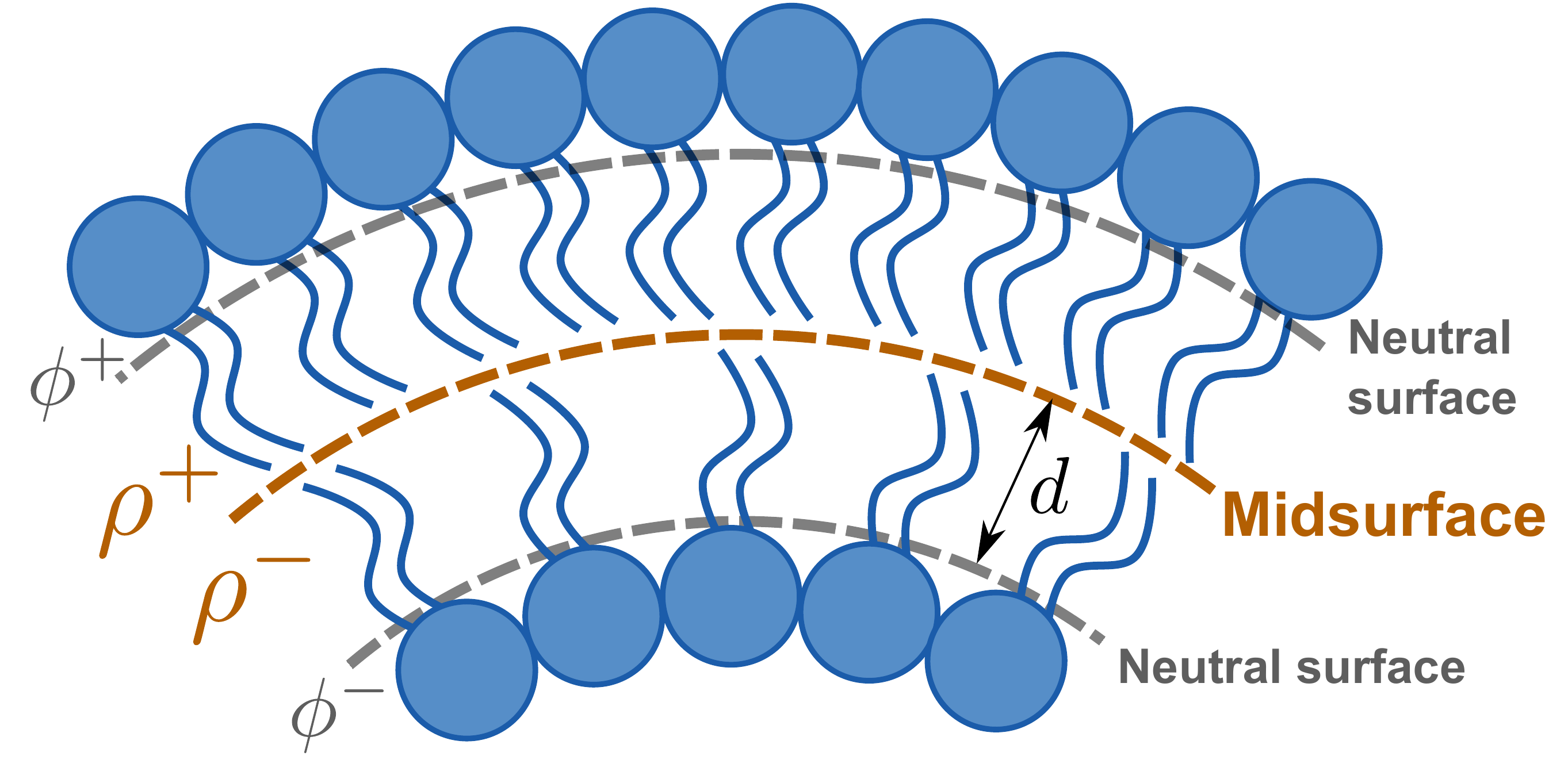}
    \caption{A curved membrane induces deviations in the packing of lipids in the bilayer. When the membrane is flat, the number densities of lipids projected by the two monolayers at the midsurface are equivalent. However, when the membrane is curved and its lipids are allowed to laterally relax to their minimum energy value, the upper ($+$) and lower ($-$) monolayers project different densities at the midsurface. The steeper the curvature, the greater the difference between the scaled densities $\rho^+$ and $\rho^-$. At steady-state, the neutral surface lipid number densities $\phi^+ =  \phi^-$. The distance between the midsurface and either neutral surface is $d$.}
    \label{fig:curvedbilayer_schematic}
\end{figure}

In addition to the height of the membrane, we also characterize the membrane by the lipid densities in each leaflet. We use the Seifert-Langer model \cite{seifert1993viscous, seifert1997configurations} to represent how the membrane's height couples to lipid densities. Due to membrane curvature, the lipid densities measured at different depths into the membrane bilayer will differ. We follow Seifert and Langer and primarily use the scaled lipid densities of the upper and lower monolayers at the midsurface,  $\rho^+$ and $\rho^-$. These are defined as $\rho^{\pm} \equiv (\psi^\pm /\phi_0 - 1)$, where $\psi^\pm$ are the densities projected onto the bilayer midsurface and $\phi_0$ is the equilibrium number density of a flat membrane. With this definition, the lipid density deviation between the upper and lower monolayers at the midsurface is given by $\rho \equiv (\rho^+ - \rho^-)/2$, and the average density is $\bar{\rho} \equiv (\rho^+ + \rho^-)/2$. As shown in Fig. \ref{fig:curvedbilayer_schematic}, when the membrane is bent to a positive curvature and the lipids allowed to laterally relax, the density projected by the upper leaflet at the midsurface is greater than that projected by the lower leaflet. (This is in contrast to the density profile when \textit{momentarily} bending the membrane, where the midsurface densities are equal and the upper and lower leaflets are stretched and compressed at the neutral surface, respectively \cite{watson2011intermediate}.) 

The membrane's total free energy $E$ consists of the sum of the Helfrich free energy due to bending the membrane \cite{helfrich1973elastic}, the energy due to lipid density deviations of the upper and lower membrane monolayers away from their ideal values, and the adhesion energy in Eq. (\ref{AdhesionEnergyReal}), such that
\begin{equation}
    \label{eqn:helfrich}
    E = \int \mathrm{d}\mathbf{r}\Big\{\frac{\kappa}{2}(2H)^2 + \frac{k}{2}[(\rho^+ - 2dH)^2 + (\rho^- + 2dH)^2]\Big\} + E_\text{adh},
\end{equation}
where $\kappa$ is the membrane bending modulus, $H$ is the mean curvature of the membrane, such that $2H = - \nabla^2 h(\mathbf{r})$, and $k$ is the monolayer area compressibility modulus. ($\rho^+ - 2dH)$ and ($\rho^- + 2dH)$ represent the deformations away from the ideal lipid density in the upper and lower membrane monolayers, respectively. $d$ is the distance between the bilayer midsurface and the neutral surfaces of the monolayers (for brevity, we refer to $d$ as the monolayer thickness). The sign conventions used for the mean curvature $H$ in Eq. (\ref{eqn:helfrich}) are as in \cite{seifert1997configurations}. Since we do not model asymmetries in lipid composition, we neglect spontaneous curvature in the curvature dependent contribution to the membrane energy \cite{hossein2020spontaneous}.

The membrane bending energy includes a term $\nabla^2 h(\rb)$, which can be more easily dealt with in Fourier space. We choose our Fourier conventions as representing a finite system size of dimensions $L \times L$, with the Fourier wave-vector $\mathbf{q} = \frac{2\pi}{L}(m,n)$ such that $-(N-1)/2 \leq (m,n) \leq (N-1)/2$ for $N \times N$ modes/lattice points, assuming $N$ is odd. The Fourier transform pair for the membrane's height is then
\begin{gather}
\label{eq:fourierTransformpair}
    h_\mathbf{q} = \int_{L^2} \mathrm{d}\mathbf{r} h(\mathbf{r}) e^{-i\mathbf{q}\cdot\mathbf{r}}, \hspace{15pt} h(\mathbf{r}) = \frac{1}{L^2}\sum_\mathbf{q}h_\mathbf{q}e^{i\mathbf{q}\cdot\mathbf{r}},
\end{gather}
and similarly for the transform pairs $\rho_\mathbf{q},  \rho(\mathbf{r})$ and $
\bar{\rho}_\mathbf{q},  \bar{\rho}(\mathbf{r})$. Additional comments on the treatment of variables in Fourier space are included in Appendix \ref{app:FourierConventions}.

The total free energy $E$ in Eq. (\ref{eqn:helfrich}) is computed by summing the contributions due to each Fourier mode as
\begin{gather}
    \label{EnergyFourierSpace}
    E = \frac{1}{L^2} \sum_\mathbf{q} \frac{1}{2}(h_\qv, \rho_\qv, \bar{\rho}_\qv) \mathbf{E} \begin{pmatrix} h_\qv \\ \rho_\qv \\ \bar{\rho}_\qv \end{pmatrix}^* + {E_\text{adh}},  \\ 
    \label{EnergyMatrixFS}
    \mathbf{E} = \begin{pmatrix} \tilde{\kappa}q^4 && -2kdq^2 && 0\\ -2kdq^2 && 2k && 0 \\ 0 && 0 && 2k \end{pmatrix}.
\end{gather}

In Eq. (\ref{EnergyMatrixFS}), $q$ is the magnitude of the Fourier wavevector $\qv$, and $h_\qv, \rho_\qv, \bar{\rho}_\qv$ are the membrane height, lipid density deviation and average density corresponding to a given mode. Here $\tilde{\kappa} = \kappa + 2d^2k$ is the renormalized bending modulus, which describes the response of the membrane over short times when lipids cannot laterally relax.  A typical value of $\kappa$ is about 20 $k_B T$, although this can be a few times larger for particularly stiff membranes. The strength of membrane substrate adhesion $\gamma$ can vary over orders of magnitude in different contexts. To best model the experiments in \cite{cannon2019amphipathic,shi2022kinetic}, we use a fairly strong $\gamma \sim$ $10^{13}$ J/m$^4$, unless otherwise stated. This is our estimate of adhesion strengths of supported lipid bilayers (SLBs) on glass substrates (see Discussion and Appendix \ref{app:justifyAdhesionStrength}). The other parameter values used in the model are included in Table \ref{tab:ModelParameters}. 

\subsection{\label{SimulationMethods}Dynamics and simulation of fluctuating membranes\protect}

A membrane that is deformed away from its equilibrium state will relax over time. The dynamics of this process are controlled by the viscosity of the fluid outside the membrane, the membrane's own viscosity, and the drag between the two leaflets \cite{seifert1993viscous,watson2011intermediate,camley2013diffusion}. To these relaxation dynamics, we add a stochastic term obeying a fluctuation-dissipation relationship, which ensures that the system will evolve into thermal equilibrium. The resulting stochastic dynamical equations for evolving $h_\qv$, $\rho_\qv$ and $\bar{\rho}_\qv$ in time are (Appendix \ref{app:simulationalgorithms})
\begin{equation}
    \label{EvolutionEquations}
    \frac{\partial}{\partial t}\left(\begin{matrix} h_\qv \\ \rho_\qv \\ \bar{\rho}_\qv\end{matrix}\right) = -L^2\left(\begin{matrix} \frac{1}{\Omega_{h}}{\partial E}/{\partial h_\qv^*} \\ \frac{1}{\Omega_\rho} {\partial E}/{\partial \rho_\qv^*} \\ \frac{1}{\Omega_{\bar{\rho}}} {\partial E}/{\partial \bar{\rho}_\qv^*} \end{matrix}\right) + \left(\begin{matrix} \xi_\qv \\ \zeta_\qv \\ \chi_\qv
    \end{matrix}\right),
\end{equation}
  where ${\Omega_h}^{-1} = {1}/{4 \eta q}$, ${\Omega_\rho}^{-1} = {q^2}/({4b + 4 \eta q + 2 \mu q^2})$, and ${\Omega_{\bar{\rho}}}^{-1} = {q^2}/({4 \eta q + 2 \mu q^2})$. These $\Omega^{-1}$ values play the role of hydrodynamic mobilities for a membrane with monolayer viscosity $\mu$ and intermonolayer friction $b$ embedded in a fluid of viscosity $\eta$, setting the time derivative of a field $\omega$ in terms of the force-like term $-L^2\partial E / \partial \omega_\qv^*$ (this is the Fourier transform of $-\delta E / \delta \omega(\mathbf{r})$ in our convention). For instance, $\Omega_h^{-1} = 1/4\eta q$ is the Oseen tensor relating the $z$-component force density on the membrane to the membrane height's velocity, as used in \cite{lin2004brownian}. Thermal fluctuations are accounted for with the stochastic terms $\xi_\qv, \zeta_\qv$ and $\chi_\qv$ (Appendix \ref{app:simulationalgorithms}). The deterministic components in the equations for $\frac{\partial h_\qv}{\partial t}$ and $\frac{\partial \rho_\qv}{\partial t}$ are consistent with the Seifert-Langer model, and we derive $\frac{\partial \bar{\rho}_\qv}{\partial t}$ from the hydrodynamic equations in \cite{seifert1993viscous} while neglecting inertial effects. While we present these equations of motion in terms of the hydrodynamics of the system for generality, our focus is on the equilibrium properties of the system, which are independent of the dynamic parameters $\eta$, $\mu$, $b$, etc. We will use this dynamical model to sample from the equilibrium thermal distributions of $h(\mathbf{r}), \rho(\mathbf{r})$ and $\bar{\rho}(\mathbf{r})$. A full understanding of the dynamics of this problem should also include the effect of the presence of the substrate near to the surface, which will alter the hydrodynamic response \cite{lin2006simulating,camley2013diffusion}. 

To simultaneously simulate the fluctuations of membrane height and lipid density, we numerically integrate Eq. (\ref{EvolutionEquations}). This essentially extends the Fourier-space Brownian Dynamics (FSBD) approach \cite{lin2004brownian}---so we will often refer to our simulations as FSBD simulations as well. The simulation algorithms, their derivations, and guidelines for choosing a manageable timestep for simulation convergence are included in Appendix \ref{app:simulationalgorithms}. To ensure that our approach creates the correct equilibrium distribution, we compared with an extension of the Fourier Monte Carlo method \cite{gouliaev1998simulations} (Appendix \ref{app:FMCsimulations}). 

\subsection{\label{ProteinSizeGaussianWeight}Modeling a protein as a perfect observer}

 To understand what will limit a protein's ability to sense membrane curvature even in ideal circumstances, we treat the protein as a perfect observer, making a precise measurement of the membrane curvature at the protein scale. The perfect observer assumption means that the protein does not affect the membrane in any way: it is a mere spectator. By a measurement ``at the protein scale,'' we describe an average over a region of the membrane of roughly the protein's size  $a$.   The local membrane curvature and local lipid density deviation sensed by the protein are then
\begin{gather}
    \label{eq:weightedcurvature}
    C_a = \int_{L^2} \mathrm{d}\mathbf{r} G(\mathbf{r}, a) \frac{-\nabla^2 h(\mathbf{r})}{2},\\
    \label{eq:weightedrho}
    \rho_a = \int_{L^2} \mathrm{d}\mathbf{r} G(\mathbf{r}, a) \rho(\mathbf{r}),
\end{gather}
where $G(\mathbf{r}, a)$ is a two-dimensional Gaussian weight centered at the protein location, such that 
\begin{equation}
    \label{eq:GaussianWeightDefinition}
    G(\mathbf{r}, a) = \frac{1}{2\pi a^2}\exp{\frac{-|\mathbf{r} - \mathbf{r}_\textrm{prot}|^2}{2a^2}}.
\end{equation}
We will always choose the protein to be located at the top of the spherical bead, $\mathbf{r}_\textrm{prot} = (L/2,L/2)$.

The integrals in Eqs. (\ref{eq:weightedcurvature})--(\ref{eq:weightedrho}) are evaluated by summing over discrete membrane lattice points. Membrane curvatures are computed from $h_\qv$ noting that the Fourier transform of the curvature is $\{-\frac{1}{2}\nabla^2 h(\mathbf{r})\}_\qv = \frac{1}{2}q^2 h_\qv$, then using the inverse Fast Fourier Transform to reconstruct the curvature field $-\frac{1}{2}\nabla^2 h(\mathbf{r})$.

\section{\label{sec:level3}Results\protect}
 
\subsection{Simulations of membrane-adhered beads}

\begin{figure}
     \centering
     \begin{subfigure}
         \centering
         \includegraphics[width=0.45\textwidth]{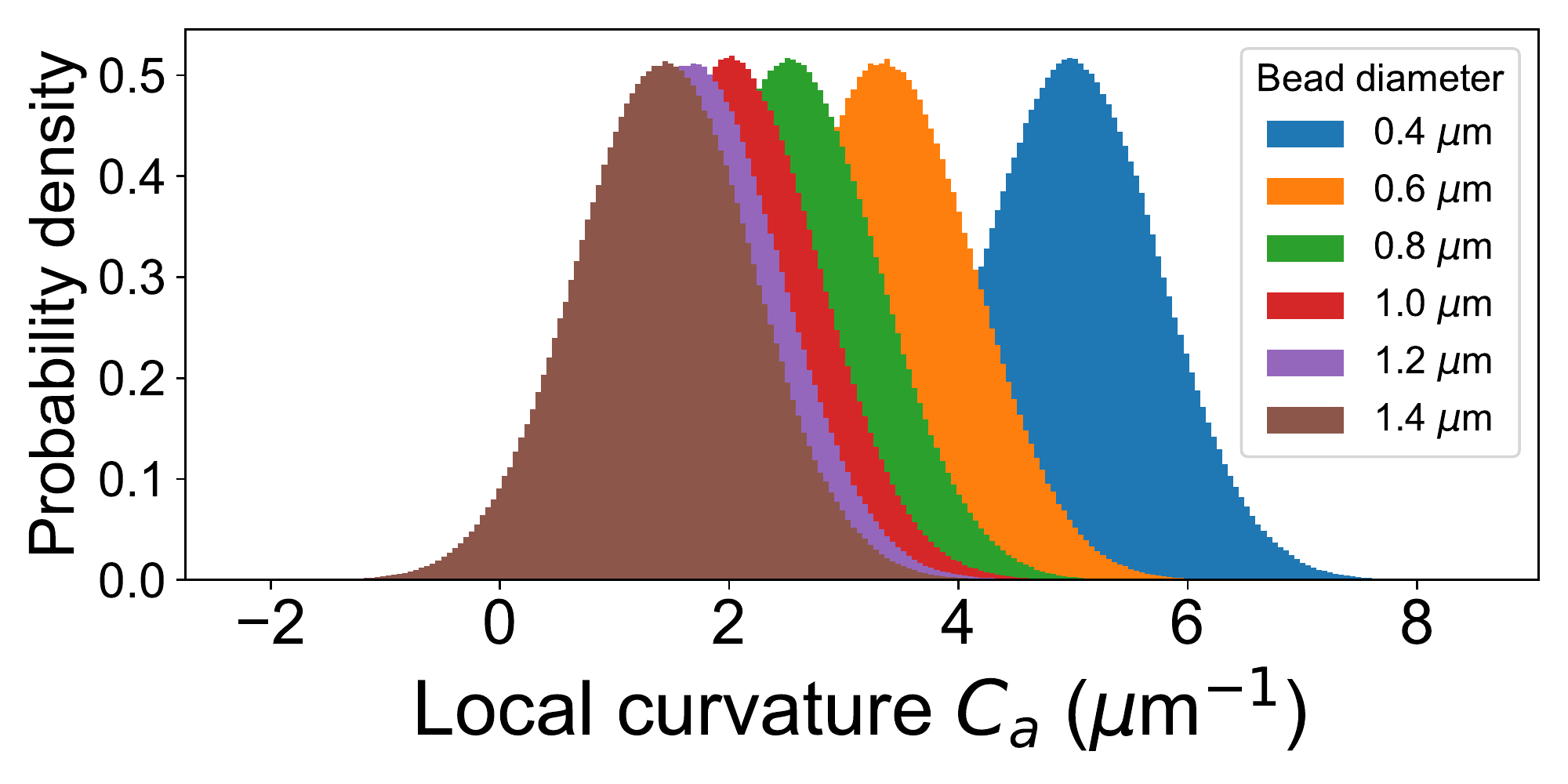}
     \end{subfigure}
     \begin{subfigure}
         \centering
         \includegraphics[width=0.45\textwidth]{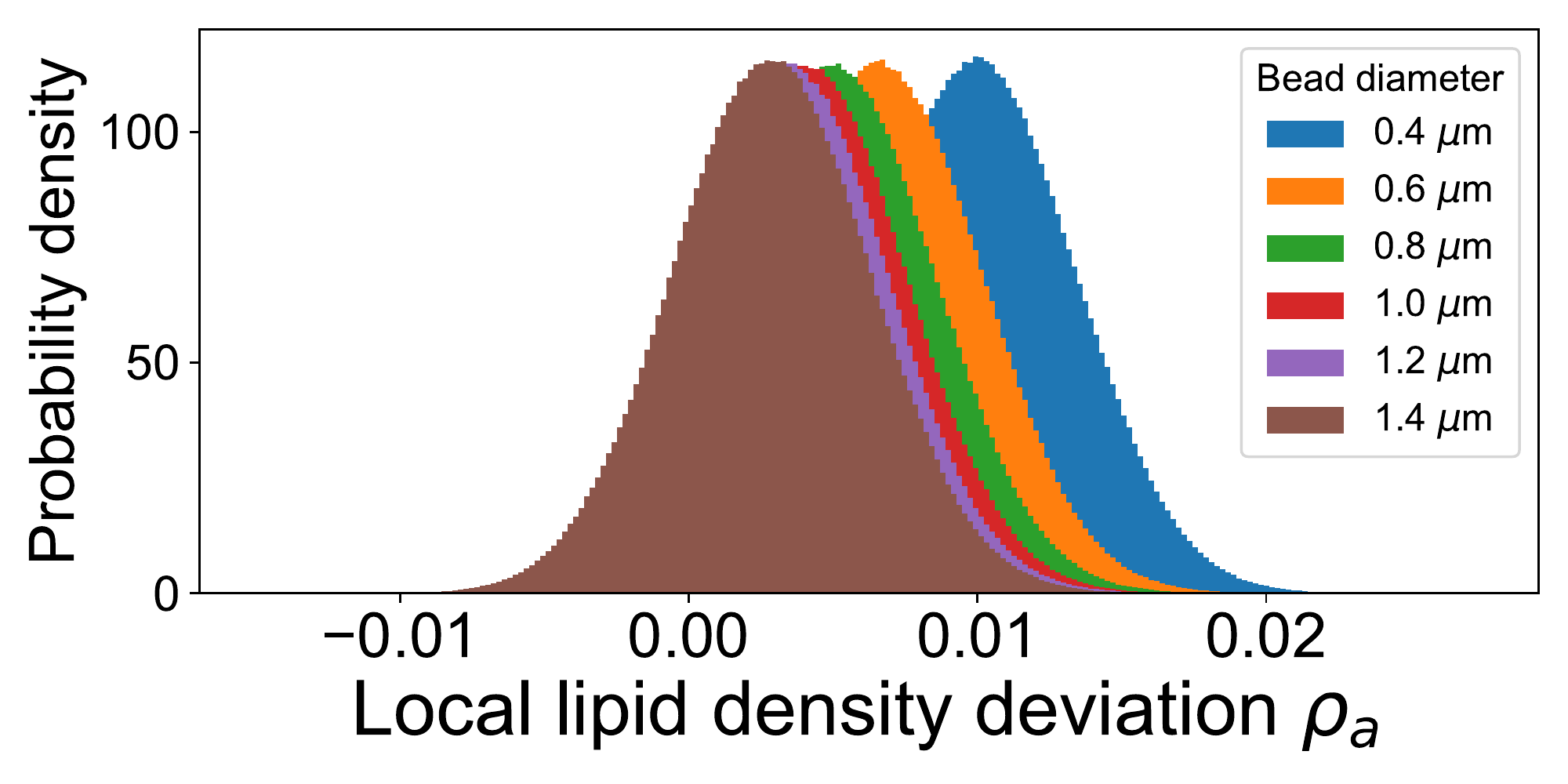}
     \end{subfigure}
     \caption{Histograms (normalized to probability densities) from FSBD simulations of local membrane curvatures and local lipid density deviations sensed by a protein of size $a = 16$ nm positioned at the top of membrane-adhered beads of different diameters. The strength of membrane-substrate adhesion is $\gamma = 10^{13}$ J/m$^4$. When the histogram distributions associated to different beads overlap considerably, there is more uncertainty about which bead curvature resulted in a particular local membrane curvature or density deviation sensed by the protein. The simulation used timesteps of $\Delta t = 3.2$ ns for total time $t_\text{sim} = 0.016$ seconds. Simulation parameters: Table \ref{tab:ModelParameters}.}
     \label{fig:CurvDensityHistogram}
\end{figure}

We simulate fluctuating membranes adhered to beads of varying sizes. In Fig. \ref{fig:CurvDensityHistogram}, we show the distribution of local curvature $C_a$ and local lipid density deviation $\rho_a$ that would be sensed over a protein scale of $a = 16$ nm such that 2$a$ roughly corresponds to the footprint of a yeast septin rod, which has an end-end length of $\sim$ 32 nm \cite{cannon2017unsolved}. We have chosen the membrane-substrate adhesion appropriate for a supported lipid bilayer, which is strongly adherent (Appendix \ref{app:justifyAdhesionStrength}). These distributions show the extent to which different beads could be distinguished by a protein: when there is significant overlap between two distributions, even a perfect detector would struggle to distinguish between beads of these radii. As the bead radius is increased, the average curvatures and density deviations sensed by the protein decrease in magnitude---as we would expect, because the bead is made locally flatter. The distributions for larger beads overlap more substantially, so a protein that measures a particular curvature or density value in this regime is subjected to more ambiguity as to which bead the measurement corresponds to.

\subsection{\label{SNRanalysis}Quantifying sensing efficacy with signal-to-noise ratio}

\begin{figure*}[htb]
     \centering
     \begin{subfigure}
         \centering \includegraphics[width=0.47\textwidth]{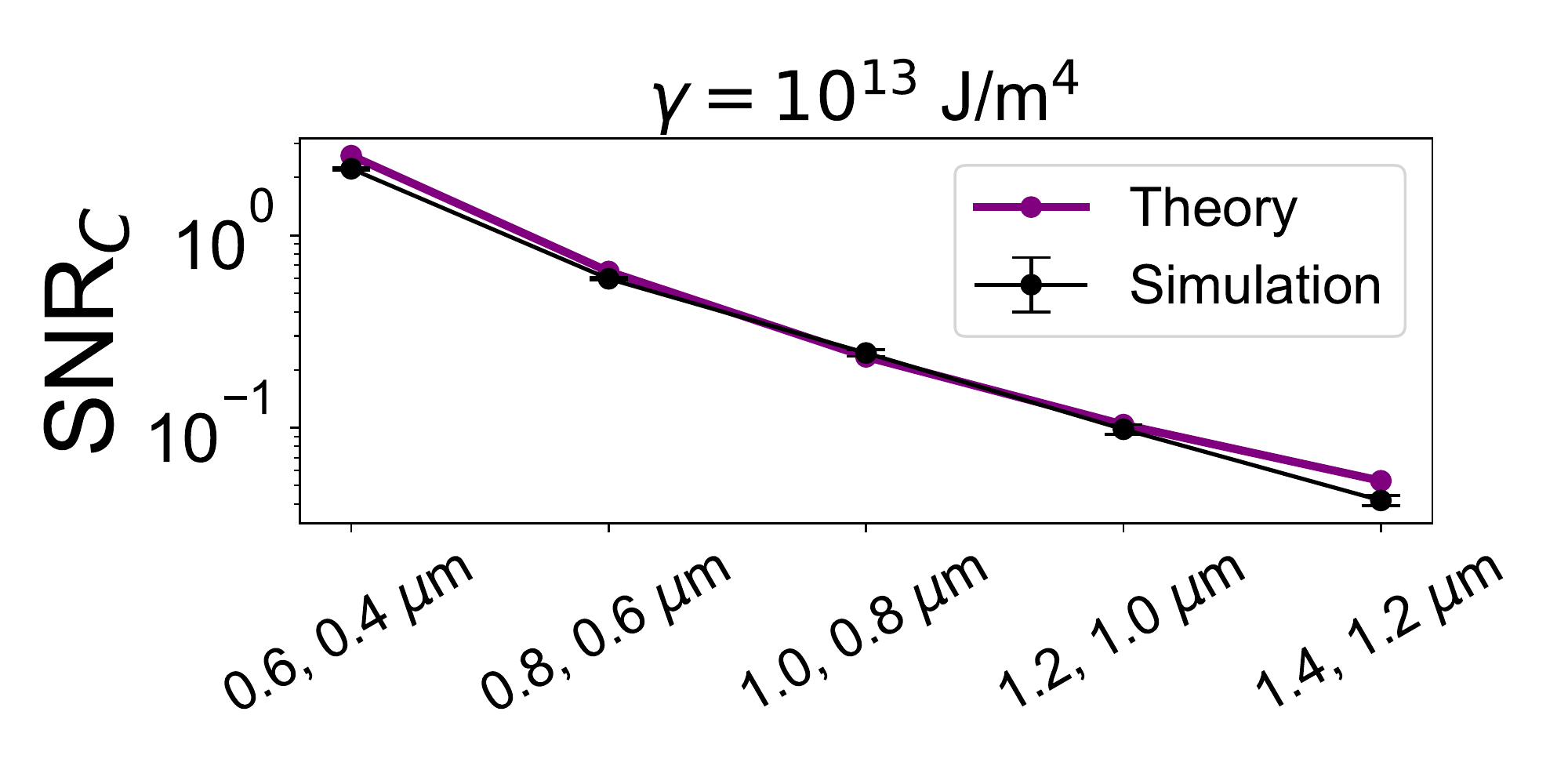}
     \end{subfigure}
     \begin{subfigure}
         \centering \includegraphics[width=0.47\textwidth]{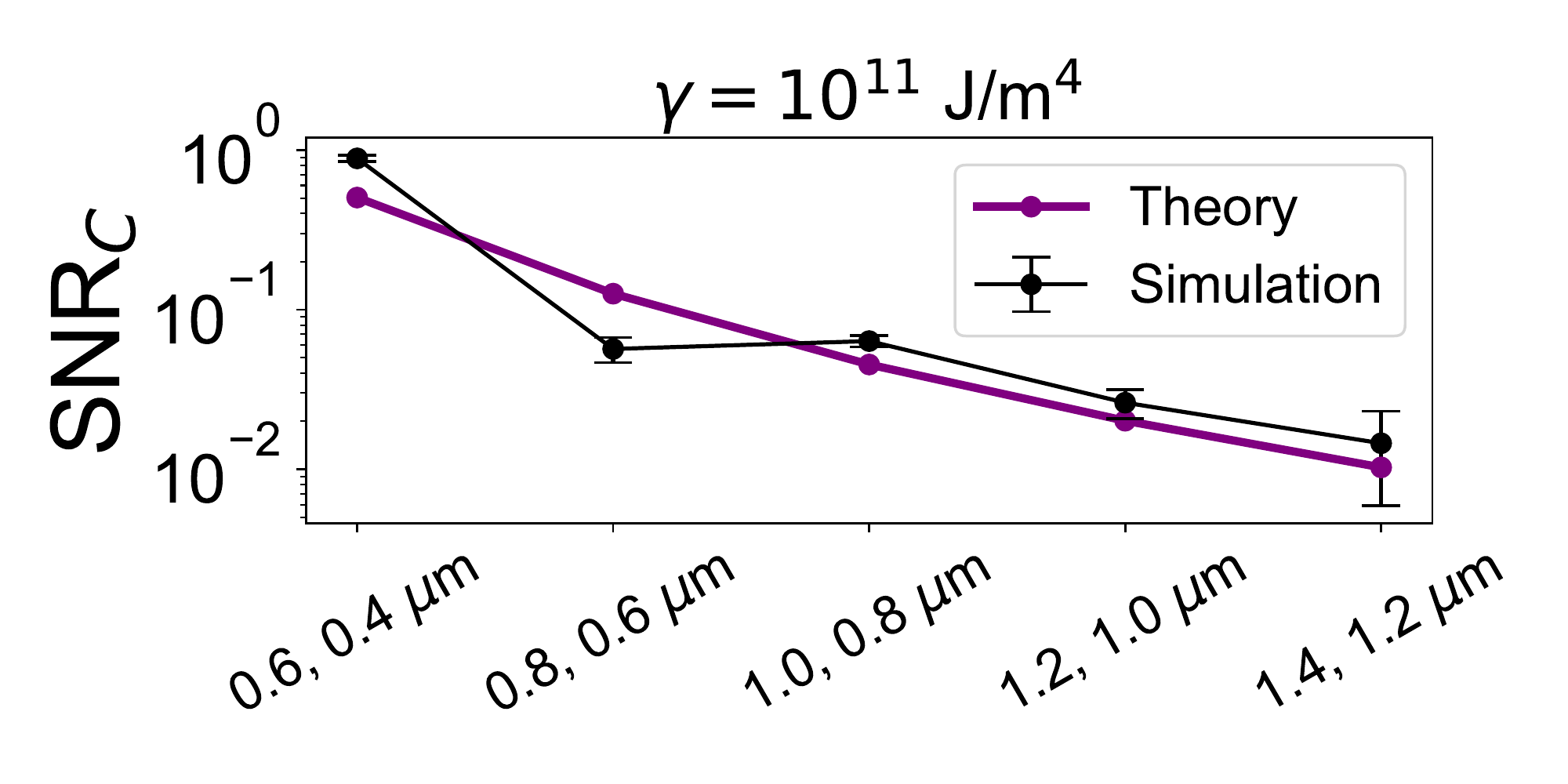}
     \end{subfigure}
     \begin{subfigure}
         \centering \includegraphics[width=0.47\textwidth]{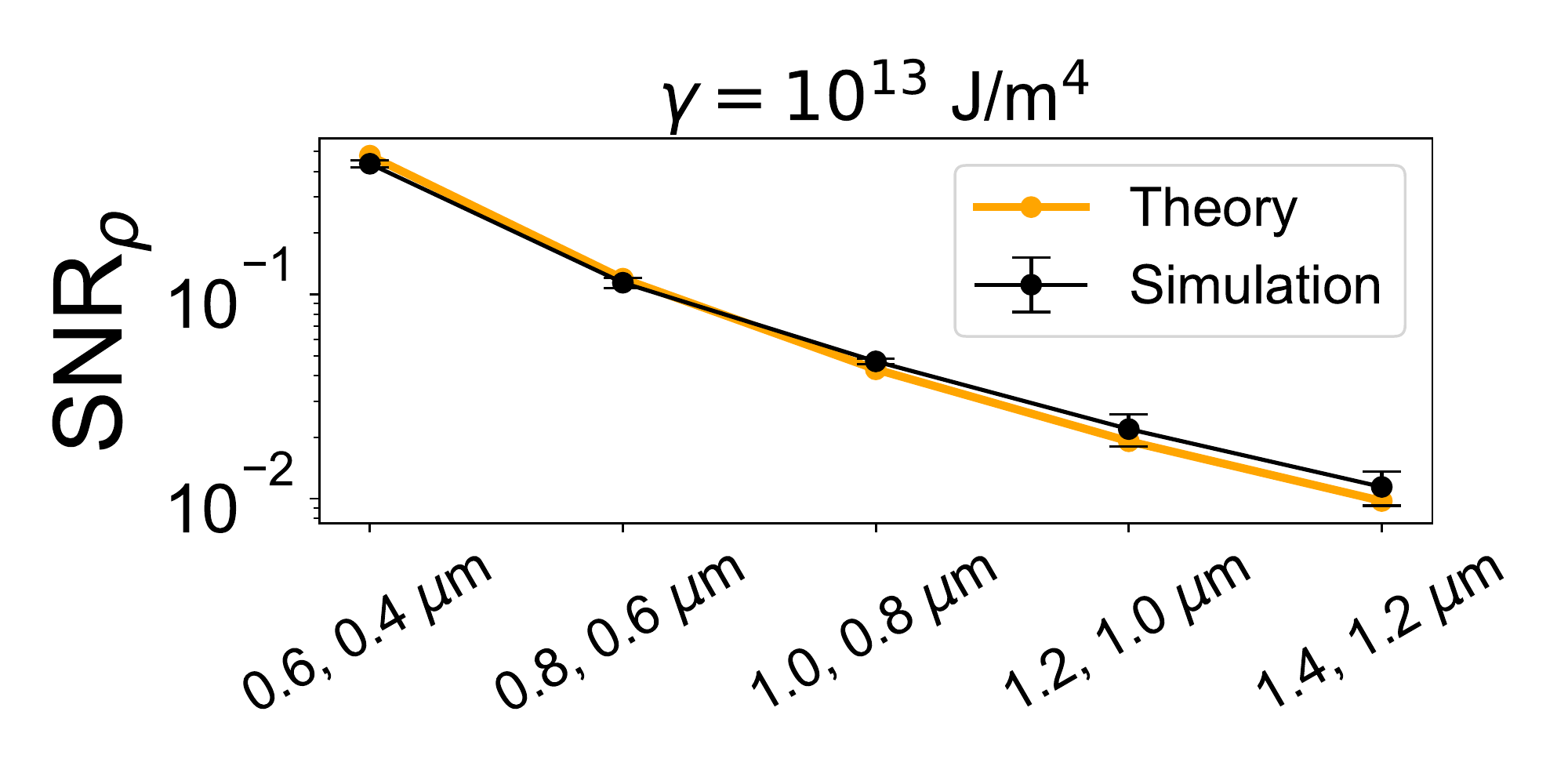}
     \end{subfigure}
     \begin{subfigure}
         \centering \includegraphics[width=0.47\textwidth]{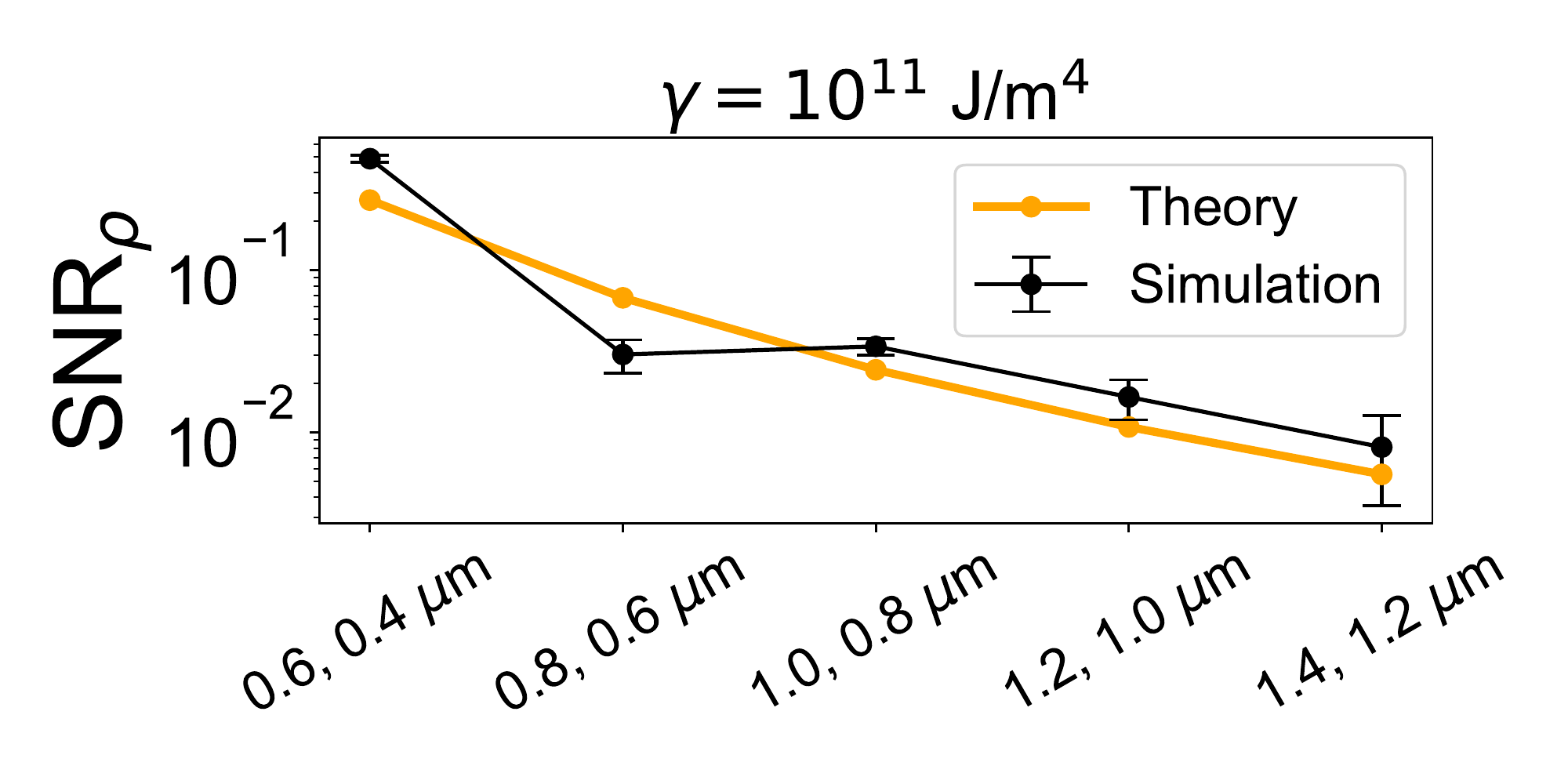}
     \end{subfigure}
     \caption{Sensing SNR for pairs of membrane-adhered beads with diameters ranging from 0.4 $\mu$m to 1.4 $\mu$m, with each pair having diameters separated by 0.2  $\mu$m. (Top-left) SNR$_C$ with $\gamma = 10^{13}$ J/m$^4$. (Top-right) SNR$_C$ with $\gamma = 10^{11}$ J/m$^4$. (Bottom-left) SNR$_\rho$ with $\gamma = 10^{13}$ J/m$^4$. (Bottom-right) SNR$_\rho$ with $\gamma = 10^{11}$ J/m$^4$. Other parameters: Table \ref{tab:ModelParameters}. Errors were computed with the block averaging method \cite{grossfield2018best}. For each pair of beads, the simulated curvature and $\rho$ trajectories were separated into $N_\textrm{block} = 4$ blocks and the mean and variance computed for each block to obtain SNR$_\textrm{block}$, which was averaged across the blocks to obtain the SNR plotted. The error bars indicate standard errors, computed by dividing the standard deviation of SNR$_\textrm{block}$ across the blocks by $\sqrt{N_\textrm{block}}$.}
    \label{fig:SNR_varyingbeads}
\end{figure*}

If a particular local curvature or density deviation is sampled from the distributions in Fig. \ref{fig:CurvDensityHistogram}, can the bead size corresponding to the sampled measurement be reliably determined? This can be challenging due to overlapping distributions, since each bead size is associated to a multiplicity of instantaneous $C_a$ and $\rho_a$ values.
We summarize the way that thermal fluctuations of the membrane could confound even a perfect detector of curvature or lipid density in distinguishing between two membrane-adhered beads of different sizes with a signal-to-noise ratio (SNR), such that
\begin{gather}
    \label{SNR_definition}
    \text{SNR} = \frac{(\mu_A - \mu_B)^2}{\sigma_A^2 + \sigma_B^2},
\end{gather}
where $\mu_A$ and $\mu_B$ are the average membrane curvature or lipid density deviation at steady-state for two beads $A$ and $B$, and $\sigma_A^2$ and $\sigma_B^2$ are the corresponding variances of the membrane curvature or density deviation sensed by the protein. The motivation for this definition is to measure the distance between the means of two histograms in Fig. \ref{fig:CurvDensityHistogram} in terms of their variance. If we define a variable $X$ which is the difference between the measured variable on bead $A$ and the measured variable on bead $B$, the SNR is $\langle X \rangle^2 / \left[ \langle X^2 \rangle - \langle X \rangle^2 \right]$---which gives Eq. (\ref{SNR_definition}) because the variance of two independent variables adds. The greater this SNR value is, the better a protein can distinguish between the two beads $A$ and $B$ in the presence of thermal fluctuations of the membrane.
We will show in Section \ref{SNR_Estimate_Experiment} that---with some additional assumptions---this SNR controls the largest possible ratio of association rates of proteins to beads of radius $R_A$ and $R_B$, and use this to estimate experimental SNR.

We use our FSBD simulation to compute the SNR for pairs of beads in Fig. \ref{fig:SNR_varyingbeads}, keeping the difference $R_A - R_B$ to be 200 nm. The smaller the beads, the better a protein can distinguish between two similarly sized beads, as indicated by the magnitude of the SNR. For beads on the micron-scale, the SNR is much smaller than $1$ when the beads are similarly sized (such as 1.2 $\mu$m and 1.4 $\mu$m diameters). This is true for both the SNR of curvature sensing, SNR$_C$, and the SNR of density sensing, SNR$_\rho$. The  decrease in SNR is largely driven by the decreasing signal: large beads have curvatures $1/R$ that are both increasingly close to zero curvature, so the term $(\mu_A - \mu_B)^2$ will shrink. As we will see later (Fig. \ref{fig:SNR_BigBeads_ProtMono}), for beads where the radii differ significantly, e.g. $1$ micron vs. $3$ micron diameters, the SNR can be appreciable.  We also see that changing the membrane-substrate adhesion energy from a weakly-adherent membrane value ($\gamma = 10^{11}$ J/m$^4$) to one appropriate to a SLB ($\gamma = 10^{13}$ J/m$^4$) increases the SNR.

To gain an understanding of how the SNR depends on the protein size, the mechanical properties of the membrane, the geometry of the bead, and the membrane-bead adhesion, we develop a theoretical model for the SNR in the next section, Sec. \ref{sec:analytical_SNR}. We plot this theoretical result against the simulation result and see good agreement, especially at strong membrane-substrate adhesion and large bead sizes (Fig. \ref{fig:SNR_varyingbeads}). This deviation at small bead sizes ($\sim 200-300$ nm radii) and weak adhesion ($\gamma \sim 10^{11}$ J/m$^4$) is expected. As we discuss in the next section, our theoretical model assumes that on average, the fluctuating membrane perfectly follows the shape of a spherical bead of any radius $R$. However, in the simulated membrane-bead systems using hemispherical beads with small $R$, weak membrane adhesion cannot overcome the substantial bending forces required at the sharp intersection between the periphery of a small hemisphere and the horizontal plane from which it is projected \cite{deserno2003wrapping}. A stronger adhesion strength such as $\gamma \sim 10^{13}$ J/m$^4$ rectifies this, and allows the membrane to follow the shape of the bead nearly perfectly. In Appendix \ref{crosssection_smallbead_adhesion}, we show a comparison of the cross-sectional profiles of a simulated membrane adhered to a small bead for both weak and strong adhesion.

\subsection{Analytical calculation of the SNR}
\label{sec:analytical_SNR}

To find an analytical form for the SNR written in Eq. (\ref{SNR_definition}), we need the average values of curvature and density deviation on a bead as well as their standard deviations. The average mean curvature and density deviation are reflected by the average shape of the membrane wrapping the bead, and we can find analytical values for them. 

If the membrane is strongly adherent to the bead, on average its shape will just be the bead's shape, $\langle h(\mathbf{r})\rangle = h_\textrm{bead}(\mathbf{r})$. The averaged mean curvature for a membrane adhered to a bead is then $-\frac{1}{2}\nabla^2 h_\textrm{bead}(\mathbf{r})$. At the top of the bead ($\mathbf{r} = \mathbf{r}_\textrm{prot}$), the curvature is then
\begin{gather}
\label{C-ss}
{\langle C_a \rangle \approx 1/R},
\end{gather}
where $R$ is the bead's radius. Our assumption that the membrane follows the shape of the bead can be checked with simulation: we see that it is reasonable at sufficiently large bead sizes and strong membrane-substrate adhesion (refer to Fig. \ref{Membraneprofile_SmallBead} in Appendix).

Given that the membrane is deformed to follow the bead, we can find the value of $\rho(\mathbf{r})$ that would minimize the energy of the membrane, solving for $\rho_\qv$ such that $\partial E/\partial \rho_\qv^* = 0$ (using Eq. (\ref{EnergyFourierSpace})). This would be the steady-state $\rho_\qv$ holding the membrane shape fixed. We find that this value is
\begin{equation}
    \label{rho-ss-q}
    \rho_{{\qv}}^{\text{ss}} = d {q}^2 h_{\qv}.
\end{equation}
Inverting the Fourier transform, we see that the density at the protein's location is
\begin{equation}
    \label{rho-ss-r}
    \rho^\textrm{ss}(\mathbf{r}_\textrm{prot}) = -d\nabla^2 h(\mathbf{r}_\textrm{prot}) = \frac{2d}{R}.
\end{equation}
To approximate the standard deviation of observed curvature and density histograms, we start by noting that in Fig. \ref{fig:CurvDensityHistogram}, the width of the histograms is broadly consistent across many different bead diameters. This suggests that $\sigma_A$ and $\sigma_B$ do not strongly depend on bead size. In fact, for a large enough bead, the variances of the observed curvature are essentially those for a membrane adherent on a flat substrate with the same adhesion strength. This approximation is warranted for the same reason that sensing micron-scale curvature is difficult: the protein scale is much smaller than the size of the bead, and locally the bead surface is nearly flat. We then expect that when the protein is small relative to the bead, the change in the variance due to the curvature will be $\sigma_A^2 \approx \sigma_\text{flat}^2 + O(a/R)$, and the terms of $O(a/R)$ can be neglected in Eq. (\ref{SNR_definition}). If this is the case, then the standard deviations in the definition of SNR in Eq. (\ref{SNR_definition}) will be the same for both beads, and approximately equal to the standard deviation for a flat membrane. Therefore, we propose as an estimate of the SNR as
\begin{align}
    \label{SNR_curvature}
    \text{SNR}_{C} &= \frac{(\frac{1}{R_A} - \frac{1}{R_B})^2}{2 \langle C_a^2 \rangle^\textrm{flat}},\\
    \label{SNR_rho}
    \text{SNR}_{\rho} &= \frac{(\frac{2d}{R_A} - \frac{2d}{R_B})^2}{2\langle \rho_a ^2 \rangle^\textrm{flat}},     
\end{align}
where $R_A$ and $R_B$ are bead radii for beads $A$ and $B$, and $\langle C_a^2 \rangle^\textrm{flat}$ and $\langle \rho_a^2 \rangle^\textrm{flat}$ are the membrane curvature and density variances sensed by a protein of size $a$ when the membrane is bound to a flat substrate. These variances can be worked out analytically in some cases, and by simple numerical quadrature in others, as we will see in the next section. When the membrane is bound to a flat substrate, $h_\textrm{bead}(\mathbf{r}) = 0$ and the height, density deviations, and curvature have zero mean, so the variance in $h_\qv$ is $\textrm{var}(h_\qv) = \langle |h_\qv|^2 \rangle - \langle h_\qv \rangle^2 = \langle |h_\qv|^2 \rangle$, and similarly for the density deviations and curvature. Therefore, we will often call $\langle C_a^2 \rangle^\textrm{flat}$ the ``curvature variance.''

\subsubsection{\label{FlatMembraneResults}Variances of membrane height and density when bound to a flat substrate}

For a planar membrane adhered to a flat substrate $h_\textrm{bead} = 0$, the adhesion energy of Eq. (\ref{AdhesionEnergyReal}) is just a simple harmonic penalty,  $E_\text{adh} = \frac{\gamma}{2} \int \mathrm{d}\mathbf{r} h(\mathbf{r})^2$. Plugging in the Fourier representation, this becomes $E_\text{adh} = \frac{\gamma}{2L^2}\sum_\qv|h_\qv|^2$.
Then, the complete energy of Eq. (\ref{EnergyFourierSpace}) is simply represented as
\begin{gather}
    \label{EnergyFourierSpaceFlat}
    E^\textrm{flat} = \frac{1}{L^2} \frac{1}{2} \sum_\mathbf{q} (h_\qv, \rho_\qv, \bar{\rho}_\qv) \mathcal{E} \begin{pmatrix} h_\qv \\ \rho_\qv \\ \bar{\rho}_\qv \end{pmatrix}^*,    \\ 
    \mathcal{E} = \begin{pmatrix} \tilde{\kappa}q^4 + \gamma && -2kdq^2 && 0\\ -2kdq^2 && 2k && 0 \\ 0 && 0 && 2k \end{pmatrix}.    \label{EnergyMatrixPlanarFS}
\end{gather}
Using Wick's theorem \cite{kardar2007statistical}, the variances of the Fourier modes of height and density are then
\begin{align}
\label{heightFlatFourier}
    \langle |h_\qv| ^2 \rangle &= L^2k_B T \mathcal{E}_{hh}^{-1} = \frac{L^2 k_B T}{\kappa q^4 + \gamma},\\
    \label{rhoFlatFourier}
    \langle |\rho_\qv| ^2 \rangle &= L^2k_B T \mathcal{E}_{\rho \rho}^{-1} = L^2 k_B T\frac{(2d^2k + \kappa )q^4 + \gamma}{2k(\kappa q^4 + \gamma)},\\
    \label{rhobarFlatFourier}
    \langle |\bar{\rho}_\qv| ^2 \rangle &= L^2k_B T \mathcal{E}_{\bar{\rho} \bar{\rho}}^{-1} = \frac{L^2 k_B T}{2k},
\end{align}
where the subscripts $hh, \rho\rho,$ and  $\bar{\rho}\bar{\rho}$ denote the corresponding diagonal elements of the matrix inverse $\mathcal{E}^{-1}$.

\subsubsection{\label{VariancesSensedProtein}Variances of curvature and densities sensed by a protein}
Assuming that the protein is a perfect sensor of the membrane curvature and density, the curvature sensed by the protein can be determined by a weighted integral over the membrane in Eq. (\ref{eq:weightedcurvature}), and similarly for the density in Eq. (\ref{eq:weightedrho}). Since these integrals are linear in the height and density fields, it is relatively simple to compute the variance of the protein-sensed curvature $C_a$ and protein-sensed densities $\rho_a$, $\bar{\rho}_a$ by substituting the Fourier transform conventions for $h(\mathbf{r})$ and $\rho(\mathbf{r})$ into Eqs. (\ref{eq:weightedcurvature}) and (\ref{eq:weightedrho}) to obtain
\begin{gather}
    \label{eq:WeightedCurvatureVariance}
    \langle C_a^2 \rangle^\textrm{flat} = \frac{1}{4L^4}\sum_\mathbf{q} q^4 \langle |h_\qv| ^2 \rangle |G(\qv)|^2,\\
    \label{eq:WeightedRhoVariance}
    \langle \rho_a^2 \rangle^\textrm{flat} = \frac{1}{L^4}\sum_\mathbf{q} \langle |\rho_\qv| ^2 \rangle |G(\qv)|^2,
\end{gather}
where $G(\qv)$ is the Fourier transform of the Gaussian weight $G(\mathbf{r},a)$, and $|G(\qv)|^2 = |G(q)|^2 = \exp{(-q^2 a^2)}$. See Appendix \ref{app:FourierSpaceDerivation_CurvatureRhoVariance} for an example derivation.

In the continuum limit, these Fourier sums can be rewritten as integrals, noting  $\sum_\mathbf{q} = \left(\frac{L}{2 \pi}\right)^2\int \mathrm{d}\mathbf{q}$ in 2D \cite{safran2018statistical}.
Since the integrands depend only on the magnitude of $\mathbf{q}$, we can further simplify $\int \mathrm{d}\mathbf{q} = 2 \pi\int_0^\infty q \mathrm{d}q$, finding
\begin{gather}
    \label{eqn:ContLimCurvatureVar}
    \langle C_a^2 \rangle^\textrm{flat} = \frac{1}{8\pi}\int_0^\infty q^5 \frac{k_B T}{\kappa q^4 + \gamma}|G(q)|^2 \mathrm{d}q,
    \\
    \label{eqn:ContLimRhoVar}
    \langle \rho_a^2 \rangle^\textrm{flat} = \frac{1}{2\pi}\int_0^\infty q \frac{k_B T(\Tilde{\kappa} q^4 + \gamma)}{2k(\kappa q^4 + \gamma)}|G(q)|^2 \mathrm{d} q,
\end{gather}
where $\Tilde{\kappa} = \kappa + 2d^2k$. 

We reformulate Eq. (\ref{eqn:ContLimCurvatureVar}) and Eq. (\ref{eqn:ContLimRhoVar}) by substituting a dimensionless parameter $u = qa$ and simplify as
\begin{gather}
    \label{ContLimCurvVar_DimLess}
    \langle C_a^2 \rangle^\textrm{flat} = \frac{k_B T}{8 \pi \kappa a^2}\int_0^\infty  u \frac{u^4}{u^4 + \frac{\gamma a^4}{\kappa}}\exp(-u^2)\mathrm{d}u,\\
    \label{ContLimRhoVar_DimLess}
    \langle \rho_a ^2 \rangle^\textrm{flat} = \frac{k_B T}{4 \pi k a^2}\int_0^\infty u \frac{u^4 + \frac{2d^2k u^4}{\kappa} + \frac{\gamma a^4}{\kappa}}{u^4 + \frac{\gamma a^4}{\kappa}}\exp(-u^2)\mathrm{d}u.
\end{gather}

\begin{figure}
    \centering \includegraphics[width=0.48\textwidth]{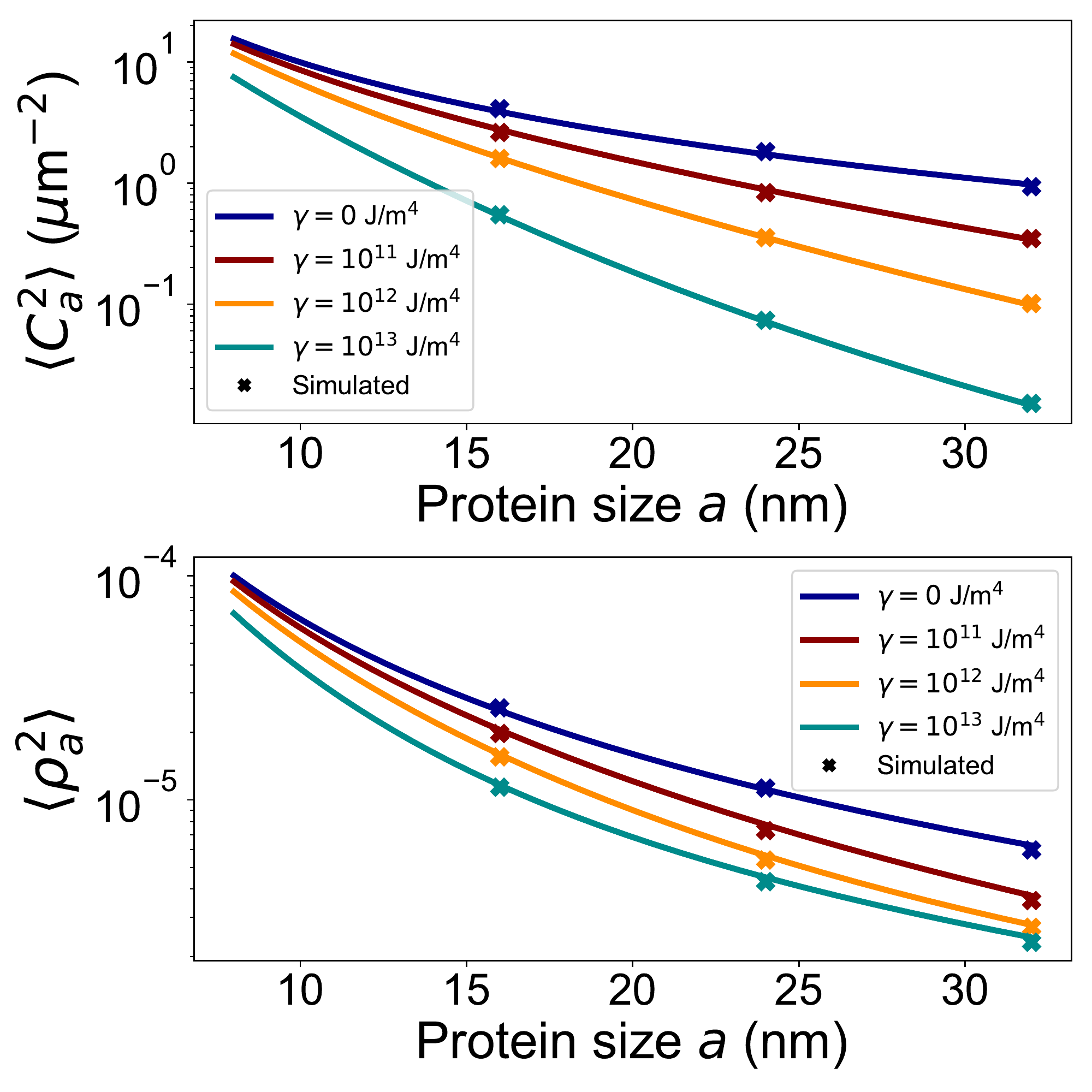}
    \caption{Theoretically predicted variances in local curvature and local lipid density deviation as sensed by proteins of varying sizes (lines) for a membrane adhered to a flat substrate with different adhesion strengths $\gamma$, as compared to variances obtained from FSBD simulations of a membrane on a flat substrate (symbols). Simulation points are $a =$ 16, 24 and 32 nm. Simulation parameters: $L = 900$ nm, $N = 49$. Other parameters: Table \ref{tab:ModelParameters}.}
    \label{fig:CurvDensityVar_ProtSize}
\end{figure}

The curvature and density variances of Eq. (\ref{ContLimCurvVar_DimLess}) and Eq. (\ref{ContLimRhoVar_DimLess}) are numerically integrated by quadrature using the \texttt{scipy.integrate.quad} \cite{2020SciPy-NMeth} method in Python. We plot the variances in Fig. \ref{fig:CurvDensityVar_ProtSize} as a function of protein size for varying membrane adhesion strengths.  We also compare these results to FSBD simulations of a fluctuating membrane bound to a flat substrate, finding excellent agreement (Fig. \ref{fig:CurvDensityVar_ProtSize}).
These variances will control the SNR through Eqs. (\ref{SNR_curvature})--(\ref{SNR_rho}), and thus the potential accuracy of sensing. How do they depend on the protein size and membrane-substrate adhesion? In general, larger protein sizes and stronger adhesion strengths allow the protein to minimize the variance in curvature and $\rho$ sensed locally. However, we note that increasing the membrane-substrate adhesion $\gamma$ seems to continually decrease the curvature variance $\langle C_a^2 \rangle^\textrm{flat}$ over orders of magnitude, while the density variance $\langle \rho_a^2 \rangle^\textrm{flat}$ seems to saturate.  We can understand these behaviors by studying some asymptotic limits of the variances of Eqs. (\ref{ContLimCurvVar_DimLess})--(\ref{ContLimRhoVar_DimLess}), which can be analytically evaluated. 

In the absence of membrane-substrate adhesion, $\gamma = 0$, and Eq. (\ref{ContLimCurvVar_DimLess}) and Eq. (\ref{ContLimRhoVar_DimLess}) evaluate to
\begin{gather}
    \label{eq:zeroadhlimCurv}
    \langle C_a^2 \rangle_{\gamma = 0}^\textrm{flat} = \frac{k_B T}{16 \pi a^2 \kappa},\\
    \label{eq:zeroadhlimRho}
    \langle \rho_a^2 \rangle_{\gamma = 0}^\textrm{flat} = \frac{k_B T \Tilde{\kappa}}{8 \pi a^2 k \kappa}.
\end{gather}
At the zero adhesion limit (freely fluctuating membrane), the curvature and density variances sensed are both inversely proportional to the protein size as $1/a^2$. However, $\langle \rho_a ^2 \rangle$ also depends on the renormalized bending modulus $\Tilde{\kappa} = \kappa + 2d^2k$, implying greater variance in the lipid density deviation sensed by the protein when the membrane is thicker. Eqs. (\ref{eq:zeroadhlimCurv})--(\ref{eq:zeroadhlimRho}) are also applicable at relatively weak adhesion; for the standard parameters values chosen, these asymptotic formulas are reasonably accurate up to adhesion strengths of $\gamma \sim 10^{9}$-$10^{10}$ J/m$^4$ (Fig. \ref{fig:CurvRhoSNRvsAdhesion}).

We can simplify the variances in Eqs. (\ref{ContLimCurvVar_DimLess})--(\ref{ContLimRhoVar_DimLess}) in the limit of high adhesion. The integrand in these equations is suppressed exponentially when $u \gg 1$, so at sufficiently high adhesion strengths, $\frac{\gamma a^4}{\kappa} \gg 1 + \frac{2d^2k}{\kappa}$, we can neglect the terms not proportional to $\gamma$ in the numerator and denominators of Eqs. (\ref{ContLimCurvVar_DimLess})--(\ref{ContLimRhoVar_DimLess}). In this limit, we find
\begin{gather}
    \label{CurvHighAdhLim}
    \langle C_a^2 \rangle_{\text{{high $\gamma$}}}^\textrm{flat} = \frac{k_B T}{8\pi a^6 \gamma},\\
    \label{RhoHighAdhLim}
    \langle \rho_a^2 \rangle_\text{{high $\gamma$}}^\textrm{flat} = \frac{k_B T}{8 \pi a^2 k}.
\end{gather}
We see in these high-adhesion limits both of the key features we observed in the numerical calculations of Fig. \ref{fig:CurvDensityVar_ProtSize}: curvature variance depends strongly on both protein size and adhesion, while the density variance does not. In the absence of adhesion, $\langle C_a ^2 \rangle \sim 1/a^2$, while in the high-adhesion regime,  $\langle C_a ^2 \rangle \sim 1/a^6$. As $\gamma$ is increased,  $\langle C_a^2 \rangle$  continues to diminish, while $\langle \rho_a^2 \rangle$ saturates asymptotically to a fixed value. 
This might be expected, as even if the membrane is effectively frozen into a flat configuration ($\gamma \to \infty$), the lipids may still diffuse in the flat membrane, leading to lipid density fluctuations. Interestingly, $\langle C_a ^2 \rangle$ and $\langle \rho_a ^2 \rangle$ lose their dependence on $\kappa$ in the high-adhesion regime. In this case, the cost for deviating from a flat height is dominated by the adhesion energy---but because bending is so strongly suppressed, the primary contribution to fluctuations in lipid density is the area compressibility modulus $k$.

\subsection{\label{ComparingSNR}Determining curvature SNR and \texorpdfstring{$\rho$}{rho} SNR for micron-sized beads}

\begin{figure}
     \centering
     \begin{subfigure}
         \centering
         \includegraphics[width=0.4\textwidth]{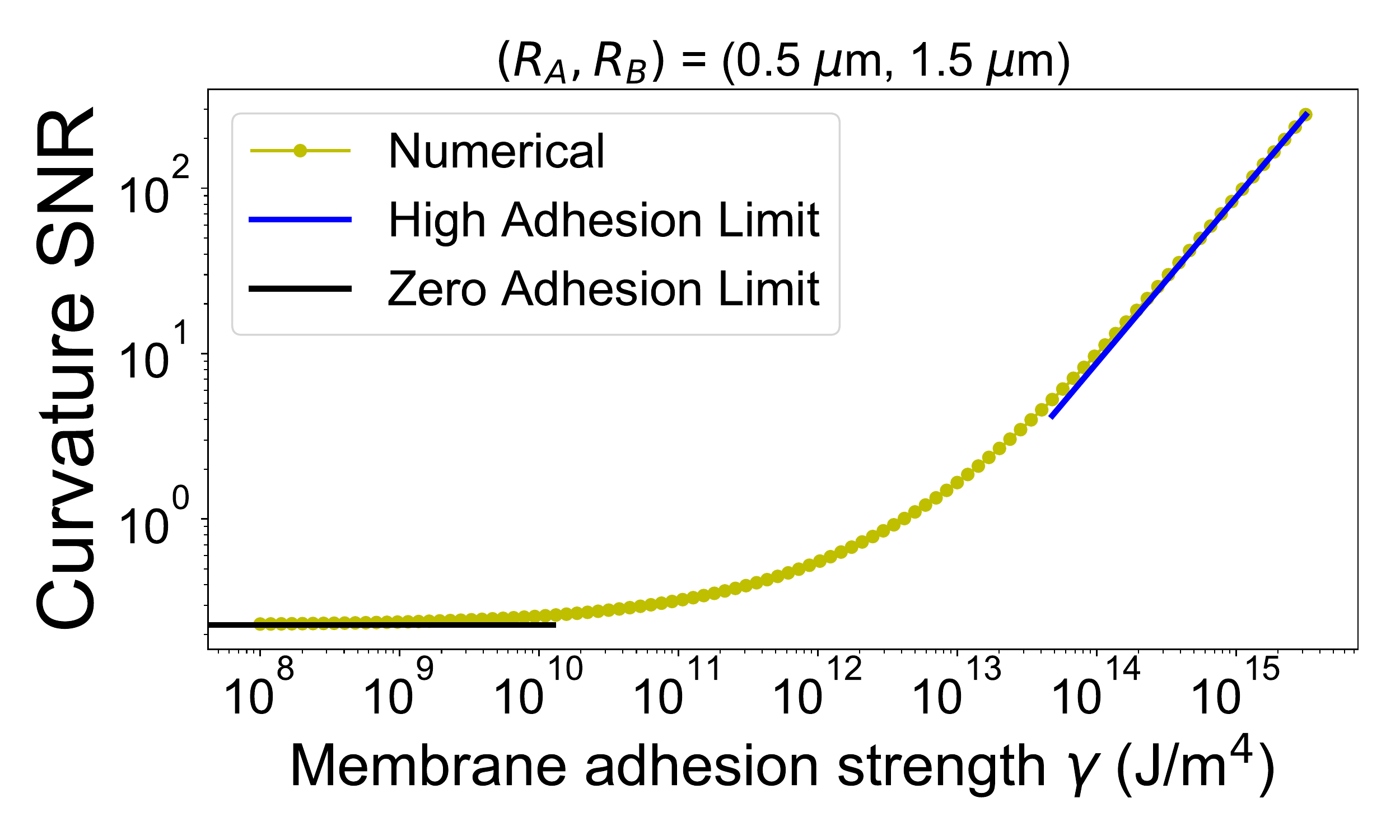}
     \end{subfigure}
     \begin{subfigure}
         \centering
         \includegraphics[width=0.4\textwidth]{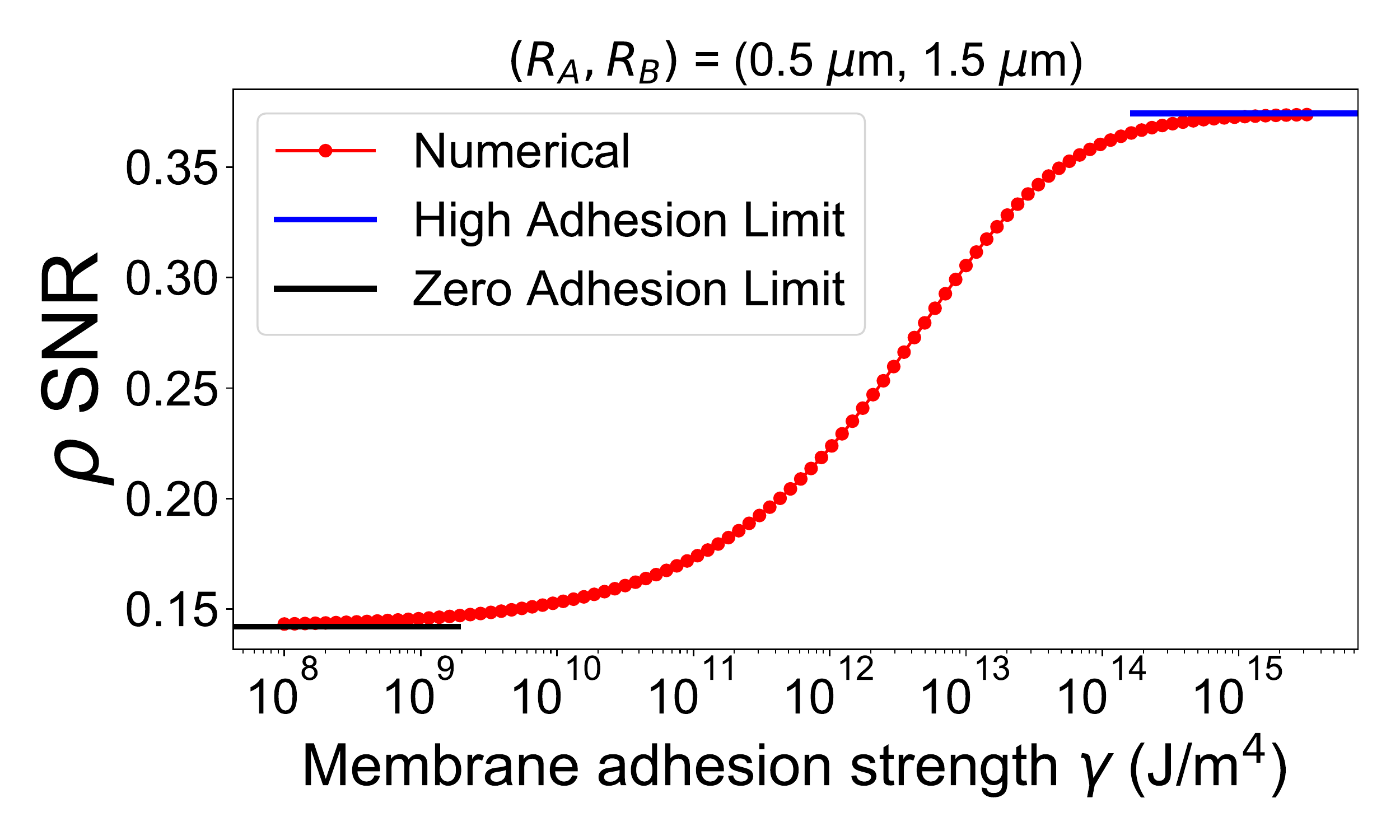}
     \end{subfigure}
     \caption{Predicted SNR$_C$ and SNR$_\rho$ for beads of radii $(R_A, R_B) = (0.5$ $\mu$m, 1.5 $\mu$m) as a function of adhesion strength $\gamma$, compared to their zero and high adhesion limits. The curvature and $\rho$ variances for the numerical  SNR values are integrated by quadrature using Eqs. (\ref{ContLimCurvVar_DimLess})--(\ref{ContLimRhoVar_DimLess}). Theory parameters: Table \ref{tab:ModelParameters}.}
     \label{fig:CurvRhoSNRvsAdhesion}
\end{figure}

With the results of the previous sections, we now have a complete theory for computing $\textrm{SNR}_C$ and $\textrm{SNR}_\rho$ using Eqs. (\ref{SNR_curvature})--(\ref{SNR_rho}) and the variances Eqs. (\ref{ContLimCurvVar_DimLess})--(\ref{ContLimRhoVar_DimLess}). We choose the radii of the beads $R_A$ = 0.5 $\mu$m and $R_B$ = 1.5 $\mu$m to correspond to typical bead sizes in the experiments of \cite{cannon2019amphipathic,shi2022kinetic}. We plot the SNR computed using numerical quadrature in Fig. \ref{fig:CurvRhoSNRvsAdhesion}. Consistent with our discussion of measured fluctuations above, the SNR for curvature increases without bound as curvature fluctuations are suppressed at high $\gamma$, while the $\rho$ SNR reaches an asymptotic limit. We can determine simple analytical forms for the SNR by using the low-$\gamma$ and high-$\gamma$ limits for the variances derived above. The closest relevant limit for understanding experiments on supported lipid bilayers on beads \cite{cannon2019amphipathic,shi2022kinetic} is the limit of strong adhesion (high $\gamma$).

In the limit of high adhesion, Eq. (\ref{SNR_curvature}) and Eq. (\ref{SNR_rho}) become
\begin{gather}
    \label{SNR_curvature_highadh}
    \text{SNR}_{C, \text{ high $\gamma$}} = \frac{4 \pi a^6 \gamma(\frac{1}{R_A} - \frac{1}{R_B})^2}{k_B T},\\
    \label{SNR_rho_highadh}
    \text{SNR}_{\rho, \text{ high $\gamma$}} = \frac{16 \pi a^2 d^2 k (\frac{1}{R_A} - \frac{1}{R_B})^2}{k_B T}.
\end{gather}

At high $\gamma$, SNR$_C$ greatly improves with protein size, varying as $a^6$. Even a small increase in protein size confers a substantial sensing advantage. In comparison, a protein's ability to sense deviations in lipid densities $\rho$ improves more modestly with protein size, varying as $a^2$. SNR$_\rho$ is also improved by increases in monolayer thickness and resistance to in-plane compression as $d^2 k$. 

\begin{figure}
     \centering
     \begin{subfigure}
         \centering
         \includegraphics[width=0.45\textwidth]{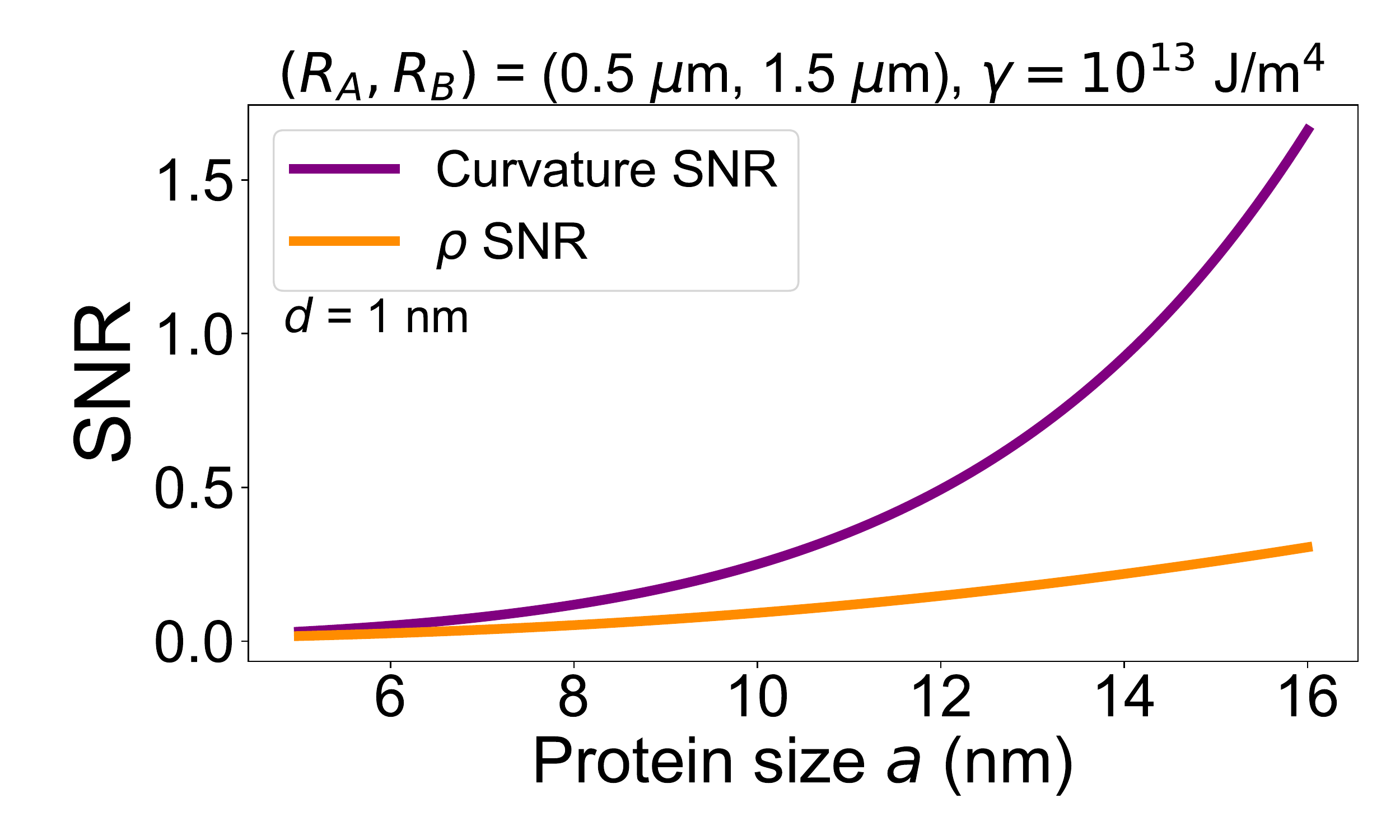}
     \end{subfigure}
     \begin{subfigure}
         \centering
         \includegraphics[width=0.45\textwidth]{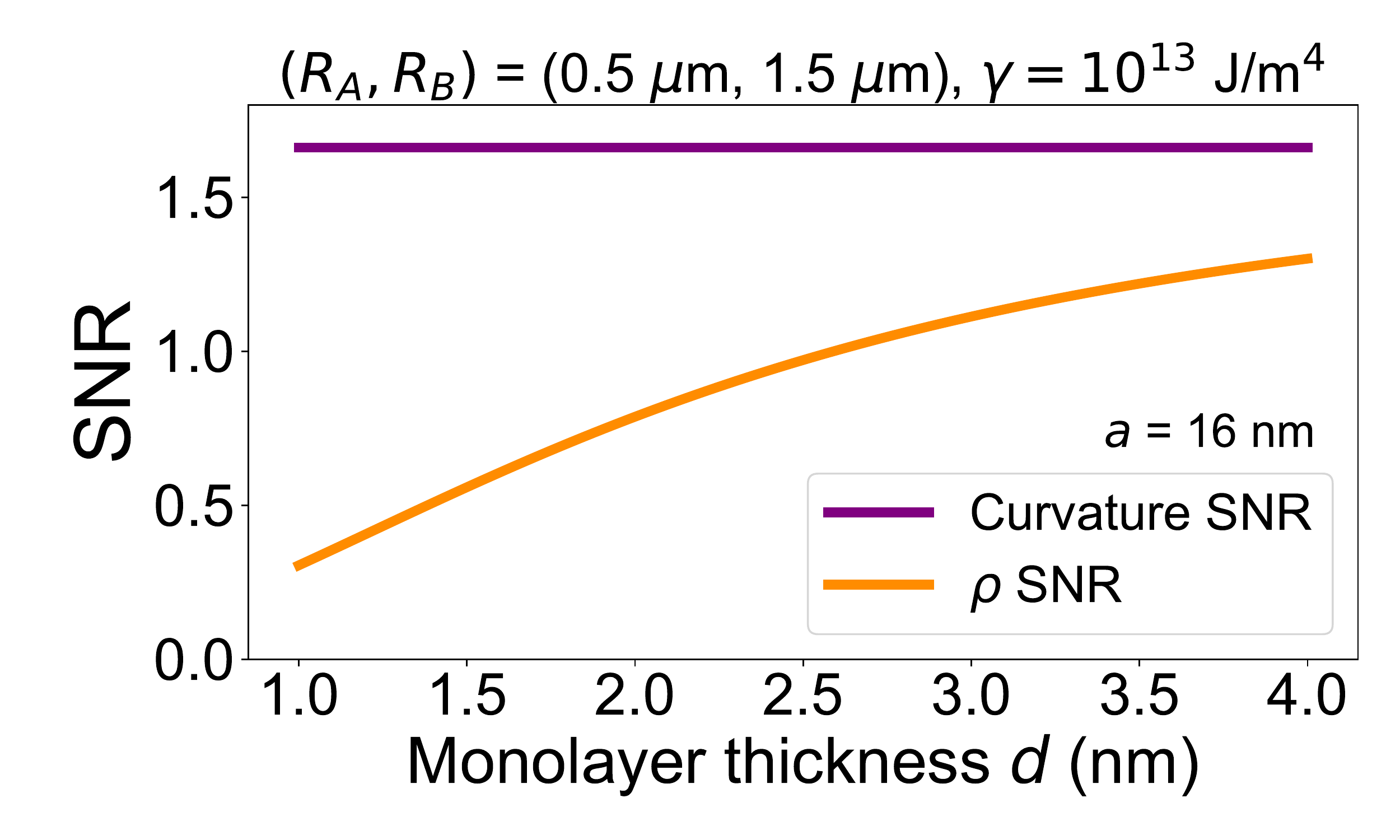}
     \end{subfigure}
     \caption{(Top) SNR vs. protein size, with $d$ = 1 nm. (Bottom) SNR vs. monolayer thickness, with $a$ = 16 nm, for experimentally relevant micron-sized pairs of beads with diameters of (1 $\mu$m, 3 $\mu$m). As the monolayer thickness is increased, SNR$_\rho$ approaches SNR$_C$ asymptotically. A physically realistic adhesion strength of $\gamma = 10^{13}$ J/m$^4$ is used to numerically compute the variances for the SNR using Eqs. (\ref{ContLimCurvVar_DimLess})--(\ref{ContLimRhoVar_DimLess}). Theory parameters: Table \ref{tab:ModelParameters}.}
     \label{fig:SNR_BigBeads_ProtMono}
\end{figure}

What is the theoretically predicted SNR when we compare two beads of experimentally relevant radii $(R_A, R_B)$ = (0.5 $\mu$m, 1.5 $\mu$m)? We choose a value of $\gamma = 10^{13}$ J/m$^4$ to represent a fairly strong SLB adhesion (see Appendix \ref{app:justifyAdhesionStrength} for calculated estimates), and plot the SNR in Fig. \ref{fig:SNR_BigBeads_ProtMono}, varying protein size and membrane monolayer thickness. (We note that our estimate for SLB adhesion does not put us in the asymptotic limit of Eqs. (\ref{SNR_curvature_highadh})--(\ref{SNR_rho_highadh}); the full form must be used.) We would like to highlight three elements of these central results. First, we see that the SNR for sensing curvature is always, over our parameter range, larger than for sensing density. This may not be surprising, since our perturbation of the membrane acts directly on the membrane height field through Eq. (\ref{AdhesionEnergyReal}), with density only correlated with this effect. Second, we see that SNR$_\rho$ can be comparable to SNR$_C$, especially for small protein size $a$ and larger membrane thickness $d$; these are the circumstances where the density difference $\rho$ is best able to act as a proxy for the membrane curvature. Third, we should comment on the overall scale: we see signal-to-noise ratios on the order of one or higher. This suggests that sensing micron-scale curvature is at least reasonably plausible. In the next section, we will ask whether this SNR is compatible with the recent experimental observations of \cite{shi2022kinetic}.

We should also note that the results of Fig. \ref{fig:SNR_BigBeads_ProtMono} vary monolayer thickness independently of other parameters. Changing lipid types to vary monolayer thickness will also potentially change the bending modulus $\kappa$ and compressibility modulus $k$. We show a variant of Fig. \ref{fig:SNR_BigBeads_ProtMono} when $\kappa$ is also changed according to phenomenological laws connecting $\kappa$ and $d$ in Appendix \ref{app:BilayerParameterDependence}.

\subsection{\label{SNR_Estimate_Experiment}Connecting SNR and membrane shape and density fluctuations to experimental protein-bead association rates}

To interpret what the SNR means, we make a correspondence between the experimental measurements and our computed distributions of curvature and lipid density. This requires a few additional assumptions. We treat two separate cases. In the first case, we assume that a single protein has a preference to bind to a specific curvature or range of curvatures. This is consistent with data showing that adsorption of septin has a clear peak at a characteristic bead curvature of 2 $\mu$m$^{-1}$ \cite{ bridges2016micron,cannon2019amphipathic}. 
In our second case, we assume that single proteins prefer to bind to beads with curvature above some threshold value, so that the association rate is enhanced above the threshold curvature, but saturates at sufficiently steep curvatures. For example, if the threshold curvature is 0.5 $\mu$m$^{-1}$, a single protein would distinguish between a flat membrane and a bead of curvature 1 $\mu$m$^{-1}$, but would not be able to distinguish between two beads both of which had curvatures considerably above the threshold, such as 4 $\mu$m$^{-1}$ and 6 $\mu$m$^{-1}$.
This is motivated by the recent work \cite{shi2022kinetic}, which notes that the single-molecule binding rates of septin increase with increasing bead curvature. These two assumptions have qualitatively distinct results. 

\subsubsection{\label{AssocPrefCurv}Proteins with a maximal association rate to a preferred curvature}

We create a simple estimate for the association rate of protein to a membrane adhered to a bead of radius $R$. We first consider the distribution of curvatures or densities that a protein would sense given the bead radius, $P(C_a|R)$ (i.e. the histograms plotted in Fig. \ref{fig:CurvDensityHistogram}). We then assume that the protein binds with a probability that is dependent on the curvature it senses. In the extreme case, binding could happen {\it only} when the protein senses its preferred curvature $C_\text{pref}$. We assume that for a given sensed curvature $C_a$, the association rate has a basal level $A_0$, and a piece that is dependent on the sensed curvature, maximal when $C_a = C_\text{pref}$, which we write as a Gaussian. Therefore,
\begin{gather}
    A(C_a) = A_0 + A_C  \exp\left({\frac{-(C_a - C_\text{pref})^2}{2 \sigma_{\text{bind}}^2}}\right). \label{eq:association_curvature}
\end{gather}

Consequently, the association rates are maximal when the protein senses its preferred curvature, and decreases when $C_a$ is steeper or shallower than $C_\text{pref}$. Here, $\sigma_\text{bind}^2$ characterizes how precisely an individual protein's binding depends on curvature and sets the range of apparent curvatures the protein binds to. In the limit $\sigma_\text{bind} \to 0$, this would correspond to proteins that observe an instantaneous curvature $C_a$ only binding if the observed curvature $C_a$ is precisely $C_\textrm{pref}$.

Eq. (\ref{eq:association_curvature}) reflects the association rate for {\it one} value of $C_a$, but proteins will sense a distribution of apparent curvatures depending on the radius of the bead, $P(C_a|R)$. The conditional  probability density of local curvatures sensed given that the membrane is adhered to a bead of radius $R$ is Gaussian,
\begin{equation}
\label{eq:localcurv_probdist}
    P(C_a|R) = \frac{1}{\sqrt{2 \pi \langle C_a^2 \rangle }}\exp\left(\frac{-(C_a - 1/R)^2}{2 \langle C_a^2 \rangle }\right),
\end{equation}
where $\langle C_a^2 \rangle $ is the curvature variance as derived in Eq. (\ref{ContLimCurvVar_DimLess}). In this section, we drop the superscript ``flat'' label for simplicity, but the variances $\langle C_a^2 \rangle$ are all computed assuming a flat substrate approximation (as explained when proposing Eq. (\ref{SNR_curvature})). Then, the average association rate to a bead of radius $R$ would average Eq. (\ref{eq:association_curvature}) over this distribution as
\begin{equation}
 A(R) \equiv \int_{-\infty}^{\infty}{\mathrm{d}C_a A(C_a) P(C_a|R)}.
\end{equation}
This integral can be evaluated analytically as
\begin{equation}
\label{eq:assocrateprefcurv}
    A(R) = A_0 +  \beta \exp\left(-\text{SNR}_C^\text{eff}\right) \; \; \; \textrm{(preferred curvature)},
\end{equation}
where $\beta = \frac{A_C \sigma_\text{bind}}{\sqrt{\langle C_a ^2 \rangle + \sigma_\text{bind}^2}}$, and the effective curvature SNR is
\begin{equation}
    \label{eq:effectiveCurvSNR}
    \text{SNR}_C^{\text{eff}} = \frac{(1/R - C_\text{pref})^2}{2(\langle C_a^2 \rangle + \sigma_\text{bind}^2)}.
\end{equation}

In the limit $\sigma_\text{bind}^2 \rightarrow 0$, SNR$_C^{\text{eff}}$ is exactly the SNR$_C$ derived in Eq. (\ref{SNR_curvature}), characterizing the ability of a perfect detector to distinguish between the bead's true curvature $C_\textrm{bead} = 1/R$ and the protein's preferred curvature $C_\textrm{pref}$.

\begin{figure}[ht]
\centering
    \includegraphics[width=0.48\textwidth]{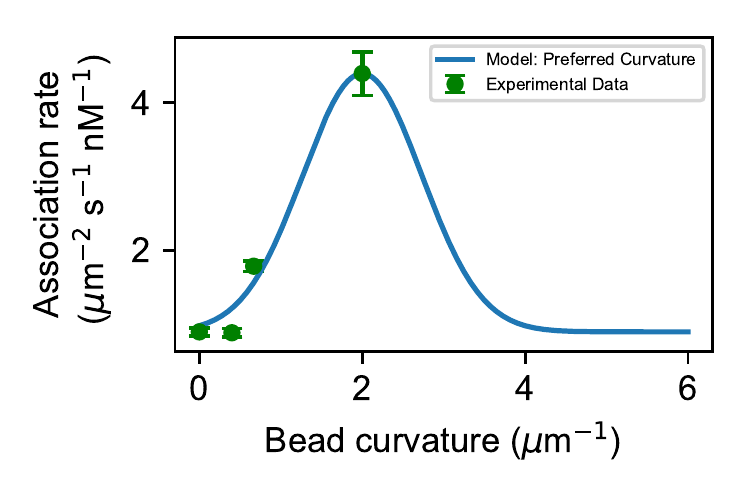}
    \caption{Preferred curvature model: association rates of a septin-sized protein to various membrane-adhered bead curvatures,  maximal at a preferred curvature of 2 $ \mu$m$^{-1}$. This is compared to the experimental data in Figure 2B of \cite{shi2022kinetic} (extracted using WebPlotDigitizer \cite{Rohatgi2022}) for single septin association rates. For an experimental basal rate of $A_0 = 0.892$ $\mu$m$^{-2}$ s$^{-1}$ nM$^{-1}$, we obtain the model parameter $\beta = 3.505$ $\mu$m$^{-2}$ s$^{-1}$ nM$^{-1}$ by subtracting $A_0$ from the maximal association rate at the preferred curvature. The binding variance is obtained as a fit parameter using non-linear least-squares fits (\texttt{lmfit} \cite{newville2016lmfit} in Python), such that $\sigma_\text{bind}^2 \approx 5\times 10^{-9} \langle C_a^2\rangle $. Parameters: Table \ref{tab:ModelParameters}.}
    \label{fig:AssocRatePrefcurvModel}
\end{figure}

Shi et al. report the association rates of a single septin oligomer to beads of different curvatures \cite{shi2022kinetic}, which we replot in Fig. \ref{fig:AssocRatePrefcurvModel}. We can extract the basal association rate $A_0$ of Eq. (\ref{eq:assocrateprefcurv}) directly from their experiments on flat surfaces (zero curvature). We assume that the preferred curvature of a septin is at 2 $\mu \textrm{m}^{-1}$, which is where the association rate is maximal, and is also the curvature of maximal adsorption \cite{bridges2016micron} (although the competition effects found by \cite{shi2022kinetic} suggest that this maximum is not straightforward when different bead sizes are present in the same assay).  This also sets the value of $\beta$, because $A(R = 1/C_\textrm{pref}) = A_0 + \beta$. We use our default parameters (Table \ref{tab:ModelParameters}) to compute $\langle C_a^2\rangle$, leaving only one fit parameter in the model, $\sigma_\textrm{bind}^2$. We fit Eq. (\ref{eq:assocrateprefcurv}) to the experimental data with $\sigma_\text{bind}^2$ as a fit parameter, and in Fig. \ref{fig:AssocRatePrefcurvModel}, we compare the data to our predicted association rates for varying bead curvatures. Although the protein associates maximally to the chosen $C_\text{pref} = 2$ $\mu$m$^{-1}$, it can also associate substantively to a fairly broad range of bead curvatures between $\sim$ 0-4 $\mu$m$^{-1}$. The width of this curve is limited by the unavoidable error in sensing the curvature, $\langle C_a^2 \rangle$, which we computed above. The best fit $\sigma_\text{bind}^2$ is negligible compared to $\langle C_a ^2 \rangle$.

In addition to the fit displayed in Fig. \ref{fig:AssocRatePrefcurvModel}, we can more directly map between the association rates we studied and the SNR we computed in the earlier sections. Using Eq. (\ref{eq:assocrateprefcurv}), the ratio between the protein's association rate to a bead with its preferred curvature and a bead of radius $R$ is
\begin{gather}
\label{eq:assocratioEqn}
    \frac{A(R = 1/C_\text{pref})}{A(R)} =  \frac{w + 1}{w + \exp({-\text{SNR}_C^{\text{eff}}})},
\end{gather}
where $w = A_0/\beta$. This ratio is maximized when $w\to0$, which would happen if the basal rate $A_0$ is negligible. The maximum possible association ratio when $w = 0$ is
\begin{gather}
    \label{maxratioSNRC}
    \left\{\frac{A(R = 1/C_\text{pref})}{A(R)}\right\}_\text{max} = \exp({\text{SNR}_C^{\text{eff}})}.
\end{gather}

Our estimates for association rate in this section let us interpret what the SNR means: a large SNR between beads of radii $R_A$ and $R_B$ indicates that there can be a large ratio of association rates between these two beads. However, if there is a large nonspecific basal level of association, or if the specific association  to curvature is very weakly dependent on curvature (large $\sigma_\textrm{bind}^2$), then the ratio of association rates could be much smaller than that predicted by the simplest SNR in Eq. (\ref{SNR_definition}). The signal-to-noise ratio of Eq. (\ref{SNR_definition}) gives the {\it best case} ability of proteins to distinguish between differing bead shapes, assuming perfect detection of the membrane shape and no non-specific binding.

What SNR does the data on septin binding to membrane-coated beads imply? In Figure 2B of \cite{shi2022kinetic} (replotted in Fig. \ref{fig:AssocRatePrefcurvModel}), the ratio between the association rates of a yeast septin to membrane-adhered beads of diameters 1 $\mu$m and 3 $\mu$m (curvatures of 2 $\mu$m$^{-1}$ and 0.67 $\mu$m$^{-1}$) is  
\begin{gather}
    \label{maxratioexperiment}
    \left\{\frac{A(R = 0.5 \text{ }\mu\text{m})}{A(R = 1.5 \text{ }\mu\text{m})}\right\}_\text{experiment} \approx 2.5 .
\end{gather}
The minimal SNR required by a protein to distinguish between these two membrane-adhered beads with a selective association ratio of 2.5 when $w = 0$ is 
\begin{gather}
\label{eq:minimumSNR}
    \text{SNR}^\text{eff}_\text{minimum}\approx \ln(2.5) \approx 0.9 .
\end{gather}

However, since the experiments indicate a basal association rate of $A_0 = 0.892$ $\mu$m$^{-2}$ s$^{-1}$ nM$^{-1}$, we obtain a non-negligible $w = A_0/\beta \approx 0.254$. Consequently, for an association ratio of 2.5, Eq. (\ref{eq:assocratioEqn}) gives the effective experimental curvature SNR:
\begin{gather}
\label{eq:exptSNR}
    \text{SNR}^\text{eff}_\text{experiment} \approx 1.4 .
\end{gather}

This SNR value is fairly close to the theoretical curvature sensing limits we have derived (Fig. \ref{fig:SNR_BigBeads_ProtMono}), suggesting that the accuracy of septin's discrimination between two curvatures may be near the limit set by stochastic fluctuations. However, as is apparent in Fig. \ref{fig:CurvRhoSNRvsAdhesion}, the theory SNR is strongly dependent on adhesion between the membrane and substrate. Our best estimate for membrane adhesion strength to a solid substrate such as a bead is $\gamma \sim 10^{13}$ J/m$^4$ (see Appendix \ref{app:justifyAdhesionStrength}). For weaker $\gamma$ values, the sensing limit would be set lower. For example, if we used parameters appropriate to membrane-cytoskeleton adhesion, where $\gamma \sim 10^{9}$ J/m$^4$ \cite{biswas2017mapping}, we obtain SNR$_C \approx 0.25$ (see Appendix \ref{app:SNRLowAdhAppendix}). An SNR of $0.25$ means that the association ratio between targets of diameters 1 $\mu$m and 3 $\mu$m could be at most about $1.3$. This is not as preferentially selective as the membrane-bead systems in \cite{cannon2019amphipathic,shi2022kinetic}. If the experiments of \cite{shi2022kinetic,cannon2019amphipathic} were repeated on a system with $\gamma \leq  10^{9}$ J/m$^4$ (e.g. a membrane attached to the cell's cortex or a giant unilamellar vesicle with no adhesion, $\gamma = 0$), then we would expect a significantly lower ratio of association rates. It is only the adhesion strength being large in the experiments of \cite{cannon2019amphipathic,shi2022kinetic} that make them consistent with our bounds. The enhancement of binding by 30\% even at weaker adhesion, though, suggests that at least some curvature sensing by single proteins may be plausible in a broader range of contexts than just strongly-adherent SLBs.

We have phrased everything so far in this section in terms of sensing membrane curvature. However, we can derive an effective lipid density sensing SNR in an exactly analogous way. We find that 
\begin{equation}
    \text{SNR}_\rho^\text{eff} = \frac{(2d/R - \rho_\text{pref})^2}{2(\langle \rho_a^2 \rangle + \sigma_{\rho\text{,bind}}^2)},
\end{equation}
and that the maximum possible ratio between a $\rho$-sensing protein's association rate to a membrane-adhered bead with its preferred lipid density deviation and a different bead is
\begin{gather}
    \label{maxratioSNRrho}
    \left\{\frac{A(R = 2d/\rho_\text{pref})}{A(R)}\right\}_\text{max} = \exp({\text{SNR}_\rho^{\text{eff}})}.
\end{gather}
Fig. \ref{fig:SNR_BigBeads_ProtMono} would suggest that the theoretical lipid density sensing limit is substantially lower than SNR$^\textrm{eff}_\textrm{experiment} \approx 1.4$ when $d = 1$ nm. However, this does not necessarily indicate that $\rho$ is an unfeasible metric to infer membrane shape, as $\rho$ SNR can be appreciable when the membrane monolayer is made thicker or more resistant to in-plane compression. Holding the other parameters constant, we find that SNR$_\rho \approx 1.4$ when $d = 4$ nm and $k = 0.1$ J/m$^2$ (a relatively small change from our default parameters in Table \ref{tab:ModelParameters}). In Appendix \ref{app:RelativeEfficacy}, we plot the ratio between SNR$\rho$ and SNR$_C$ for different sets of physical parameters and compare their relative sensing efficacy.

\subsubsection{\label{AssocThreshCurv}Proteins with enhanced binding above a threshold curvature}

If, instead of the protein binding when it measures a particular curvature, we can loosen our requirements and assume that the protein binds to the bead when it measures a local membrane curvature $C_a$ greater than a threshold curvature $C_\textrm{thresh}$. 
This is motivated by the idea that nm-scale curvature can induce defects in the lipid order \cite{cui2011mechanism}, so rare fluctuations to very steep curvatures could induce local defects in the lipid order, allowing for easier insertion of the amphipathic helices of the protein. Under this threshold assumption, the protein's association rate to the bead will be proportional to the probability that $C_a>C_\textrm{thresh}$, i.e. $P(C_a > C_\textrm{thresh}) = \int_{C_\textrm{thresh}}^{\infty} \mathrm{d}C_a P(C_a | R)$. We define an association rate that depends on $C_\textrm{thresh}$ as
\begin{equation}
    A(R) = A_0 + A_C \int_{C_\textrm{thresh}}^{\infty}\mathrm{d}C_a P(C_a|R),
\end{equation}
where $P(C_a|R)$ is the conditional probability density of local curvatures for a membrane adhered to a bead of radius $R$, as in Eq. (\ref{eq:localcurv_probdist}). 

Evaluating this analytically, we obtain
\begin{equation}
\label{eq:associationR_threshold}
    A(R) = A_0 +  \frac{A_C}{2}\mathrm{erfc}\left(\frac{C_\textrm{thresh} - 1/R}{\sqrt{2 \langle C_a^2 \rangle}}\right) \; \; \; \textrm{(threshold)},
\end{equation}
where the complementary error function is defined as $\mathrm{erfc}(x) = \frac{2}{\sqrt{\pi}}\int_{x}^{\infty}e^{-t^2}\mathrm{d}t$ \cite{oldham2008error}.

We can view Eq. (\ref{eq:associationR_threshold}) as the association rate as a function of the bead curvature $C_\mathrm{bead} = 1/R$. When $C_\textrm{bead} = C_\mathrm{thresh}$, $\mathrm{erfc}(0) = 1$, and $A(C_\mathrm{bead}) = A_0 + A_C/2$. Therefore, $C_\mathrm{thresh}$ indicates the curvature at which the protein association rate's increase above the basal level is half-maximal.

\begin{figure}[ht]
\centering
    \includegraphics[width=0.47\textwidth]{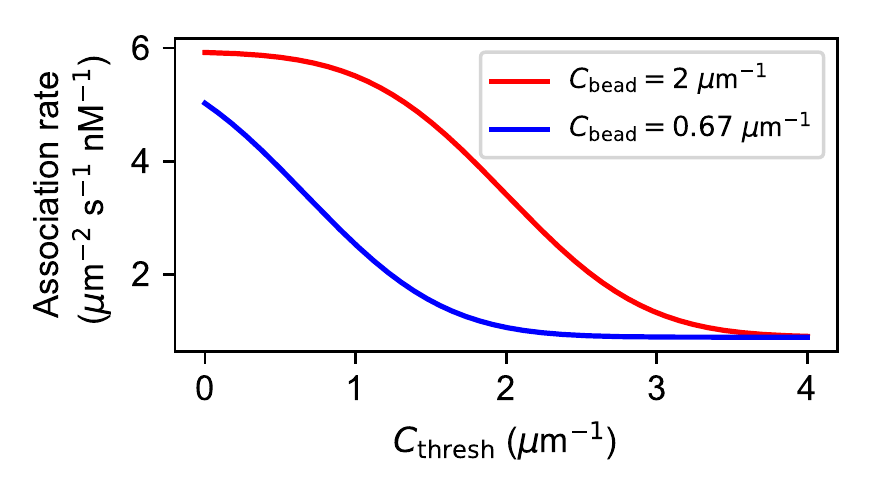}
    \caption{Predicted association rates of a septin-sized protein to membrane-adhered beads of curvature 2 $\mu$m$^{-1}$ and 0.67 $\mu$m$^{-1}$, as a function of the protein's threshold curvature $C_\textrm{thresh}$, for a fixed $A_0$ and $A_C$. Theory parameters: Table \ref{tab:ModelParameters}. The values used for $A_0$ and $A_C$ are taken from the fit in Fig. \ref{fig:AssocRateThresholdModel}.} 
    \label{fig:AssocRate_Cthresh}
\end{figure}

In the simulated histograms in Fig. \ref{fig:CurvDensityHistogram}, we showed how beads can have local curvature distributions that overlap considerably, making discriminating between these two beads more difficult. If the protein binds only at sufficiently large curvature, $C_a > C_\textrm{thresh}$, this means that proteins are probing the tails of these histograms. This has two effects. First, looking at the tail of the distribution can highlight a small difference between the means---a higher $C_\textrm{thresh}$ makes it less likely that a protein incorrectly attributes a steep local membrane curvature to a shallow bead's curvature distribution. However, as $C_\textrm{thresh}$ is increased above $C_\textrm{bead}$, it is rarer and rarer that a curvature this high is observed, so the curvature-dependent association rate decreases, and eventually any specificity is lost because the curvature-dependent association rate is smaller than the basal rate. In Fig. \ref{fig:AssocRate_Cthresh}, we compare the predicted association rates of a septin-sized protein to beads of curvature 2 $\mu$m$^{-1}$ and 0.67 $\mu$m$^{-1}$. The association rates for both these beads are suppressed at high $C_\textrm{thresh}$ values; however, the association to the bead with steeper curvature ($C_\textrm{bead} = 2$ $\mu$m$^{-1}$) is considerably higher over a broad range of $C_\textrm{thresh}$. 

\begin{figure}[ht]
\centering
    \includegraphics[width=0.48\textwidth]{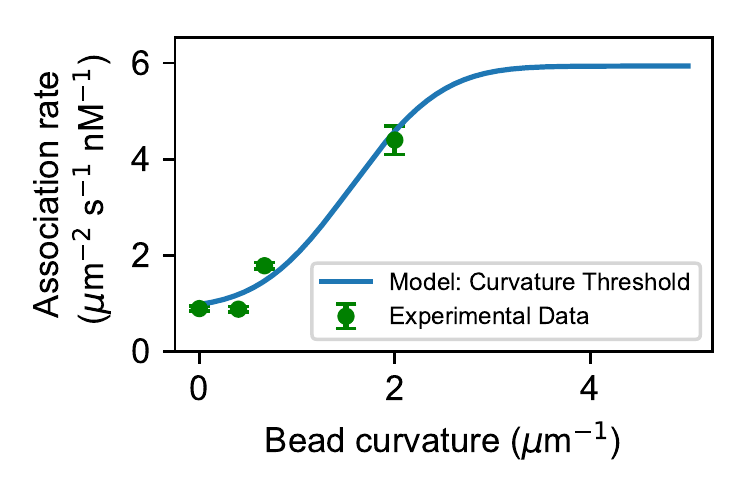}
    \caption{Protein association rates predicted by the curvature threshold model, as compared to the experimental single septin association rates in Figure 2B of \cite{shi2022kinetic} for varying bead curvatures.  The basal rate $A_0$ is 0.892 $\mu$m$^{-2}$ s$^{-1}$ nM$^{-1}$. The model parameters, fitted using a non-linear least-squares method (\texttt{lmfit} \cite{newville2016lmfit} in Python), are $A_C \approx 5.05$ $\mu$m$^{-2}$ s$^{-1}$ nM$^{-1}$ and $C_\textrm{thresh} \approx 1.55$ $\mu$m$^{-1}$. Parameters: Table \ref{tab:ModelParameters}.}
    \label{fig:AssocRateThresholdModel}.
\end{figure}

How does the threshold model compare to experiment? We fit Eq. (\ref{eq:associationR_threshold}) to the experimental data (Figure 2B, \cite{shi2022kinetic}) using their basal rate for $A_0$. The variance $\langle C_a ^2 \rangle$ is computed using our default physical parameters (Table \ref{tab:ModelParameters}), and we obtain $C_\textrm{thresh}$ as a fit parameter. The data indicate that $C_\textrm{thresh} \approx 1.55$ $\mu$m$^{-1}$ when the adhesion strength $\gamma = 10^{13}$ J/m$^4$.  As plotted in Fig. \ref{fig:AssocRateThresholdModel}, the association rate increases sigmoidally as a function of bead curvature, saturating when the bead curvature is sufficiently greater than $C_\textrm{thresh}$. Therefore, a protein that is described by the threshold model exhibits increased association to curvatures that are above a threshold, but loses the ability to discriminate between two bead curvatures that are substantially more than the threshold curvature.

In the model of Section \ref{AssocPrefCurv}, where individual proteins have a preferred curvature, the SNR controls the sharpness of how association rates can depend on bead curvature. In a similar fashion, in the threshold model, the sharpness of the transition in association rates is determined by the variance in membrane curvatures sensed by the protein. In Eq. (\ref{eq:associationR_threshold}), we see that the predicted association rate increases from its basal level to its maximal level as the bead curvature $1/R$ is increased---the value $\sqrt{\langle C_a^2 \rangle}$ sets the scale of this transition. The fit between experiment and data in Fig. \ref{fig:AssocRateThresholdModel} could be improved if we made the transition sharper. If we increase the adhesion $\gamma$---corresponding to fluctuations that are more suppressed---we do find a better fit (Appendix \ref{app:ImprovedFitsCurvatureThreshold}). However, this takes the adhesion beyond what we think is the likely range (Appendix \ref{app:justifyAdhesionStrength}).

\section{\label{sec:level4}Discussion\protect}
Our results identify the key elements that determine whether micron-scale curvature sensing via sensing local membrane shape or lipid density is physically plausible. These limits occur because even a perfect detection of the membrane's shape or density is subject to unavoidable thermal fluctuations. We identify a signal-to-noise ratio that describes the accuracy with which a protein can discriminate between two different large-scale curvatures solely by sampling the local properties of the membrane. This SNR can be related to the relative association rates of the protein to two membrane-coated beads of different curvatures. The SNR naturally depends on the curvature of the bead---larger bead radii lead to shallower curvatures, which are more difficult to distinguish---but also on the properties of the membrane and the protein. Of particular importance are the membrane-bead adhesion strength and membrane bending stiffness, which suppress membrane fluctuations, and the protein's size.  In addition, when sensing lipid densities, the membrane's thickness and area compressibility modulus (indicating its resistance to in-plane compression) play an important role. Our estimates of SNR suggest that the ability of septin proteins to distinguish between micron-scale curvatures  \cite{cannon2019amphipathic,shi2022kinetic} is consistent with the idea that septin is measuring the local shape of the membrane or the local changes in density. However, for consistency, we must assume both that the measurements are near-perfect and that the membrane is strongly adherent to its bead ($\gamma \approx 10^{13}$ $ \textrm{J}/\textrm{m}^4$); lower adhesion strengths lead to insufficient SNR to explain the experiments (Appendix \ref{app:SNRLowAdhAppendix}). We would then expect that preferential binding as a function of curvature would be much lower for giant unilamellar vesicles that are not attached to a bead, though we note that there may be confounding issues when changing vesicle size \cite{jin2022curvature}. Although the importance of membrane-substrate adhesion in biological processes is widely acknowledged \cite{sengupta2018adhesion, gordon2015membrane, hynes1999cell, sheetz2001cell}, it remains challenging to ascertain the range of adhesion strengths that are applicable to any given system. Evidence suggests that large vesicles exhibit weak adhesion to glass substrates, while supported lipid bilayers with direct lipid-glass binding can have stronger adhesion \cite{ursell2011lipid,schmidt2014signature}. Membranes supported on glass beads generally exhibit strong adhesion; the hydration layer between the membrane and bead is only a few nanometers thick \cite{anderson2009formation,zwang2010quantification, israelachvili2011intermolecular}. Given the orders-of-magnitude-broad range of reported values, we have generally tried to show how our SNR depends on adhesion. The value $\gamma \approx 10^{13}$ J/m$^4$ was estimated based on van der Waals interactions and hydration forces (Appendix \ref{app:justifyAdhesionStrength}). 

We have computed SNR$_C$ under the assumption of perfect local detection of the curvature. However, there may be strong biophysical constraints on curvature sensing beyond the statistical ones we have raised here. In particular, as noted by \cite{huang2010macromolecules}, if a perfectly straight rod-shaped protein is placed on top of a spherical membrane of micron-scale diameter, the gap distance between the protein and the membrane is below the angstrom scale for a protein of length $\sim$ 4 nm, such as SpoVM. It is extremely biophysically implausible that any underlying binding could be sensitive to this subangstrom gap. A simple curvature sensing mechanism is not as immediately ruled out for proteins that are larger, such as septin. For yeast septins with an end-end length of $\sim 32$ nm (emulated by our ``sensing radius'' $a \sim 16$ nm), this gap is about 1 nm even for a bead that is a micron in diameter. Both SpoVM and septin, though, are likely to be able to sense local lipid densities to some extent. Despite SpoVM being only 4 nm in length, its amphipathic helix has membrane insertion depths of $\sim$ 1 nm \cite{gill2015structural}, indicating that even small proteins may sense leaflet lipid densities. The clear relevance of the amphipathic helices of septins in curvature sensing strongly suggests that curvature sensing arises from sensing some aspect of lipid membrane structure \cite{cannon2017unsolved,cannon2019amphipathic}. We have primarily focused on proteins sensing the local value of $\rho$, the difference in lipid densities between the upper and lower leaflets at the midsurface. While $\rho$ is a lipid feature that clearly reflects the curvature, it is not the only possibility. We can also generalize our results to describing proteins that simply sense the projected density of the upper leaflet $\rho^+$---which might be more appropriate for proteins that only embed amphipathic helices shallowly into the membrane. We performed simulations and derived the theory for $\langle \rho^{+^2} \rangle$ and SNR$_{\rho^+}$ (Appendix \ref{app:rhoplusSensing}). We find that SNR$_\rho$ is greater than SNR$_{\rho^+}$, and interestingly, in the high-adhesion limit, the variance in $\rho^+$ is exactly twice the variance in $\rho$. It may also be possible for proteins to effectively sense other membrane properties, such as lipid tilt \cite{watson2011thermal}, which can be a more reliable readout of bending moduli at small scales \cite{watson2012determining}. 

To compute the best possible accuracy of membrane curvature sensing by proteins, we assumed that the protein is a perfect sensor of local curvature or $\rho$, representing the measurement by a weighted integral (Eqs. (\ref{eq:weightedcurvature})--(\ref{eq:weightedrho})). There are several important caveats to this approach. While we think that the local average used here is the most natural measurement of curvature, it is possible that a more complex observable could be less noisy. For example, past modeling of concentration sensing by a single receptor has shown that the best-achievable accuracy is twice that of a naive average \cite{endres2009maximum}, and this difference can be even larger if there are multiple receptor types \cite{hopkins2020chemotaxis}. This is a clear area for future research. We have also neglected the anisotropy of the septin by choosing an isotropic weight function $G(\mathbf{r})$, but expect this to play a small role in setting SNR. We also note that binding uses effectively only a single snapshot of the membrane state, neglecting potential time-averaging. This differs from the Berg-Purcell approach and generalizations, where noisy measurements are integrated over time in order to better resolve them \cite{berg1977physics,bialek2005physical,endres2009maximum}. We are motivated in this by results suggesting that these time-averaging schemes require energy dissipation and cannot be carried out in equilibrium \cite{mehta2012energetic,govern2014energy,lang2014thermodynamics}; time-averaging is then likely not relevant to understanding {\it in vitro} experiments of curvature sensing, though it is an intriguing possibility within a living cell.

In Fig. \ref{fig:AssocRatePrefcurvModel} and Fig. \ref{fig:AssocRateThresholdModel}, we have made predictions for how we would expect association rates to depend on bead curvature. These are somewhat speculative, because they depend strongly on assumptions of how single proteins bind based on their instantaneous measurement of curvature. We have made two physically plausible assumptions, which make qualitatively different predictions on how association rate will depend on increasing bead curvature. These could be distinguished by measuring single-molecule association rates at smaller bead sizes, extending the results of \cite{cannon2019amphipathic,shi2022kinetic}---though this would be experimentally difficult due to the small patch of membrane resolved with these beads \cite{shi2022kinetic}. In addition, while both models are roughly consistent with the existing experimental data, neither is a perfect fit. It is possible to improve the fit quality if the membrane is more adherent than our expectations or there is another reason why fluctuations are suppressed (Appendix \ref{app:ImprovedFitsCurvatureThreshold}).

Another potentially important factor in improving the quantitative comparison between experiment and theory is to understand the extent to which diffusion of septin to the bead influences binding. If there was no selectivity in binding, and binding occurred immediately upon contact with a sphere of radius $R$, the rate of binding would be $4\pi D R$  \cite{berg1977physics}---so the association rate (rate per surface area) computed by \cite{shi2022kinetic,cannon2019amphipathic} would decrease for increasing radius $R$, as observed by \cite{shi2022kinetic}. However, the association rate observed is smaller than we would expect from a diffusion-limited rate, so we have neglected these factors. In future work we will consider complications arising from competing diffusion and adsorption timescales \cite{alvarez2010diffusion}.

What if the membranes are under tension? We expect that a probe of local membrane tension \cite{pinot2018feedback,colom2018fluorescent} may also be constrained by related fluctuation results, as probes of tension are related to probing lipid structure and packing \cite{colom2018fluorescent}. Added tension on the membrane will suppress thermal fluctuations \cite{brown2008elastic}. Added tension on the membrane due to osmotic effects may play several other roles.  As pointed out by Wasnik et al. in analyzing SpoVM localization, in the presence of an osmotic pressure difference across the membrane, the tension will be different for different-sized vesicles due to the Young-Laplace equation  \cite{wasnik2015modeling}, with the tension increasing linearly with vesicle radius. If this is the case, then the relevant distinction between vesicles of different sizes may not be the shape, but the tension. In our view, then, we would expect that the averaged lipid densities in the two leaflets $\rho^{\pm}$ could vary systematically with radius in a more complex way than that given by our simple $\rho = 2d/R$. If so, the osmotic pressure could lead to systematic shifts between the histograms in Fig. \ref{fig:CurvDensityHistogram}, increasing the SNR beyond our predictions here. 

Basic considerations of physical and statistical bounds limit the accuracy of a vast number of sensing processes across biology, from chemotaxis to pattern formation and differentiation. Our results suggest that similar physical constraints may be relevant for curvature sensing by single proteins---proteins like septin may be performing nearly as well as possible, given the inevitable thermal fluctuations of the membrane. These predictions, though, must be tested against experiments, e.g. varying the membrane-bead adhesion, membrane compressibility modulus, or bending modulus, to be viewed confidently. Our results also have broader implications for sensors of related properties, e.g. fluorescent probes that reflect membrane structure or tension \cite{colom2018fluorescent}. We would predict that the distribution of signals arising from these fluorescent probes of membrane structure are limited by the thermal fluctuations in lipid density and membrane shape, and could be fit to models extending our work.  
These results may also provide inputs into probe design for  curvature or stress sensors. It is more advantageous to use larger probe sizes for curvature sensors ($\sim a^6$ dependence) than for lipid density sensors ($\sim a^2$ dependence) in the high-adhesion limit, while density sensors can benefit from greater membrane insertion depths (probing $\rho$ instead of $\rho^+$). As we have studied here, probe accuracies would also depend on membrane-substrate adhesion strength, suggesting that substrate types and preparations \cite{scomparin2009diffusion,gunderson2018liquid,amjad2017membrane} may play a role in curvature sensing experiments.

\begin{acknowledgments}
We thank Amy Gladfelter, Ehssan Nazockdast, and all the authors of \cite{shi2022kinetic} for useful conversations. We thank Yongtian Luo, Pedrom Zadeh, Yiben Fu, and Margaret Johnson for a close reading of the manuscript. This work was supported by the National Science Foundation under Grant No. DMR 1945141. This research project was conducted using computational resources at the Maryland Advanced Research Computing Center (MARCC).
\end{acknowledgments}

\onecolumngrid

\section*{APPENDIX}
\appendix

\section{\label{app:FourierConventions}Fourier space conventions}

Our Fourier modes are $\mathbf{q} = \frac{2\pi}{L}(m,n)$ with $m,n$ in the range $-(N-1)/2,\cdots,0,\cdots,(N-1)/2$. Since $h(\mathbf{r}), \rho(\mathbf{r})$ and $\bar{\rho}(\mathbf{r})$ must necessarily be real quantities, the modes must fulfill the condition $h_\mathbf{q}^* = h_\mathbf{-q}$, $\rho_\mathbf{q}^* = \rho_\mathbf{-q}$, and $\bar{\rho}_\mathbf{q}^* = \bar{\rho}_\mathbf{-q}$. Therefore, only half the modes are independently evolved as a function of time, and the dependent modes are computed as complex conjugates of the independent modes. We choose the independent modes analogously to \cite{sigurdsson2013hybrid}. In general, we would expect Fourier modes to be complex, but because we are performing Fourier transforms with a finite set of modes, some modes are forced to be their own complex conjugate, requiring them to be real. Aside from these explicitly real modes, other modes have both real and imaginary components. The specific modes that are explicitly real depend on whether $N$ is chosen to be even \cite{lin2006simulating} or odd \cite{sigurdsson2013hybrid,camley2014fluctuating}. For convenience, we choose $N$ to be an odd number, such that only a single mode corresponding to $(m = 0, n = 0)$ of $h_q, \rho_q$ and $\bar{\rho}_q$ is explicitly real. As the membrane's center of mass is not evolved in our system \cite{lin2004dynamics}, these zeroth modes are not evolved in time after their initial values are set (see Appendix \ref{app:sec:numericalEvaluationMotion} for details).  To perform Fourier transforms and their inverses (as in Eq. (\ref{eq:fourierTransformpair})), we used the two-dimensional Fast Fourier Transform (FFT2) and Inverse Fast Fourier Transform (IFFT2) methods in Python's $\texttt{numpy}$ package \cite{harris2020array}, scaling by $1/N^2$ and $N^2$ as appropriate to the respective transform conventions. 

\section{\label{app:modelparameters}Model parameters}
The following parameters are applicable unless otherwise stated for a particular result or figure.

\begin{table}[ht]
\begin{tabular}{|c|c|c|}
\hline
Description                               & Param.         & Value                  \\ \hline
Approximate protein size (sensing radius)                 & $a$            & 16 nm                  \\ \hline
Membrane-substrate adhesion strength      & $\gamma$       & $10^{13}$ J/m$^4$      \\ \hline
Membrane bending modulus                           & $\kappa$       & 20 $k_B T$             \\ \hline
Monolayer area compressibility modulus    & $k$            & 0.07 J/m$^2$           \\ \hline
Monolayer thickness                       & $d$            & 1 nm                   \\ \hline
Temperature                               & $T$            & 310 K                  \\ \hline
Monolayer viscosity                       & $\mu$          & $10^{-8}$ kg/s         \\ \hline
Solvent fluid viscosity                   & $\eta$         & 0.02 Pa$\cdot$s        \\ \hline
Intermonolayer friction                   & $b$            & $10^7$ J$\cdot$s/m$^4$ \\ \hline
Simulation timestep                       & $\Delta t$     & $3.2 \times 10^{-9}$ s            \\ \hline
Total simulation time                     & $t_\text{sim}$ & 0.016 s                 \\ \hline
Edge length of simulated membrane         & $L$            & 1600 nm                \\ \hline
Number of membrane lattice points/Fourier modes on a side& $N$            & 73                     \\ \hline
\end{tabular}
\caption{\label{tab:ModelParameters} Parameters used for theory and FSBD simulations unless otherwise stated.}
\end{table}

The values used for $k$ and $\mu$ are consistent with typical values in \cite{boal2012mechanics} and \cite{faizi2021viscosity}, respectively. 
The solvent fluid viscosity $\eta$, the membrane monolayer viscosity $\mu$, and the intermonolayer friction $b$ are dynamical parameters that represent dissipative mechanisms, and do not influence the equilibrium distribution, which only depends on the energy of a particular state. (This is why we can reproduce our SNR values with Monte Carlo methods in Section \ref{app:FMCsimulations}.) However, the dynamic parameters do determine the rate at which disturbances relax and the magnitude of thermal fluctuations---so they influence the stability properties of the numerical algorithm and the equilibration time required. Our approach has been to begin with somewhat-realistic dynamic parameters, and then tune them to allow for easier convergence (see Section \ref{app:choosing_parameters}).

\section{\label{app:FourierSpaceDerivation_CurvatureRhoVariance}Variances in local curvature and \texorpdfstring{$\rho$}{rho} derived analytically in Fourier-space}

Consider the local membrane curvature sensed by a protein of size $a$, as in Eq. (\ref{eq:weightedcurvature}):
\begin{equation}
    C_a = \int_{L^2} \mathrm{d}\mathbf{r}  \frac{-\nabla^2 h(\mathbf{r})}{2} G(\mathbf{r}, a).
\end{equation}

Using the Fourier-space representation of $h(\mathbf{r})$ and the relation $G(\rb,a) = \frac{1}{L^2}\sum_\qv G(\qv)e^{i\qv \cdot \rb}$, we obtain
\begin{align}
    C_a &= \int_{L^2} \mathrm{d}\mathbf{r}\frac{1}{2L^2} \sum_\mathbf{q} q^2 h_{\qv} e^{i\qv \cdot \rb}  \frac{1}{L^2} \sum_\mathbf{q'} G(\qv') e^{i\qv' \cdot \rb}  \notag\\
    &= \frac{1}{2L^4}\sum_\mathbf{q, q'}{q}^2 h_{\qv}G(\qv') \int_{L^2} \mathrm{d}\mathbf{r} e^{i\qv \cdot \rb}e^{i\qv' \cdot \rb} \notag\\
    &= \frac{1}{2L^4}\sum_\mathbf{q, q'}{q}^2 h_{\qv}G(\qv') L^2 \delta_{\qv',-\qv} \notag\\
    &= \frac{1}{2L^2}\sum_\mathbf{q} q^2 h_{\qv}G(-\qv),
\end{align}

where $\delta_{\qv',-\qv}$ is a Kronecker delta function.

The variance in $C_a$ for a flat, fluctuating membrane is then derived as
\begin{align}
    \langle {C_a}^2 \rangle^\text{flat} &= \left\langle \frac{1}{2L^2}\sum_\mathbf{q} q^2 h_\qv G(-\qv)  \frac{1}{2L^2} \sum_\mathbf{q'} {q'}^2 h_{\qv'} G(-{\qv'})   \right\rangle  \notag\\ 
    &=  \frac{1}{4L^4}\sum_{\mathbf{q},\qv'} q^2 q'^2\left\langle h_\qv  h_{\qv'}\right\rangle   G(-\qv)   G(-{\qv'})    \notag\\
    &=  \frac{1}{4L^4}\sum_\mathbf{q} q^4 \left\langle h_\qv h_{-\qv}\right\rangle G(-\qv) G(\qv),
\end{align}
where in the last step we have noted that $\langle h_\qv  h_{\qv'}\rangle = \delta_{\qv,-\qv'} \langle h_\qv  h_{-\qv}\rangle$. Since $h_{-\qv} = h_{\qv}^*$ and $G(-\qv) = G(\qv)^*$, this simplifies to
\begin{equation}
    \langle {C_a}^2 \rangle^\text{flat} = \frac{1}{4L^4}\sum_\mathbf{q} q^4 \langle |h_\qv| ^2 \rangle |G(\qv)|^2.
\end{equation}

Similarly, the lipid density deviation sensed by a protein of size $a$ in Eq. (\ref{eq:weightedrho}) can be used to derive the variance in density deviations for a flat membrane, as shown in Eq. (\ref{eq:WeightedRhoVariance}).

\section{\label{app:simulationalgorithms}Simulation algorithms}

\subsection{Derivation of equation of motion}

\subsubsection{Choosing thermal noises to obey detailed balance}

We have written our equations of motion as
\begin{equation}
\label{app:eq:eqnsofmotion}
 \frac{\partial}{\partial t}\left(\begin{matrix} h_\qv \\ \rho_\qv \\ \bar{\rho}_\qv\end{matrix}\right) = -L^2\left(\begin{matrix} \frac{1}{\Omega_{h}}{\partial E}/{\partial h_\qv^*} \\ \frac{1}{\Omega_\rho} {\partial E}/{\partial \rho_\qv^*} \\ \frac{1}{\Omega_{\bar{\rho}}} {\partial E}/{\partial \bar{\rho}_\qv^*} \end{matrix}\right) + \left(\begin{matrix} \xi_\qv \\ \zeta_\qv \\ \chi_\qv
    \end{matrix}\right).  
\end{equation}

The correlations of the Gaussian Langevin noises can be written as
\begin{align}
\langle \xi_\mathbf{q}(t) \xi_\mathbf{q'}(t') \rangle &= 2D_h  \delta_\mathbf{q,-q'}\delta(t - t'), \\
\langle \zeta_\mathbf{q}(t) \zeta_\mathbf{q'}(t') \rangle &= 2D_\rho  \delta_\mathbf{q,-q'}\delta(t - t'),\\
    \langle \chi_\mathbf{q}(t) \chi_\mathbf{q'}(t') \rangle &= 2D_{\bar{\rho}}  \delta_\mathbf{q,-q'}\delta(t - t').
\end{align}
This serves as a definition for $D_{h,\rho,\bar{\rho}}$.

The amplitudes of the noises $D_h$, $D_\rho$, $D_{\bar{\rho}}$, which are analogous to diffusion coefficients in simple Brownian dynamics \cite{doi1988theory}, must obey a fluctuation-dissipation relationship. This can be found by writing down the Fokker-Planck equation \cite{gardiner1985handbook} for the time evolution of the probability distribution of the fields $P(\{h_\qv\},\{\rho_\qv\},\{\bar{\rho}_\qv\})$ as
\begin{equation}
    \frac{\partial P}{\partial t} = \sum_\qv \frac{\partial}{\partial h_\qv}\left[\frac{L^2}{\Omega_h(q)}\frac{\partial E}{\partial h_{-\qv}}P + D_h\frac{\partial P}{\partial h_{-\qv}}\right] +
    \sum_\qv \frac{\partial}{\partial \rho_\qv}\left[\frac{L^2}{\Omega_\rho(q)}\frac{\partial E}{\partial \rho_{-\qv}}P + D_\rho\frac{\partial P}{\partial \rho_{-\qv}}\right] + \sum_\qv \frac{\partial}{\partial \bar{\rho}_\qv}\left[\frac{L^2}{\Omega_{\bar{\rho}}(q)}\frac{\partial E}{\partial \bar{\rho}_{-\qv}}P + D_{\bar{\rho}}\frac{\partial P}{\partial \bar{\rho}_{-\qv}}\right],
\end{equation}
where $E$ is the total membrane free energy defined in Eq. (\ref{EnergyFourierSpace}), and we have noted that $h_\qv^* = h_{-\qv}$, $\rho_\qv^* = \rho_{-\qv}$, $\bar{\rho}_\qv^* = \bar{\rho}_{-\qv}$ for the real-valued functions $h(\rb), \rho(\rb), \bar{\rho}(\rb)$. 

For the steady-state probability to have the Gibbs-Boltzmann form, $P^{GB}(\{h_\qv\},\{\rho_\qv\},\{\bar{\rho}_\qv\}) = \frac{1}{Z}\exp(-E/k_B T)$, it must set the right hand side of this Fokker-Planck equation to zero. We note that $\frac{\partial}{\partial h_\qv} P^{GB} = -\frac{1}{k_B T}\frac{\partial E}{\partial h_\qv} P^{GB}$. Plugging in the Gibbs-Boltzmann solution to the Fokker-Planck equation, we find:
\begin{equation}
\begin{split}
\frac{\partial P^{GB}}{\partial t} = &\sum_\qv \frac{\partial}{\partial h_\qv}\left[\frac{L^2}{\Omega_h(q)}\frac{\partial E}{\partial h_{-\qv}}P^{GB} - \frac{D_h}{k_B T}\frac{\partial E}{\partial h_{-\qv}} P^{GB}\right] + \\
    &\sum_\qv \frac{\partial}{\partial \rho_\qv}\left[\frac{L^2}{\Omega_\rho(q)}\frac{\partial E}{\partial \rho_{-\qv}}P^{GB} - \frac{D_\rho}{k_B T}\frac{\partial E}{\partial \rho_{-\qv}} P^{GB}\right] + \\
    &\sum_\qv \frac{\partial}{\partial \bar{\rho}_\qv}\left[\frac{L^2}{\Omega_{\bar{\rho}}(q)}\frac{\partial E}{\partial \bar{\rho}_{-\qv}}P^{GB} - \frac{D_{\bar{\rho}}}{k_B T}\frac{\partial E}{\partial {\bar{\rho}}_{-\qv}} P^{GB}\right].
\end{split}
\end{equation}
For the equation to be at steady-state at the Gibbs Boltzmann distribution, and the right hand side to be zero, we then need
\begin{align}
    D_h &= \frac{k_B T L^2}{\Omega_h(q)}, \\
        D_\rho &= \frac{k_B T L^2}{\Omega_\rho(q)}, \\
            D_{\bar{\rho}} &= \frac{k_B T L^2}{\Omega_{\bar{\rho}}(q)}.
\end{align}
These are the Einstein relations for our system.

\subsubsection{Deriving hydrodynamic mobilities for \texorpdfstring{$h$}{h}, \texorpdfstring{$\rho$}{rho}, and \texorpdfstring{$\bar{\rho}$}{rhobar}}

To obtain the mobilities $\Omega_h^{-1}$, $\Omega_{\rho}^{-1}$ and $\Omega_{\bar{\rho}}^{-1}$, we derive the dynamical equations for $\partial {h_\qv}/\partial t$, $\partial {\rho_\qv}/\partial t$, and $\partial \bar{\rho}_\qv/\partial t$ from the hydrodynamic equations in the Seifert-Langer model \cite{seifert1993viscous} while neglecting inertial effects with the Stokes approximation. The model assumes that the membrane is surrounded by fluid above and below the membrane. Our derivation here is a variant of that presented in \cite{seifert1993viscous}, to highlight how the Seifert-Langer results can be generalized to an arbitrary Hamiltonian. 

We describe the fluid flow above and below the membrane using the incompressible Stokes equations with a fluid velocity $\mathbf{v}_f^\pm(x,y,z)$, where $\pm$ indicates whether we are above ($z>0$) or below ($z<0$) the membrane. These equations are
\begin{align}
    \nabla \cdot \mathbf{v}_f^\pm &= 0,\\
    \eta \nabla^2 \mathbf{v}_f^\pm &= \nabla p^\pm,
\end{align}
where $p^\pm$ is the pressure above/below the membrane. 

The two monolayers of the membrane have in-plane velocity fields $\Tilde{\mathbf{v}}^\pm(x,y)$---these are treated as completely two-dimensional. The Stokes equations for the velocity fields of the monolayers are
\begin{align}
    -\Tilde{\nabla} \sigma^{+} + T^+\cdot \hat{\mathbf{e}}_z + \mu \Tilde{\nabla}^2 \Tilde{\mathbf{v}}^+ - b(\Tilde{\mathbf{v}}^+ - \Tilde{\mathbf{v}}^-) = 0,\\
    -\Tilde{\nabla} \sigma^{-} - T^-\cdot \hat{\mathbf{e}}_z + \mu \Tilde{\nabla}^2 \Tilde{\mathbf{v}}^- + b(\Tilde{\mathbf{v}}^+ - \Tilde{\mathbf{v}}^-) = 0,
\end{align}
where a tilde denotes quantities in two dimensions, $\sigma^\pm(x,y) = -\delta E/\delta \rho^\pm(x,y)$ is the surface pressure due to varying densities in the two leaflets, $T^{\pm}$ is the stress tensor of the surrounding fluid, i.e. $T^{\pm}\cdot(\pm \hat{\mathbf{e}}_z)$ is the force per unit area exerted by the outside fluid onto the monolayers, $\mu$ is the monolayer viscosity, and $b$ is the intermonolayer friction. The components of the stress tensor $T^\pm$ are
\begin{equation}
    T_{ij}^\pm = -p^\pm\delta_{ij} + \eta(\partial_i v_{f,j}^\pm + \partial_j v_{f,i}^\pm).
\end{equation}

There is also a force balance equation in the vertical direction, written in real space as
\begin{equation}
\label{eq:vertical_force_balance}
    -T_{zz}^+(x,y,z=0) + T_{zz}^-(x,y,z=0) = - \frac{\delta E}{\delta h}.
\end{equation}

We assume a no-slip boundary condition between the membrane and the outside fluid---the velocity of the membrane must match the external fluid velocity. In the plane of the membrane, this requires that the monolayer velocities match the in-plane components of $\mathbf{v}_f^\pm$ at $z = 0$:
\begin{align} \label{eq:velocity_x_bc}
    \Tilde{v}_{x}^\pm(x,y) &= v_{f,x}^\pm(x,y,z=0), \\
    \Tilde{v}_{y}^\pm(x,y) &= v_{f,y}^\pm(x,y,z=0).
\end{align}
In addition, the $z$ velocity of the membrane $\partial_t h(x,y)$ must match the external fluid flow in the $z$ direction, assuming that the fluid does not penetrate the membrane. Therefore,
\begin{equation}
    v_{f,z}^\pm(x,y,z=0) = \partial_t h(x,y,t). \label{eq:velocity_z_bc}
\end{equation}
The leaflet densities obey (approximately; see \cite{seifert1993viscous}) an in-plane continuity equation,
\begin{equation} \label{eq:inplane_continuity}
\frac{\partial \rho^{\pm}}{\partial t}(x,y,t) = -\Tilde{\nabla} \cdot \Tilde{\mathbf{v}}^{\pm}.
\end{equation}

We want to determine, from these hydrodynamic equations, what the equation of motion for the rescaled densities in the top and bottom leaflets $\rho^\pm(x,y,t)$ and the membrane height $h(x,y,t)$ are. This requires us to simultaneously solve for the fluid flow in-plane and out-of-plane. This is easier to do in Fourier space. We also follow Seifert-Langer by using an Ansatz that in-plane flows are only in the $\mathbf{e}_x$ direction. 

We can then write the in-plane monolayer velocities in Fourier space as
\begin{equation}
    \Tilde{v}_x^\pm = \frac{1}{L^2} \sum_q \Tilde{v}_q^\pm e^{i q x}.
\end{equation}

Given this form, the Stokes equations for the $x$ component of the in-plane velocity fields of the monolayers are, in real space,
\begin{align}
    \label{eq:lateralforcebalance1}
    -\partial_x \sigma^{+} + T^+_{xz} + \mu \Tilde{\nabla}^2 \Tilde{v}^+_x - b(\Tilde{v}^+_x - \Tilde{v}_x^-) = 0,\\
    \label{eq:lateralforcebalance2}
    -\partial_x \sigma^{-} - T^-_{xz} + \mu \Tilde{\nabla}^2 \Tilde{v}_x^- + b(\Tilde{v}^+_x - \Tilde{v}_x^-) = 0.
\end{align}
Then, in Fourier space, these can be written as
\begin{align}
    \label{eq:lateralforcebalance1_fourier}
    -i q \sigma^{+}(q) + T^+_{xz}(q) - \mu q^2 \Tilde{v}^+_q - b(\Tilde{v}^+_q - \Tilde{v}_q^-) = 0,\\
    \label{eq:lateralforcebalance2_fourier}
    -i q \sigma^{-}(q) - T^-_{xz}(q) - \mu q^2 \Tilde{v}_q^- + b(\Tilde{v}^+_q - \Tilde{v}_q^-) = 0,
\end{align}
where we have defined $\sigma^{\pm} = 1/L^2 \sum_q \sigma^{\pm}(q) e^{iqx}$. 
The Fourier transforms of the surface pressure are:
\begin{equation}
    \sigma^\pm(q) = \left\{-\frac{\delta E}{\delta \rho^\pm(x,y)}\right\}_q = - L^2 \frac{\partial E}{\partial \rho^\pm_{-q}},
\end{equation}

where $\{\cdots\}_q$ is the Fourier transform, and the second equation can be derived from applying the chain rule on functional derivatives to our convention for Fourier transforms. (Note $\rho_{-q} = (\rho_q)^*$ because $\rho(x,y)$ is a real function.)

The fluid velocity and pressure above and below the membrane can be written in the form
\begin{align} \label{eq:fluid_velocity_ansatz}
    \mathbf{v}_f^\pm(x,y,z) &= \frac{1}{L^2}\sum_q [w^\pm(z)\mathbf{e}_x + u^\pm(z)\mathbf{e}_z]\mathrm{exp}[iqx],\\
    p^\pm(x,y,z) &= \frac{1}{L^2}\sum_q B^\pm(z)\exp[iqx],
\end{align}
where $\mathbf{e}_x$ and $\mathbf{e}_z$ are unit vectors in the $x$ and $z$ directions, and
\begin{align}
    w^\pm(z) &= [((\pm \bar{w} - w) - iu)qz + \bar{w} \pm w]\exp[\mp qz],\\
    u^\pm(z) &= [(-i(\bar{w} \pm w) \pm u)qz + u]\exp[\mp qz],\\
    B^\pm(z) &= 2 \eta q[-i(\bar{w} \pm w) \pm u]\exp[\mp qz],
\end{align}
where $w$, $\bar{w}$, and $u$ are constants to be solved for. Note that these constants will depend on $q$. 

The boundary condition of Eq. (\ref{eq:velocity_x_bc}) then reduces to
\begin{align}
    \Tilde{v}_x^{\pm}(x,y) &= v_{f,x}^\pm(x,y,z=0) \\
   \implies \frac{1}{L^2}\sum_q \tilde{v}_q^\pm e^{iqx} &= \frac{1}{L^2} \sum_q (\bar{w} \pm w)e^{iqx} \\
   \implies \Tilde{v}_q^\pm &= \bar{w} \pm w.
\end{align}
Similarly, the boundary condition of Eq. (\ref{eq:velocity_z_bc}) gives
\begin{align}
    \partial_t h(x,y,t) &= v_{f,z}^\pm(x,y,z=0) \\
    \implies \frac{1}{L^2} \sum_q \partial_t h_q(t) e^{iqx} &= \frac{1}{L^2} \sum_q u^\pm(z=0) e^{iqx} \\
    \implies \partial_t h_q(t) &= u,
\end{align}
where in the last equation, $u = u^\pm(z=0)$ is a constant.

We will now simplify the in-plane force balance equations (Eqs. (\ref{eq:lateralforcebalance1_fourier})--(\ref{eq:lateralforcebalance2_fourier})). To compute the surface pressure gradients, we use the change of variables
\begin{align}
    \sigma^{\pm}(q) &=  -L^2\left(\frac{\partial E}{\partial \rho_q^{{\pm}^*}}\right)
    = -L^2\left(\frac{\partial E}{\partial \rho_q^*} \frac{\partial \rho_q^*}{\partial \rho_q^{{\pm}^*}} +  \frac{\partial E}{\partial \bar{\rho}_q^*}\frac{\partial \bar{\rho}_q^*}{\partial \rho_q^{{\pm}^*}}\right)
    = -L^2\left(\pm \frac{1}{2}\frac{\partial E}{\partial \rho_q^*} + \frac{1}{2}\frac{\partial E}{\partial \bar{\rho}_q^*} \right).
\end{align}

This means that the difference of surface pressures depends on the derivative of energy with the density difference $\rho$, i.e. $\sigma^+(q) - \sigma^-(q) = -L^2 \frac{\partial E}{\partial \rho_q^*}$, and relatedly the sum of the surface pressures will be related to the derivative with respect to $\bar{\rho}$. 
The next term in the force balance requires $T_{xz}^\pm(q)$---the fluid's stress tensor in Fourier space, evaluated at $z = 0$. We will start by evaluating $T_{xz}(x,y,z)$ in real space, plugging in our Ansatz for the fluid velocity (Eq. (\ref{eq:fluid_velocity_ansatz})). The pressure, which only contributes to the diagonal component of the stress tensor, does not show up in the $xz$ component, and so $T^\pm_{xz} =  \eta(\partial_x v_{f,z}^\pm + \partial_z v_{f,x}^\pm)$. We find, then,
\begin{align}
    T^\pm_{xz}(x,y,z) &= \frac{1}{L^2} \sum_q \eta \left( \frac{\partial [u^{\pm}(z)e^{iqx}]}{\partial x} + \frac{\partial[w^{\pm}(z)e^{iqx}]}{\partial z}\right) \notag\\
    \implies T^\pm_{xz}(x,y,z) &= \frac{1}{L^2} \sum_q \eta\left([(-i(\bar{w} \pm w) \pm u)qz + u]\exp[\mp qz]iq \exp[iqx] \notag \right. \\
     & + [((\mp \bar{w} -w)-iu)qz + \bar{w} \pm w](\mp q)\exp[\mp qz]\exp[iqx]  \notag\\
     & + \left.[((\mp \bar{w} -w)-iu)q]\exp[\mp qz]\exp[iqx]\right).
\end{align}
Evaluating this stress tensor at the membrane location, $z = 0$, 
\begin{align}
    T_{xz}^{\pm}(x,y,z=0) &= \frac{1}{L^2} \sum_q \eta(uiq e^{iqx} + (\bar{w} \pm w)(\mp q)e^{iqx} + ((\mp \bar{w} - w) - iu)q e^{iqx}) \\
    &= \frac{1}{L^2} \sum_q  {\mp 2\eta q(\bar{w} \pm w)e^{iqx}} \\
    \implies T^\pm_{xz}(q) &= \mp 2\eta q(\bar{w} \pm w).
\end{align}

Adding the Fourier-space lateral force balance equations, Eq.  (\ref{eq:lateralforcebalance1_fourier}) and Eq. (\ref{eq:lateralforcebalance2_fourier}), we can solve for $\bar{w}$ by plugging in the expressions for $\sigma^\pm(q)$, $T_{xz}^\pm(q)$, and $\tilde{v}_q^\pm$ derived above. Therefore,
\begin{align}
    & -iq \sigma^{+}(q)  -iq \sigma^{-}(q) + T_{xz}^+(q)  - T_{xz}^-(q) - \mu q^2 (\Tilde{v}_q^+ +  \Tilde{v}_q^-) = 0 \\
    &\implies iq L^2 \frac{\partial E}{\partial \rho_q^*} - 4 \eta q \bar{w} - 2\mu q^2 \bar{w} = 0 \\
    &\implies \bar{w} = \frac{iq}{4 \eta q + 2 \mu q^2} L^2 \frac{\partial E}{\partial \rho_q^*}.
\end{align}

Subtracting Eq. (\ref{eq:lateralforcebalance2_fourier}) from Eq. (\ref{eq:lateralforcebalance1_fourier}), we can solve for $w$ as
\begin{align}
 & -iq \sigma^{+}(q)  +iq \sigma^{-}(q) + T_{xz}^+(q)  + T_{xz}^-(q) - \mu q^2 (\Tilde{v}_q^+ -  \Tilde{v}_q^-) -2b(\Tilde{v}_q^+ -  \Tilde{v}_q^-) = 0 \\
 &\implies iq L^2 \frac{\partial E}{\partial \bar{\rho}_q^*} - 4 \eta q w - 2\mu q^2 w -4 b w= 0 \\
 &\implies w = \frac{iq}{4 \eta q + 2 \mu q^2 + 4b} L^2 \frac{\partial E}{\partial \bar{\rho}_q^*}.
\end{align}

We can find the remaining parameter, $u$, and the corresponding dynamics of the height field, from the vertical force balance equation (Eq. (\ref{eq:vertical_force_balance})). In Fourier space, this equation is
\begin{equation}
    -T_{zz}^+(q) + T_{zz}^-(q) = - L^2 \frac{\partial E}{\partial h_q^*}.
\end{equation}

We obtain the $(z,z)$ component of the stress tensor as 
\begin{align}
T_{zz}^\pm(x,y,z) &= -p^\pm\delta_{zz} +         \eta(\partial_z v^\pm_{f,z} +         \partial_z v^\pm_{f,z}) \\
    &= -p^\pm(x,y,z) + 2\eta \partial_z v^\pm_{f,z} \\
    &= -\frac{1}{L^2}\sum_q B^\pm(z) e^{iqx} + 2 \eta \frac{1}{L^2} \sum_q \frac{\partial [u^\pm(z) e^{iqx}]}{\partial z}.
\end{align}
At $z = 0$, plugging in the formulas for $B^\pm(z)$ and $u^\pm(z)$, 
\begin{align}
\label{app:eq:zz_tensors_z0}
    T_{zz}^{\pm}(x,y,z = 0) &= 2\eta \frac{1}{L^2} \sum_q q[-i(\bar{w} \pm w) \pm u]\exp[iqx] + 2\eta\frac{1}{L^2} \sum_q q(-i(\bar{w} \pm w)\exp[iqx]) \\
    &= \mp \frac{1}{L^2} \sum_q 2 \eta q u \exp[iqx]\\
    \implies T_{zz}^\pm(q) &= \mp 2 \eta q u.
\end{align}

The force balance equations in the vertical direction then become
\begin{align}
\label{app:eq:verticalforcebalance1}
    -T_{zz}^+(q) + T_{zz}^-(q) &= -L^2 \frac{\partial E}{\partial h_q^*} \\  
    \implies 4\eta q u  &= -L^2 \frac{\partial E}{\partial h_q^*} \\
\label{app:eq:solving_u}
    \implies u  &= \frac{-1}{4\eta q} L^2\frac{\partial E}{\partial h_q^*}.
\end{align}

Now that we have values for $u$, $w$, and $\bar{w}$, we can find the equations of motion for $h_q$, $\rho_q$, and $\bar{\rho}_q$. We already know that $\partial_t h_q = u$ from the boundary condition at the membrane. The other equations arise from applying the in-plane continuity equation, Eq. (\ref{eq:inplane_continuity}). If we plug in our Fourier transform representation of the functions, we see that
\begin{align}
\frac{1}{L^2} \sum_q \frac{\partial \rho^{\pm}_q}{\partial t} e^{iqx} = -\frac{1}{L^2}\sum_q i q \Tilde{v}_q^{\pm} \\
\implies \frac{\partial \rho^{\pm}_q}{\partial t} = - iq \Tilde{v}_q^{\pm} = -iq(\bar{w}\pm w).
\end{align}

Using the definitions  $\rho_q = (\rho_q^+ - \rho_q^-)/2$, $\bar{\rho}_q = (\rho_q^+ + \rho_q^-)/2$, we then get
\begin{align}
\label{app:eq:h_evolutioneqn}
    \frac{\partial h_q}{\partial t} &= u = -\frac{1}{4\eta q}L^2\frac{\partial E}{\partial h_q^*},\\
\label{app:eq:rhoevolutioneqn}
    \frac{\partial {\rho}_q}{\partial t} &= \frac{1}{2}\left(\frac{\partial \rho_q^+}{\partial t} - \frac{\partial \rho_q^-}{\partial t}\right) = -iq w  = \frac{q^2}{4 \eta q + 2 \mu q^2 + 4b} L^2\frac{\partial E}{\partial \rho_q^*},\\
\label{app:eq:rhobarevolutioneqn}
    \frac{\partial \bar{\rho}_q}{\partial t} &= \frac{1}{2}\left(\frac{\partial \rho_q^+}{\partial t} + \frac{\partial \rho_q^-}{\partial t}\right) = -iq \bar{w} = \frac{q^2}{4 \eta q + 2 \mu q^2} L^2\frac{\partial E}{\partial \bar{\rho}_q^*}.
\end{align}

Equating Eqs. (\ref{app:eq:h_evolutioneqn})--(\ref{app:eq:rhobarevolutioneqn}) to the deterministic part of our equations of motion in Eq. (\ref{app:eq:eqnsofmotion}) gives us the hydrodynamic mobilities, 
\begin{align}
\label{app:eq:hmobility}
    \frac{1}{\Omega_h} &= \frac{1}{4 \eta q},\\
\label{app:eq:rhomobility}
    \frac{1}{\Omega_\rho} &= \frac{q^2}{4 \eta q + 2 \mu q^2 + 4b},\\
\label{app:eqn:rhobarmobility}
    \frac{1}{\Omega_{\bar{\rho}}} &= \frac{q^2}{4 \eta q + 2 \mu q^2}.
\end{align}

The first two of these mobilities could be derived simply by requiring that the deterministic equations of motion matched those of Seifert and Langer. The third is not quite the result of Seifert and Langer, as we have neglected inertia in the $\bar{\rho}$ mode. Although this assumption influences the dynamics, the resultant thermal equilibrium distribution of $\bar{\rho}$ does not change.

\subsubsection{Deriving the equations of motion using functional derivatives of the membrane energy \texorpdfstring{$E$}{E}}

The membrane energy $E$ in Eq. (\ref{EnergyFourierSpace}) may be expressed as 
\begin{equation*}
    E = \sum_{\mathbf{q}} \frac{1}{2}\frac{1}{L^2}(\Tilde{\kappa}q^4h_\qv h_{-\qv} - 2kdq^2\rho_{-\qv}h_\qv - 2kdq^2h_{-\qv}\rho_\qv + 2k\rho_\qv \rho_{-\qv} + 2k\bar{\rho}_\qv \bar{\rho}_{-\qv}) + E_\text{adh}.
\end{equation*}

We show here that we get the Seifert-Langer equations of motion back in the limit of zero adhesion ($E_\textrm{adh} = 0$). Differentiating $E$ with respect to $\rho_{-\jv}$, such that $\jv$ is an arbitrary Fourier index,
\begin{equation}
        \frac{\partial E}{\partial \rho_{-\jv}} = \sum_\mathbf{q} \frac{1}{2}\frac{1}{L^2}(-2kdq^2h_\qv \delta_{-\jv, -\qv} - 2kdq^2h_{-\qv} \delta_{-\jv, \qv} + 2k\rho_\qv \delta_{-\jv,-\qv} + 2k\rho_{-\qv}\delta_{-\jv,\qv}),
\end{equation}
where $\delta_{-\jv, -\qv}$ and $\delta_{-\jv, \qv}$ are Kronecker delta terms. 

Consequently, 
\begin{equation}
    \frac{\partial E}{\partial \rho_{-\jv}} =  \frac{1}{2L^2}(-2kdj^2h_\jv - 2kdj^2h_\jv + 2k\rho_\jv + 2k\rho_\jv).
\end{equation}

Since $\jv$ is an arbitrary Fourier index, we may equivalently reformulate in terms of $\qv$ (noting that $\frac{\partial E}{\partial \rho_{-\jv}} = \frac{\partial E}{\partial \rho_\jv^*})$ as

\begin{equation}
\label{app:eq:E_rhoq_funcderiv}
    \frac{\partial E}{\partial \rho_\qv^*} =  \frac{1}{L^2}(-2kdq^2h_\qv + 2k\rho_\qv).
\end{equation}

Similarly, it can be shown that
\begin{align}
\label{app:eq:E_rhobarq_funcderiv}
    \frac{\partial E}{\partial \bar{\rho}_\qv^*} &= \frac{1}{L^2}(2k\bar{\rho}_\qv),\\
\label{app:eq:E_hq_funcderiv}
    \frac{\partial E}{\partial h_\qv^*} &= \frac{1}{L^2}(\Tilde{\kappa}q^4h_\qv + 2kdq^2\rho_\qv).
\end{align}

Substituting Eqs. (\ref{app:eq:E_rhoq_funcderiv})--(\ref{app:eq:E_hq_funcderiv}) as appropriate to the dynamical equations in Eqs. (\ref{app:eq:h_evolutioneqn})--(\ref{app:eq:rhobarevolutioneqn}), we have
\begin{equation}
\label{app:eq:dynamicalequations_full}
    \frac{\partial}{\partial t}\begin{pmatrix} h_\qv \\ \rho_\qv \\ \bar{\rho}_\qv \end{pmatrix} \equiv -\mathbf{M}\begin{pmatrix} h_\qv \\ \rho_\qv \\ \bar{\rho}_\qv \end{pmatrix} = -\begin{pmatrix} \frac{\tilde{\kappa}q^4 }{4\eta q} && \frac{-qkd}{2\eta} && 0 \\ \frac{-kdq^4}{2b + 2\eta q + \mu q^2} && \frac{kq^2}{2b + 2\eta q + \mu q^2} && 0 \\ 0 && 0 && \frac{kq^2}{2 \eta q + \mu q^2} \end{pmatrix}\begin{pmatrix}h_\qv \\ \rho_\qv \\ \bar{\rho}_\qv \end{pmatrix}.
\end{equation}

These are (except for the $\bar{\rho}$ mode, as noted above) the equations of motion from \cite{seifert1993viscous}, who have neglected membrane-substrate adhesion ($\gamma = 0$). It is also straightforward to show that if we have adhesion to a flat surface, then $E_\textrm{adh} = \frac{\gamma}{2}\int \mathrm{d}\rb h^2(\rb) = \frac{\gamma}{2 L^2} \sum_\qv |h_\qv|^2$, which will lead to an added term $\gamma/4 \eta q$ to the $\mathbf{M}_{hh}$ term as shown in Eq. (\ref{app:eq:DynamicalMatrix_h_rho}).

\subsection{Numerical evaluation of the equation of motion\label{app:sec:numericalEvaluationMotion}}

 To numerically solve the stochastic equations of motion in Eq. (\ref{app:eq:eqnsofmotion}), we take the simplest approach, the Euler-Maruyama method \cite{kloeden2012numerical}. Let us take the dynamics of the height variable $h_\mathbf{q}$ as an example. Integrating from a time $t$ to $t+\Delta t$, this equation becomes
\begin{equation}
    h_{\mathbf{q}}(t+\Delta t)-    h_{\mathbf{q}}(t) = - \int_{t}^{t+\Delta t} \mathrm{d}t' \frac{L^2}{\Omega_h} \frac{\partial E}{\partial h_{\qv}^*} + \int_{t}^{t+\Delta t} \mathrm{d}t' \xi_\qv(t').
\end{equation}
The term without the Langevin noise can be approximated simply with the usual Euler rule, and we define a new function $\Xi_\qv(\Delta t) \equiv \int_{t}^{t+\Delta t} \mathrm{d}t' \xi_\qv(t')$, so we have
\begin{equation}
    h_{\mathbf{q}}(t+\Delta t)     = h_{\mathbf{q}}(t) - \Delta t \frac{L^2}{\Omega_h} \frac{\partial E}{\partial h_{\qv}^*} + \Xi_\qv(\Delta t).
    \label{HeightEvolutionAlgorithm}
\end{equation}
Here, $\Xi(\Delta t)$ is a Gaussian random variable with mean zero and a variance that will depend on the timestep $\Delta t$. We can compute its variance straightforwardly by using the correlation of $\xi_{\qv}(t)$ given above, 
    $\langle \xi_\mathbf{q}(t) \xi_\mathbf{q'}(t') \rangle = 2D_h \delta_\mathbf{q,-q'}\delta(t - t')$, such that 
\begin{align}
\label{app:eq:XiVariance}
\langle |\Xi_\qv(\Delta{t})|^2\rangle    =\langle \Xi_\qv(\Delta{t})\Xi_{-\qv}(\Delta{t})\rangle &= \left\langle   \int_t^{t+\Delta{t}} \mathrm{d}t \xi_\qv(t) \int_{t'}^{t' + \Delta{t'}} \mathrm{d}t' \xi_{-\qv}(t')  \right\rangle \notag\\
     &= \int_{t}^{t+ \Delta t} \int_{t'}^{t' + \Delta t'}  \left \langle {\xi_\qv(t) \xi_{-\qv}(t')}  \right \rangle \mathrm{d}t \mathrm{d}t' \notag\\
    &=  \int_{t}^{t + \Delta t}\int_{t'}^{t' + \Delta t} {2 D_h \delta(t - t')} \mathrm{d}t \mathrm{d}t'
    \notag\\
     &= \frac{2 k_B T L^2}{\Omega_h} \int_{t}^{t + \Delta {t}}  \mathrm{d}t \notag\\
    &=  \frac{2 k_B T L^2}{\Omega_h} \Delta t.
\end{align}

Similarly, to integrate over the Langevin noises associated to fluctuations in lipid densities, we define $\Theta_{\qv}(\Delta t) \equiv \int_t^{t+\Delta t} \mathrm{d}t' \zeta_{\qv}(t')$ and $\Upsilon_{\qv}(\Delta t) \equiv \int_t^{t+\Delta t} \mathrm{d}t' \chi_{\qv}(t')$ and derive
\begin{align}
\label{app:eq:ThetaVariance}
    \langle |\Theta_\qv(\Delta{t})|^2\rangle &= \frac{2 k_B T L^2}{\Omega_\rho} \Delta t,\\
    \langle
\label{app:eq:UpsilonVariance}
|\Upsilon_\qv(\Delta{t})|^2\rangle &= \frac{2 k_B T L^2}{\Omega_{\bar{\rho}}} \Delta t.
\end{align}

In ordinary Brownian dynamics, we would generate a real random variable with a variance given by Eqs. (\ref{app:eq:XiVariance})--(\ref{app:eq:UpsilonVariance}) in order to evolve the equations of motion. However, our Fourier modes are complex, except for $\qv = (0,0)$ (see Appendix \ref{app:FourierConventions}).

We sample the real and imaginary parts of our Fourier modes separately, with variances so that their absolute value obeys Eqs. (\ref{app:eq:XiVariance})--(\ref{app:eq:UpsilonVariance}), such that
\begin{align}
\label{app:eq:samplingXi}
    \Re[\Xi_{\qv}(\Delta t)]; \Im[\Xi_{\qv}(\Delta t)] &\sim \mathcal{N}\left(0, \frac{k_B T L^2}{\Omega_h}\Delta t\right),\\
\label{app:eq:samplingTheta}
    \Re[\Theta_{\qv}(\Delta t)]; \Im[\Theta_{\qv}(\Delta t)] &\sim \mathcal{N}\left(0, \frac{k_B T L^2}{\Omega_\rho}\Delta t\right),\\
\label{app:eq:samplingUpsilon}
    \Re[\Upsilon_{\qv}(\Delta t)]; \Im[\Upsilon_{\qv}(\Delta t)] &\sim \mathcal{N}\left(0, \frac{k_B T L^2}{\Omega_{\bar{\rho}}}\Delta t\right),
\end{align}
where $\mathcal{N}(\mu,\sigma^2)$ is a Gaussian distribution with mean $\mu$ and variance $\sigma^2$. This is essentially the approach used by \cite{lin2004brownian,camley2010dynamic,camley2014fluctuating}, etc.

The equation evolving the membrane's height in Fourier space, Eq. (\ref{HeightEvolutionAlgorithm}), includes a term $\frac{-\partial E}{\partial h_{\qv}^*}$, which includes the forces acting on the membrane from both deformation forces (bending and monolayer compression) in addition to the forces due to membrane-bead adhesion. The deformation forces are as shown earlier in Eq. (\ref{app:eq:E_hq_funcderiv}). When we simulate a membrane-bead system, we explicitly compute the force due to membrane-bead adhesion by Fast Fourier transforming the functional derivative $-\delta E_\text{adh}/\delta h(\mathbf{r})$, which is computed in real space. Therefore at time $t$, 
\begin{equation}
\label{ForceHeightEvolution}
 \frac{-\partial E}{\partial h_{\qv}^*}
= \frac{1}{L^2}(-\tilde{\kappa}q^4 h_\mathbf{q}(t) + 2kd q^2 \rho_\mathbf{q}(t)) + \left\{ \frac{-\delta E_\text{adh}(\mathbf{r})}{\delta h(\mathbf{r})}\right\}_\qv , 
\end{equation}

where $\left\{\cdots\right\}_\qv$ indicates the Fourier transform. The numerical algorithm for evolving $h_\mathbf{q}$ is obtained by writing Eq. (\ref{HeightEvolutionAlgorithm}) explicitly as
\begin{align}
     h_{\mathbf{q}}(t+\Delta t)     &= h_{\mathbf{q}}(t)    +  \Delta t \frac{L^2}{\Omega_h} \left[ \frac{1}{L^2}(-\tilde{\kappa}q^4 h_\mathbf{q}(t) + 2kd q^2 \rho_\mathbf{q}(t)) + \left\{ \frac{-\delta E_\text{adh}(\mathbf{r})}{\delta h(\mathbf{r})}\right\}_\qv \right] + \Xi_\qv(\Delta t)   \\
    &= h_{\mathbf{q}}(t)    +  \frac{\Delta t}{4 \eta q} \left[ -\tilde{\kappa}q^4 h_\mathbf{q}(t) + 2kd q^2 \rho_\mathbf{q}(t) + L^2\left\{ \frac{-\delta E_\text{adh}(\mathbf{r})}{\delta h(\mathbf{r})}\right\}_\qv \right] + \Xi_\qv(\Delta t). \label{HeightEvolutionAlgorithmwithFFT}
\end{align}

The corresponding numerical algorithms for evolving $\rho_\mathbf{q}$ and $\bar{\rho}_\mathbf{q}$ are
\begin{equation}
    \label{DensityFluctuationAlgorithmRho}
    \rho_{\mathbf{q}}(t + \Delta t) = \rho_{\mathbf{q}}(t) + \Delta{t}\frac{kd q^4h_{\mathbf{q}}(t) - k q^2 \rho_{\mathbf{q}}(t)}{2b + 2\eta q + \mu q^2} + \Theta_\mathbf{q}(\Delta t),
\end{equation}
\begin{equation}
    \label{DensityFluctuationAlgorithmRhoBar}
    \bar{\rho}_\mathbf{q}(t + \Delta t) = \bar{\rho}_\mathbf{q}(t) + \Delta{t}\frac{-kq^2 \bar{\rho}_\mathbf{q}(t)}{2 \eta q + \mu q^2} + \Upsilon_\mathbf{q}(\Delta t).
\end{equation}

With the exception of $\mathbf{q} = (0,0)$, for both the real and imaginary components of the remaining independent modes, the thermal noise $\Xi_\mathbf{q}({\Delta t})$ is sampled from a Gaussian distribution with a mean of zero and variance $\frac{L^2k_B T \Delta t}{4 \eta q}$, $\Theta_\mathbf{q}(\Delta t)$ is sampled from a Gaussian distribution with mean zero and variance $\frac{L^2 k_B T q^2 \Delta t}{4b + 4\eta q + 2 \mu q^2}$, and $\Upsilon_\mathbf{q}(\Delta t)$ is sampled from a Gaussian distribution with mean zero and variance $\frac{L^2 k_B T q^2 \Delta t}{4 \eta q + 2 \mu q^2}$. 

 To allow for faster simulation equilibration, we set the membrane's initial height field $h(\mathbf{r})$ at $t = 0$ to be equal to the bead's height $h_\textrm{bead}(\mathbf{r})$, and then Fast Fourier Transform this to obtain $h_\qv(t = 0)$. The lipid density $\rho_\qv$ is initialized at zero and allowed to evolve due to its coupling with membrane height; $\bar{\rho}_\qv$ is also initialized at zero. After initializing $h_\mathbf{q}$, $\rho_\mathbf{q}$ and $\bar{\rho}_\mathbf{q}$, the zeroth modes of each of these variables are not evolved further either due to the deterministic or stochastic contributions. To avoid division by zero when $q = 0$, the zeroth modes of the mobilities $1/\Omega_h = 1/4\eta q$ and $1/\Omega_{\bar{\rho}} = q^2/(2\eta q + \mu q^2)$ must be set to zero when evaluating these algorithms. After each successive $\Delta t$, the arrays computed for $h_\mathbf{q}(t + \Delta t), \rho_\mathbf{q}(t + \Delta t)$, and $\bar{\rho}_\mathbf{q}(t + \Delta t)$ are inverse Fast Fourier transformed to store their corresponding real-space values.

\subsubsection{Choosing parameters appropriately for simulation convergence \label{app:choosing_parameters}}

 Choosing a small $\Delta t$ is necessary for simulation convergence, but a $\Delta t$ that is too small prolongs the computation time required since a greater number of timesteps must be simulated for the same $t_\text{sim}$. The following are useful guidelines when assessing whether a chosen set of dynamical parameters are practically feasible for the desired simulation.

 Consider the dynamical equations for evolving $h_q$, $\rho_q$, and $\bar{\rho}$ for a membrane adhered to a planar substrate,
\begin{equation}
    \label{app:eq:DynamicalEqn_Convergence}
     \frac{\partial}{\partial t}\begin{pmatrix} h_\qv \\ \rho_\qv \\ \bar{\rho}_\qv \end{pmatrix} \equiv -\mathcal{M}\begin{pmatrix} h_\qv \\ \rho_\qv \\ \bar{\rho}_\qv \end{pmatrix}
\end{equation}
\begin{equation}
    \label{app:eq:DynamicalMatrix_h_rho}
    \mathcal{M} \equiv \begin{pmatrix} \frac{\tilde{\kappa}q^4 + \gamma }{4\eta q} && \frac{-qkd}{2\eta} && 0 \\ \frac{-kdq^4}{2b + 2\eta q + \mu q^2} && \frac{kq^2}{2b + 2\eta q + \mu q^2} && 0 \\ 0 && 0 && \frac{kq^2}{2 \eta q + \mu q^2} \end{pmatrix},
\end{equation}
where we note the inclusion of adhesion strength $\gamma$ in $\mathcal{M}_{hh}$.

\begin{figure}[!htbp]
    \centering
    \subfigure[]{\includegraphics[width=0.32\textwidth]{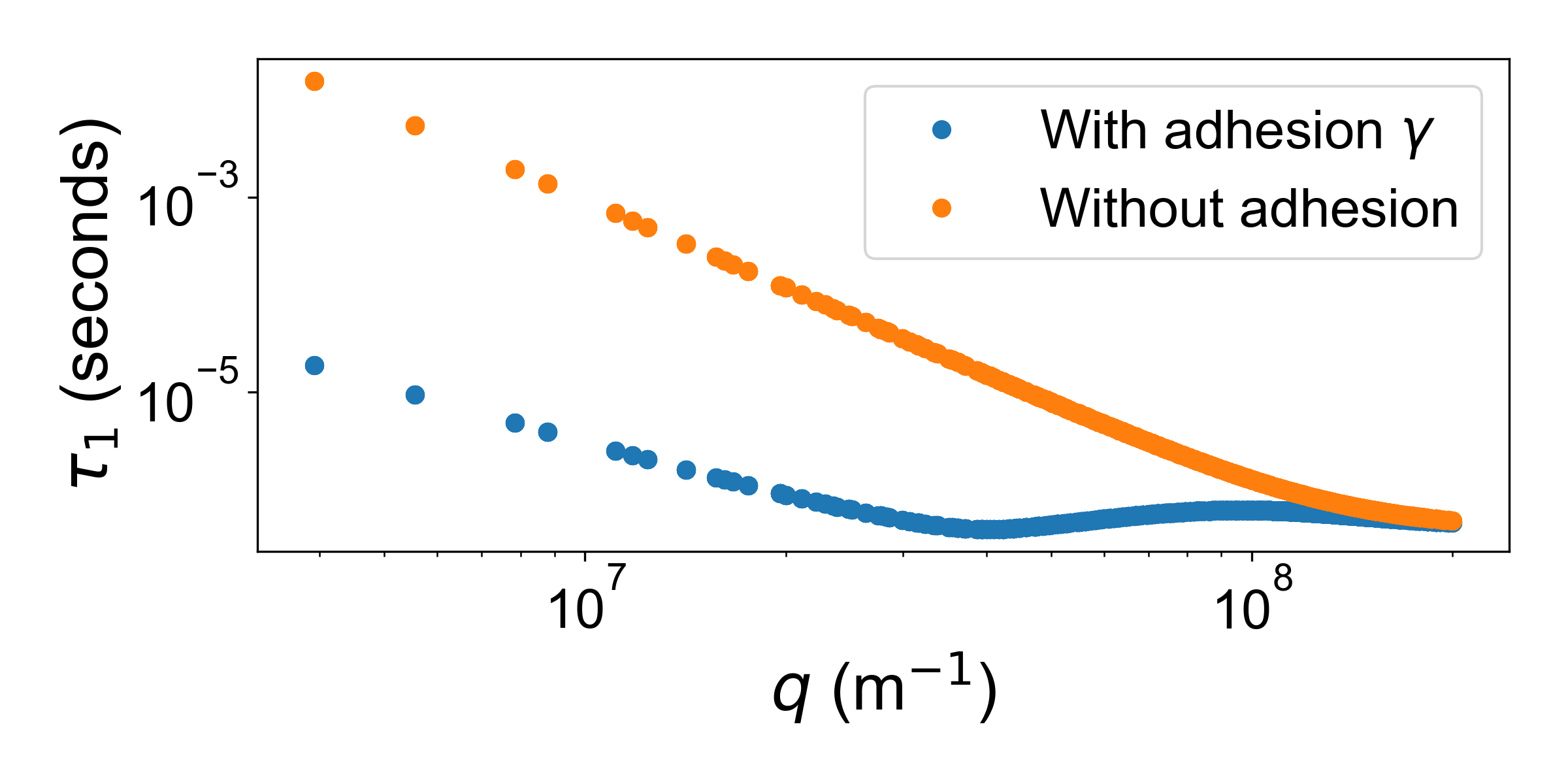}}
    \subfigure[]{\includegraphics[width=0.32\textwidth]{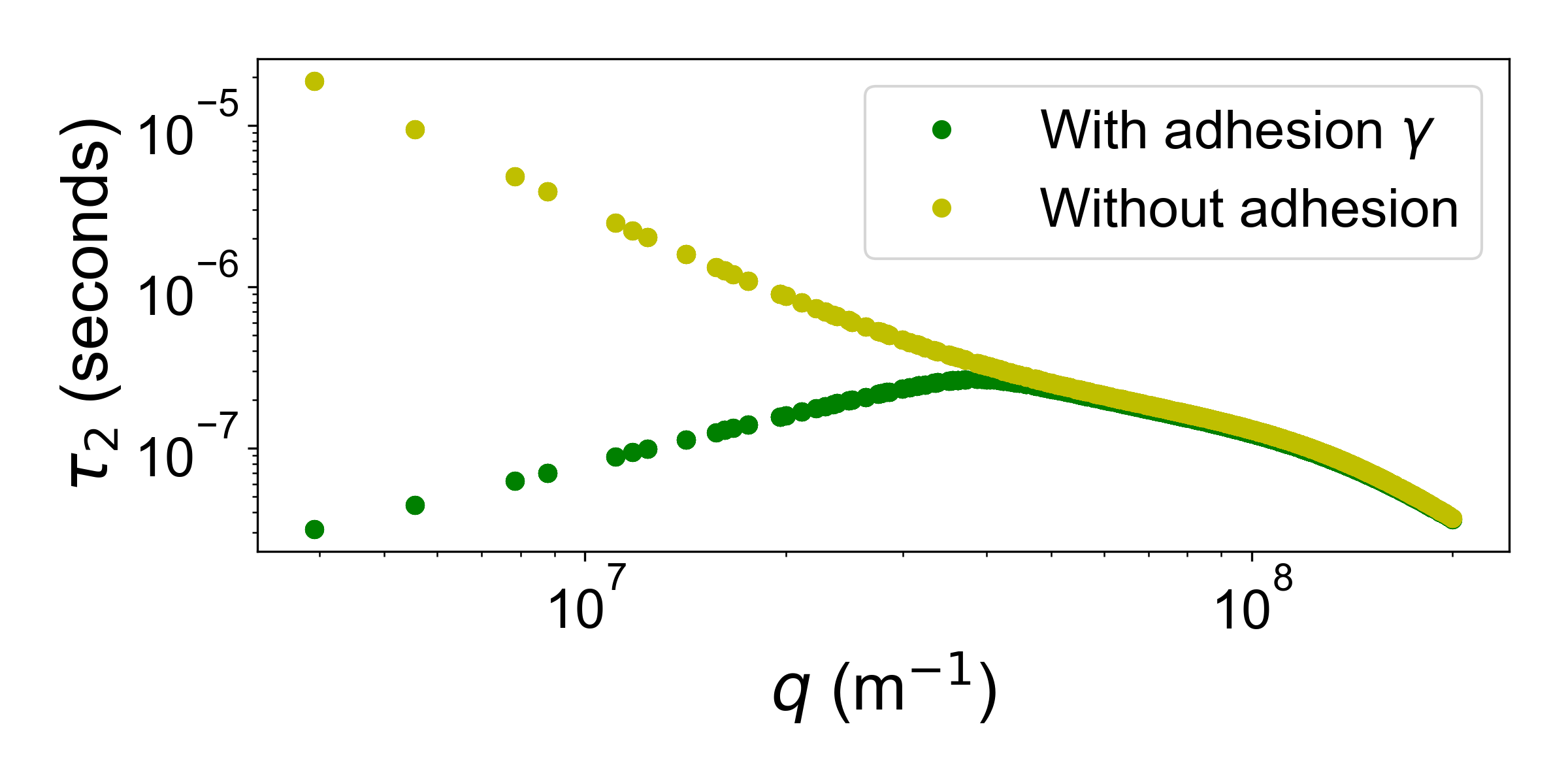}}
    \subfigure[]{\includegraphics[width=0.32\textwidth]{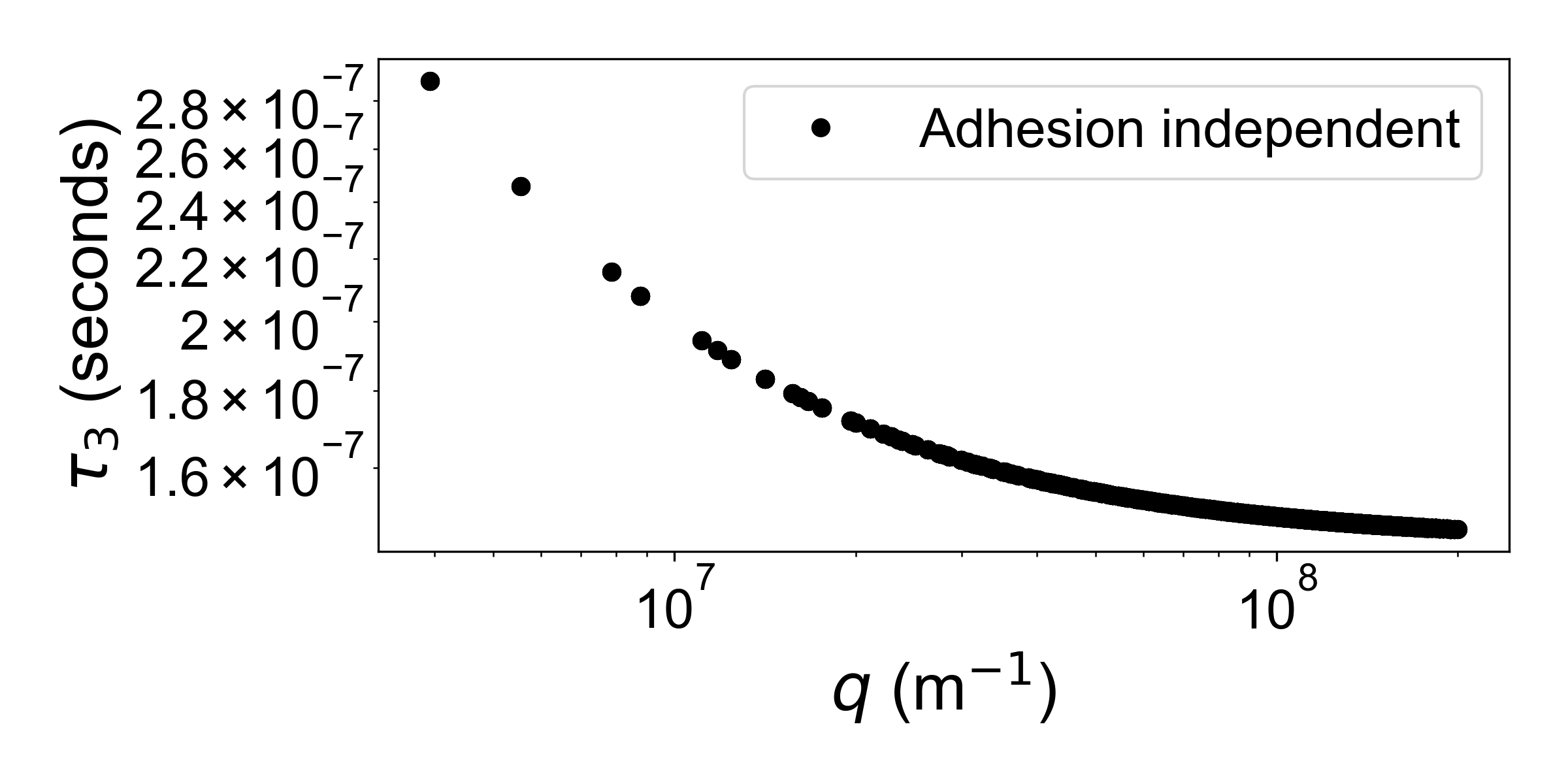}}
    \caption{A comparison of the relaxation times $\tau = 1/\lambda(q)$ for $\lambda_1(q)$, $\lambda_2(q)$, and $\lambda_3(q)$, demonstrating the effects of adhesion on the relaxation of modes. The presence of membrane-substrate adhesion ($\gamma = 10^{13}$ J/m$^4$) results in faster relaxation of low-$q$ modes for two of the eigenvalues, while the third eigenvalue is independent of adhesion.}
    \label{RelaxationModesAdhesion}
\end{figure}

The eigenvalues for $\mathcal{M}$ can be obtained symbolically by matrix diagonalization (we used the \texttt{sympy} package in Python). This results in three sets of eigenvalues, which we denote as $\lambda_1(q)$, $\lambda_2(q)$, and $\lambda_3(q)$. These eigenvalues correspond to the relaxation frequencies of the modes $q$, and depend on their relaxation times $\tau_q$ as $\lambda(q) \sim 1/\tau_q$. In Fig. \ref{RelaxationModesAdhesion}, we plot the relaxation times for each set of eigenvalues. For $\tau_1$ and $\tau_2$, the presence of strong adhesion can allow the low-$q$ (large wavelength) modes to relax orders of magnitude more quickly.

For simulation convergence, the total simulation time $t_\text{sim}$ must be at least a few times longer than the relaxation time of the slowest relaxing mode. Also, the timestep $\Delta t$ must be a fraction of the relaxation time of the fastest relaxing mode for numerical stability of the integration algorithm. Therefore,
\begin{gather*}
    \lambda_\text{slowest} t_\text{sim} \gtrsim \textrm{5--10},\\
    \lambda_\text{fastest} \Delta t \ll 1.
\end{gather*}
Practically, it suffices to have $\lambda_\text{fastest} \Delta t \approx 0.2$. $\lambda_\text{slowest}$ and $\lambda_\text{fastest}$ can be obtained as the minimum and maximum values of $\lambda_i(q)$.

\section{Cross-sectional profiles of simulated membranes adhered to a small bead with different adhesion strengths\label{crosssection_smallbead_adhesion}}

\begin{figure}[ht]
     \centering
     \begin{subfigure}
         \centering
         \includegraphics[width=0.42\textwidth]{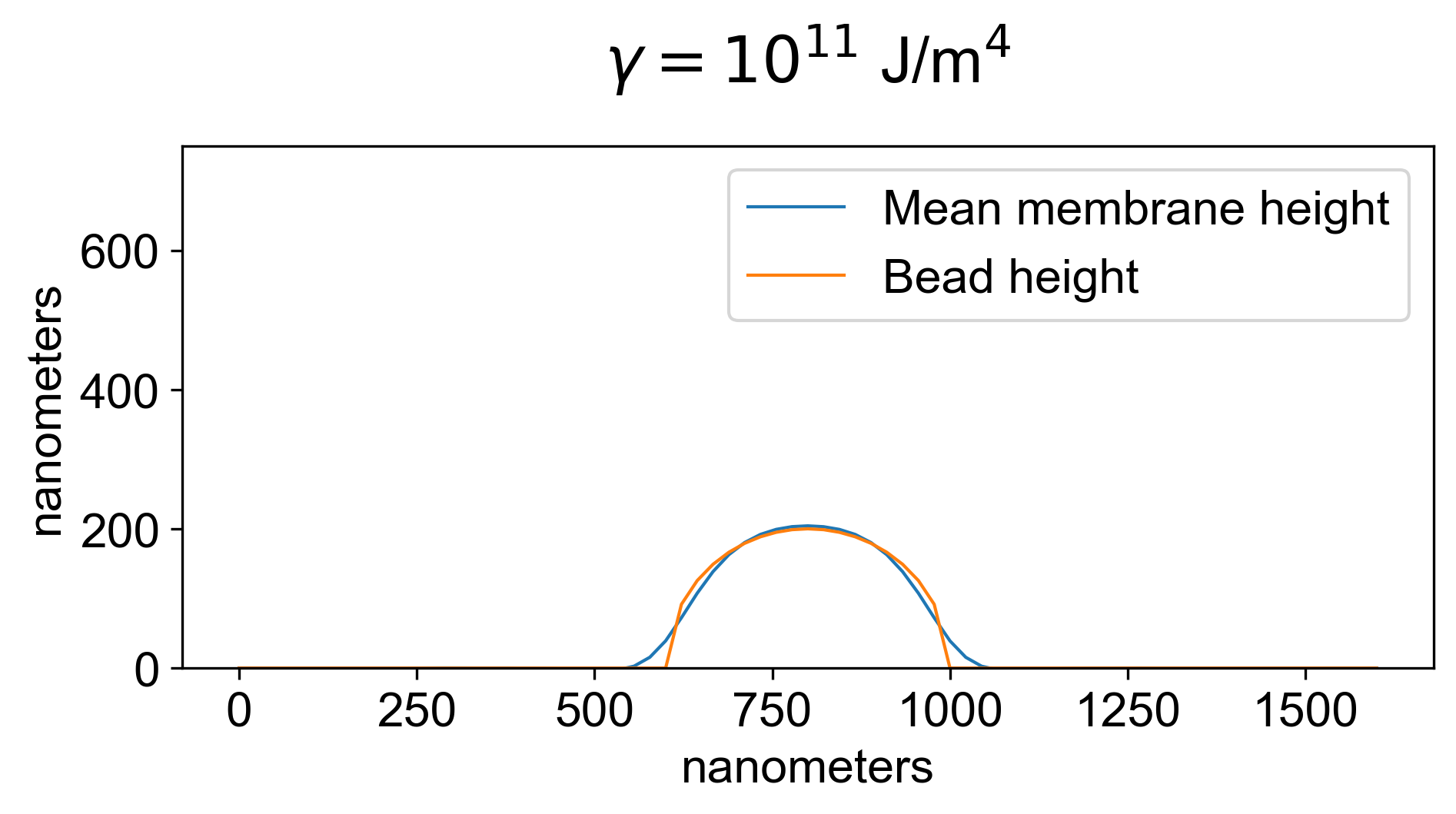}
     \end{subfigure}
     \begin{subfigure}
         \centering
         \includegraphics[width=0.42\textwidth]{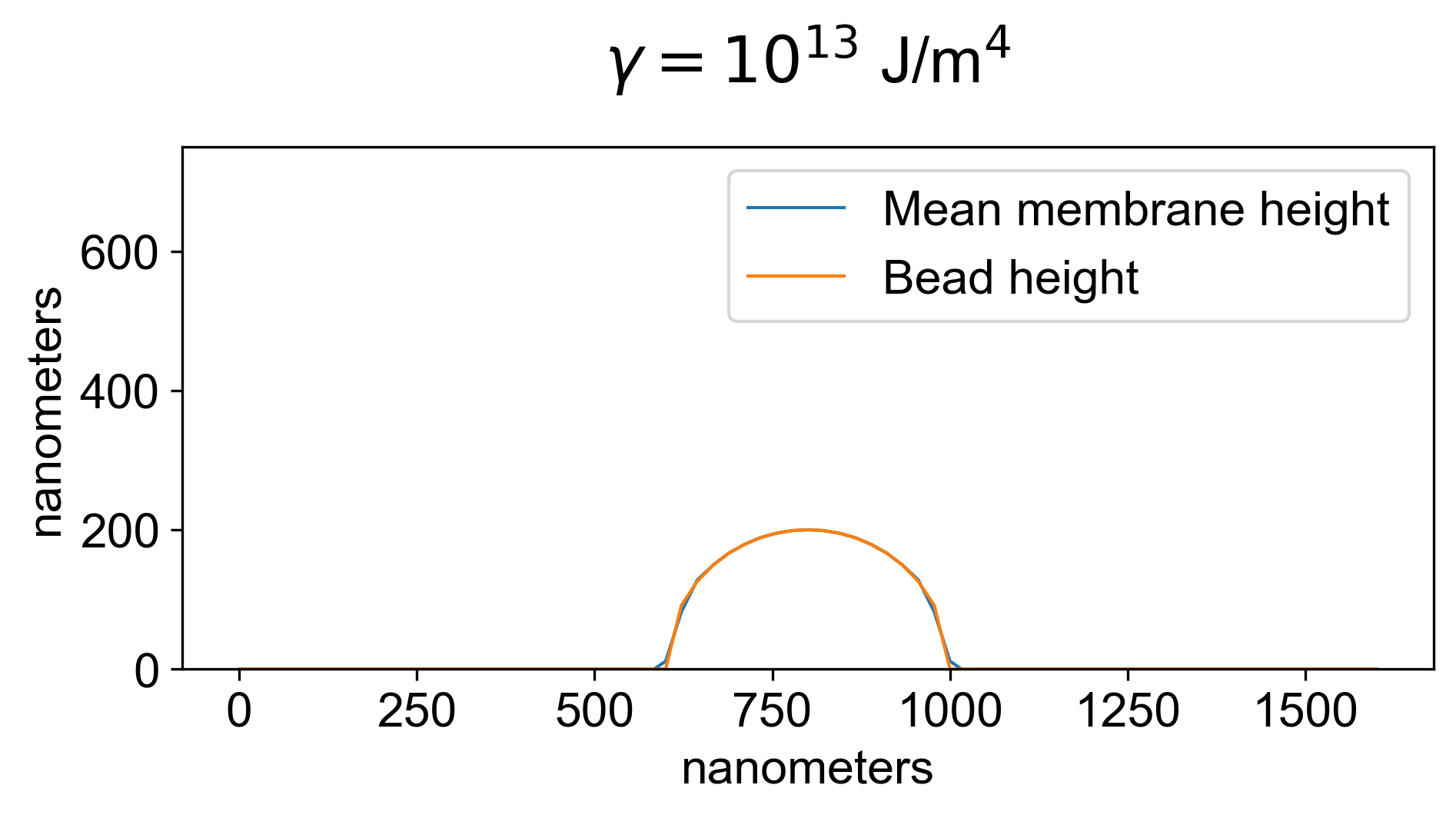}
     \end{subfigure}
     \caption{Cross-sectional profile of bead height for a bead of diameter 0.4 $\mu$m and the mean height of a simulated membrane adhered to this bead. (Top): with weak adhesion ($\gamma = 10^{11}$ J/m$^4$), the simulated membrane overestimates the curvature at the center of the bead due to a height discrepancy of 4.25 nm at the bead center (Bottom): with stronger adhesion ($\gamma = 10^{13}$ J/m$^4$), the membrane wraps around the bead with a negligible 0.11 nm central overhang.}
     \label{Membraneprofile_SmallBead}
\end{figure}

To perfectly adhere a membrane onto a hemispherical bead and the flat substrate around it would require large bending forces at the periphery of the bead due to the sharp curve. If the membrane-substrate adhesion strength is too weak, the membrane does not exactly follow the bead shape, even on average. We show in Fig. \ref{Membraneprofile_SmallBead} that, for a small bead of $R = 200$ nm, the simulated membrane's curvature at the center of the bead deviates from $C_\textrm{bead} = 1/R$ when $\gamma$ is weak. This discrepancy between bead shape and the average membrane shape is most relevant at small bead sizes and at weak adhesion strengths, and leads to the deviation between theory and simulated SNR in Fig. \ref{fig:SNR_varyingbeads}. We note that the membrane’s average profile can be below the ``bead height'' line at $\gamma = 10^{11}$ J/m$^4$ in Fig. \ref{Membraneprofile_SmallBead}. The bead height line indicates the energy minimum of the harmonic potential---this does not necessarily indicate that the membrane is crossing the bead itself. At these low adhesions, using a more complex potential with a hard core might be necessary in order to prevent the membrane from penetrating the bead. However, we expect that the distributions of height from the harmonic potential are a good approximation to distributions for fluctuations in the vicinity of the substrate \cite{schmidt2014signature}.

\section{\label{app:SNRLowAdhAppendix}SNR\texorpdfstring{$_C$}{C} and SNR\texorpdfstring{$_\rho$}{rho} in the low-adhesion regime}

\begin{figure}[ht]
     \centering
     \begin{subfigure}
         \centering
         \includegraphics[width=0.35\textwidth]{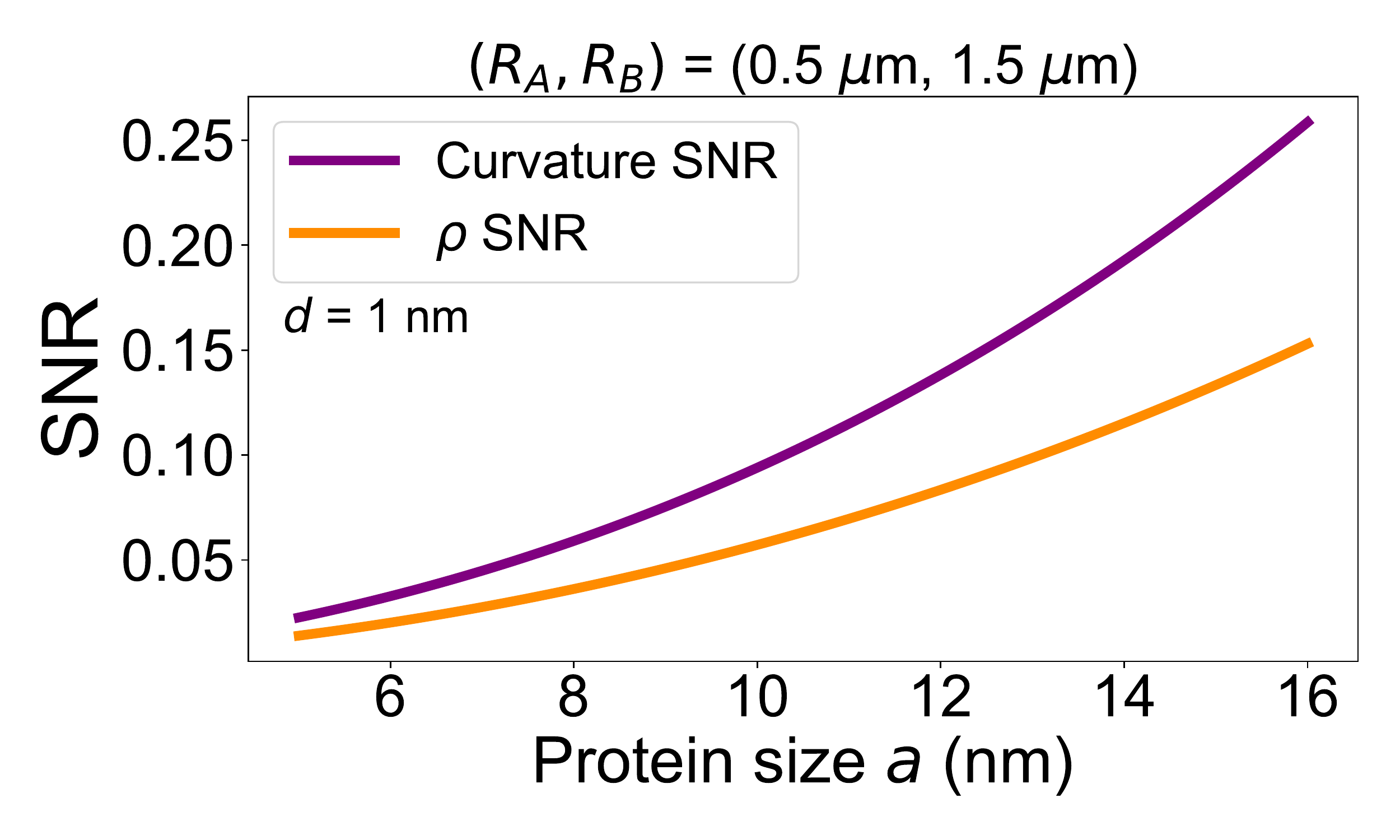}
     \end{subfigure}
     \begin{subfigure}
         \centering
         \includegraphics[width=0.35\textwidth]{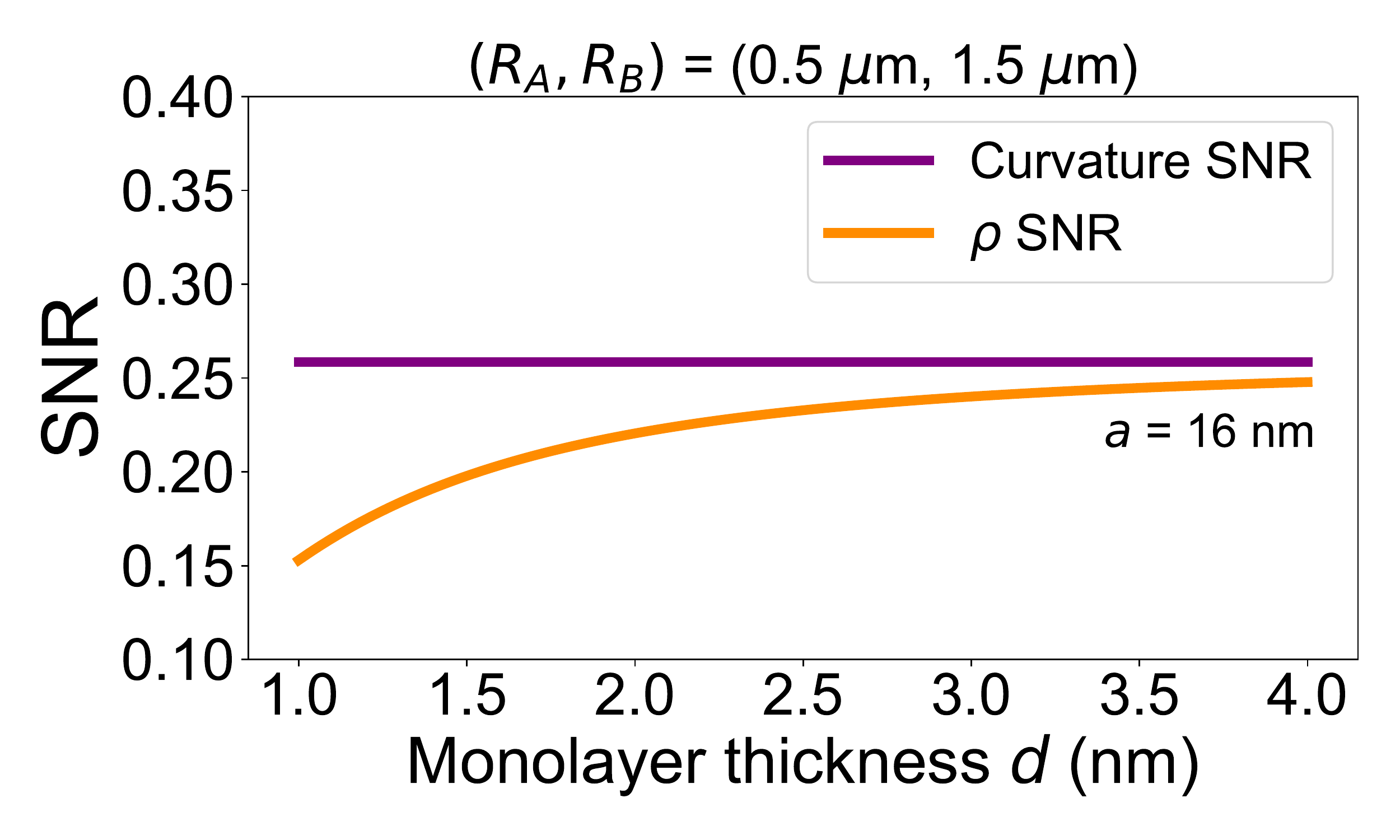}
     \end{subfigure}
     \caption{Sensing SNR in the low-adhesion regime for $\gamma = 10^9$ J/m$^4$. (Top) SNR vs. protein size (when $d$ = 1 nm) and (Bottom) SNR vs. monolayer thickness (when $a$ = 16 nm) for  $(R_A, R_B)$ = (0.5 $\mu$m, 1.5 $\mu$m). The curvature and $\rho$ variances for these SNR values are computed with Eq. (\ref{ContLimCurvVar_DimLess}) and Eq. (\ref{ContLimRhoVar_DimLess}).}
     \label{fig:SNR_BigBeads_ProtMono_LowG}
\end{figure}

In the main text, we primarily use our best estimate of $\gamma$ for the supported lipid bilayer systems of  \cite{cannon2019amphipathic,shi2022kinetic}. Here, we show some corresponding plots of SNR in the low-adhesion regime. Membrane-cytoskeleton confinement in cells have reported $\gamma \sim 10^{9}$ J/m$^4$ \cite{biswas2017mapping}. This corresponds to SNR$_C \approx 0.25$ (as shown in Fig. \ref{fig:SNR_BigBeads_ProtMono_LowG}). Assuming a negligible basal association rate, Eq. (\ref{maxratioSNRC}) implies a maximal association ratio of $ \exp(0.25) = 1.3$ to the preferred radius when distinguishing between cells of radii $(R_A, R_B) = (0.5, 1.5)$ $\mu$m.

\section{\label{app:justifyAdhesionStrength}Justification for adhesion strength \texorpdfstring{$\gamma$}{gamma}}

We estimate the membrane-substrate adhesion parameter $\gamma$ in our model using the approach of \cite{swain1999influence}. They model the energy per unit area of membrane at height $h$ as 
\begin{equation}
    V = -\frac{A}{12 \pi} \left( \frac{1}{h^2} - \frac{1}{(h+\delta)^2} \right) + \beta e^{-\alpha h},
\end{equation}
where the first term is the van der Waals interaction between a bilayer of thickness $\delta = 3.8$ nm and the substrate, with $A \approx 2.6 \times 10^{-21}$ J as the Hamaker constant \cite{israelachvili1994strength, israelachvili2011intermolecular}. The second term is a phenomenological form for the hydration force with $\beta \approx 0.93 $ J/m$^2$ and $\alpha^{-1} \approx 0.22$ nm. Instead of this complex potential, we have used a harmonic approximation to it about an equilibrium height $h_0$,
\begin{equation}
    V(h) \approx V_0 + \frac{1}{2} V''(h_0) (h-h_0)^2,
\end{equation}
where $V''(h_0) = \gamma$, corresponding to our adhesion strength. Note, again, that $V$ here is an energy per unit area, so $\gamma$ has units of J/m$^4$. 
Using the parameter values of  \cite{swain1999influence}, stated above, we find that the minimum energy distance is $h_0 \approx 3.02$ nm, and find $V''(h_0) \approx 1.6 \times 10^{13} \text{ J/m}^4$. We view this as the roughly correct order of magnitude for a supported lipid bilayer, which is strongly adherent to the substrate. However, it is possible that this adhesion energy could be a little higher in some SLBs. Experimental data indicates that the hydration layer can be as thin as $1$ nm \cite{zwang2010quantification}; if this arose from a larger Hamaker constant or lower repulsion energy, that would increase the value of $\gamma$. There is also evidence suggesting that adhesion strengths can vary over orders of magnitude in different contexts. Large membrane vesicles adhere to glass substrates relatively weakly, with corresponding $\gamma$ values of $10^7$ J/m$^4$ \cite{schmidt2014signature}, and whole-cell experiments have reported membrane-cytoskeleton adhesion strengths in the order of $10^{9} - 10^{10}$ J/m$^4$ \cite{biswas2017mapping}.

\section{\label{app:ImprovedFitsCurvatureThreshold}Curvature threshold model at higher membrane adhesion strengths}

\begin{figure}[ht]
    \includegraphics[width=0.47\textwidth]{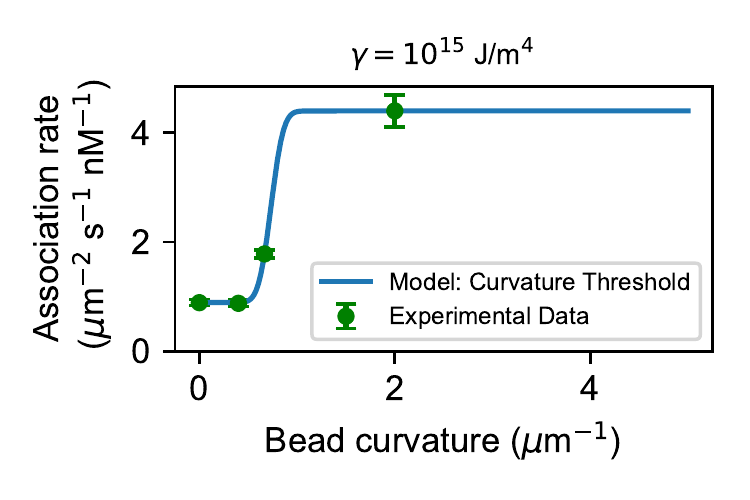}
    \caption{Curvature threshold model fits to experimental data in \cite{shi2022kinetic} for varying bead curvatures when $\gamma = 10^{15}$ J/m$^4$. With $A_0 = 0.892$ $\mu$m$^{-2}$ s$^{-1}$ nM$^{-1}$, the fit parameters obtained with non-linear least squares fits are: $A_C \approx 3.505$ $\mu$m$^{-2}$ s$^{-1}$ nM$^{-1}$ and $C_\textrm{thresh} \approx 0.733$ $\mu$m$^{-1}$. Other physical parameters: Table \ref{tab:ModelParameters}.}
    \label{app:fig:thresholdmodel_highgamma}
\end{figure}

Although we choose $\gamma = 10^{13}$ J/m$^4$ as a realistic estimate of the adhesion strength relevant to the experimental membrane-bead system, it is useful to examine the model's fit to the data for higher adhesion strengths. In Fig. \ref{app:fig:thresholdmodel_highgamma}, we choose $\gamma = 10^{15}$ J/m$^4$ and observe a nearly perfect fit to the data in \cite{shi2022kinetic} with a lower $C_\textrm{thresh} = 0.733$ $\mu$m$^{-1}$ (compared to $C_\textrm{thresh} = 1.55$ $\mu$m$^{-1}$ at $\gamma = 10^{13}$ J/m$^4$, Fig. \ref{fig:AssocRateThresholdModel}). This does not necessarily indicate that the experimental system is subject to such strong adhesion strengths, but only that minimizing the curvature variance improves the fit to the data.  Therefore, sources of membrane fluctuation suppression other than adhesion (such as membrane tension; see Discussion) may also contribute to improved fits to the data.

\section{\label{app:rhoplusSensing}Sensing lipid densities projected by the upper monolayer}

Instead of sensing the lipid density deviation $\rho$ between the upper and lower monolayer at the midsurface, we investigate here whether the protein might comparably infer differences in bead sizes by sensing the density $\rho^+$ projected solely by the upper monolayer at the midsurface. Using the definitions for $\rho$ and $\bar{\rho}$, we have
\begin{equation}
    \rho + \bar{\rho} = \left(\frac{\rho^+ - \rho^-}{2}\right) +  \left(\frac{\rho^+ + \rho^-}{2}\right)  = \rho^+.
\end{equation}

The mean squared value of $\rho_\qv^+$ is then derived as
\begin{align}
    \rho^+_\qv &= \rho_\qv + \bar{\rho}_\qv\\
    \implies \langle {|\rho_\qv^+|}^2 \rangle &= \langle |\rho_\qv|^2 \rangle + \langle |\bar{\rho}_\qv|^2 \rangle + \langle \rho_\qv \bar{\rho}_{-\qv} \rangle + \langle \rho_{-\qv} \bar{\rho}_{\qv} \rangle
    \\
    &= \langle |\rho_\qv|^2 \rangle + \langle |\bar{\rho}_\qv|^2 \rangle,
\end{align}
where the last step is true because  $\langle \rho_\qv \rangle =  \langle \bar{\rho}_\qv \rangle = 0$ for a membrane associated to a flat substrate, and $\rho_\qv$ and $\bar{\rho}_{-\qv}$ are independent given the flat-membrane energy of Eq. (\ref{EnergyMatrixPlanarFS}).

From Eq. (\ref{rhoFlatFourier}) and Eq. (\ref{rhobarFlatFourier}), we obtain, for a flat membrane,
\begin{align}
    \langle {|\rho_\qv^+|}^2 \rangle = L^2 k_B T \left(\frac{\Tilde{\kappa}q^4 + \gamma}{2k(\kappa q^4 + \gamma)} + \frac{1}{2k}\right).
\end{align}

In the continuum limit, the variance in $\rho^+$ sensed by a protein of size $a$ is
\begin{align}
    \langle {\rho_a^+}^2 \rangle &= \frac{1}{L^4}\sum_\qv \langle {|\rho_\qv^+|}^2 \rangle |G(\qv)|^2  \notag\\
    &= \frac{1}{2\pi}\int_0^\infty q k_B T \left(\frac{\Tilde{\kappa}q^4 + \gamma}{2k(\kappa q^4 + \gamma)} + \frac{1}{2k}\right)|G(q)|^2 \mathrm{d}q.
\end{align}

Substituting a dimensionless parameter $u = qa$, and since $|G(q)|^2 = \exp(-q^2 a^2)$, it can be shown that
\begin{align}
    \label{app:eq:rhoplusquadrature}
    \langle {\rho_a^+}^2 \rangle &= \frac{k_B T}{4 \pi k a^2}\int_0^\infty u \left(\frac{u^4 + \frac{2d^2ku^4}{\kappa} + \frac{\gamma a^4}{\kappa}}{u^4 + \frac{\gamma a^4}{\kappa}} + 1 \right)\exp(-u^2) \mathrm{d}u
    \\
    &= \langle \rho_a^2 \rangle + \frac{k_B T}{4 \pi k a^2}\int_0^\infty{u  \exp(-u^2)} \mathrm{d}u \\
        &= \langle \rho_a^2 \rangle + \frac{k_B T}{8 \pi k a^2},
\end{align}

where $\langle \rho_a^2 \rangle$ is as in Eq. (\ref{ContLimRhoVar_DimLess}). We see that the variance in $\rho_a^{+}$ is always greater than the variance in $\rho_a$ by the simple additive factor $k_B T / 8 \pi k a^2$.

In the absence of membrane-substrate adhesion, 
\begin{equation}
    \langle \rho_a^{+^2} \rangle_{\gamma = 0} = \frac{k_B T (d^2k + \kappa)}{4 \pi a^2 k \kappa}.
\end{equation}

In the high-adhesion limit, 
\begin{align}
    \langle {\rho_a^+}^2 \rangle_\text{high $\gamma$} &= \frac{k_B T}{4 \pi a^2 k} = 2\langle \rho_a ^2 \rangle_\text{high $\gamma$}\\
    \implies \text{SNR}_{\rho^+, \text{ high} \gamma} &= \frac{1}{2}\text{SNR}_{\rho, \text{ high} \gamma}. \label{eq:SNR_rhoplus_highadh}
\end{align}
In deriving the SNR for $\rho^+$, we have used the result that the mean value of $\rho^+$ is the same as the mean value for $\rho$. We can see that the steady state solution for the average density is $\bar{\rho}_\qv = 0$, as obtained by solving $\frac{\partial E}{\partial \rho_\qv^*} = 0$. Therefore, $\rho^+$ has the same mean value as $\rho$, but a greater variance in its distributions.

\begin{figure}[ht]
\centering
    \includegraphics[width=0.47\textwidth]{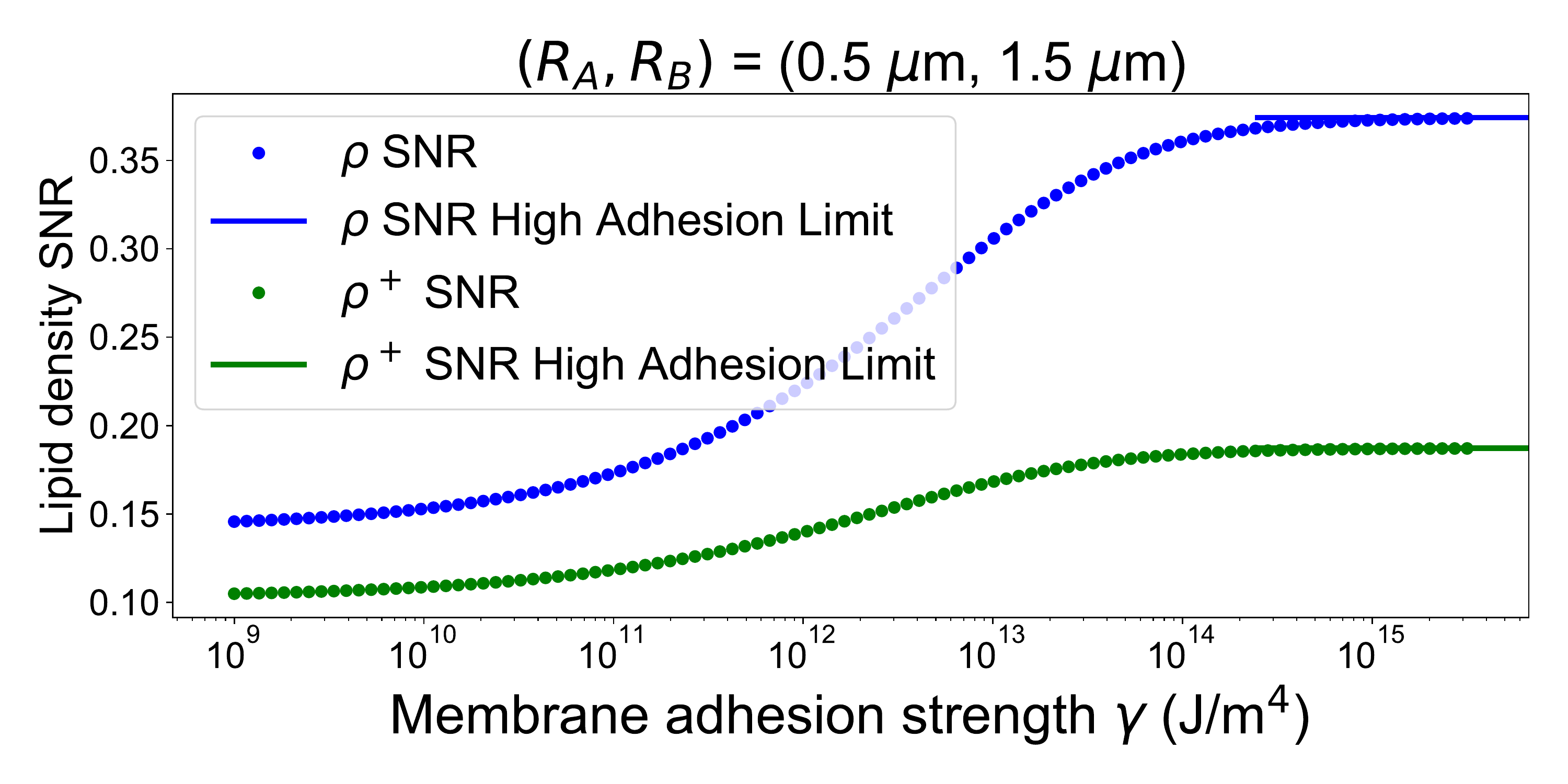}
    \caption{Lipid density SNR for $\rho$ and $\rho^+$ for varying membrane adhesion strengths, in comparison to their high-adhesion limits. The SNR is computed as in Eq. (\ref{SNR_rho}) and the corresponding flat membrane variances for $\langle \rho_{a}^{2} \rangle$ and $\langle \rho_{a}^{+^2} \rangle$ are computed by numerical quadrature. Theory parameters: Table \ref{tab:ModelParameters}.}
    \label{rhorhoplusVarComp}
\end{figure}

We plot the theoretical SNR resulting from a protein probing $\rho$ or $\rho^+$ in Fig. \ref{rhorhoplusVarComp}. We see that the difference between probing $\rho$ and $\rho^+$ becomes largest at high adhesion, where the SNR of probing $\rho$ is twice that of probing $\rho^+$, as discussed in the main text and seen in Eq. (\ref{eq:SNR_rhoplus_highadh}).

\begin{figure}[ht]
\centering
    \includegraphics[width=0.7\textwidth]{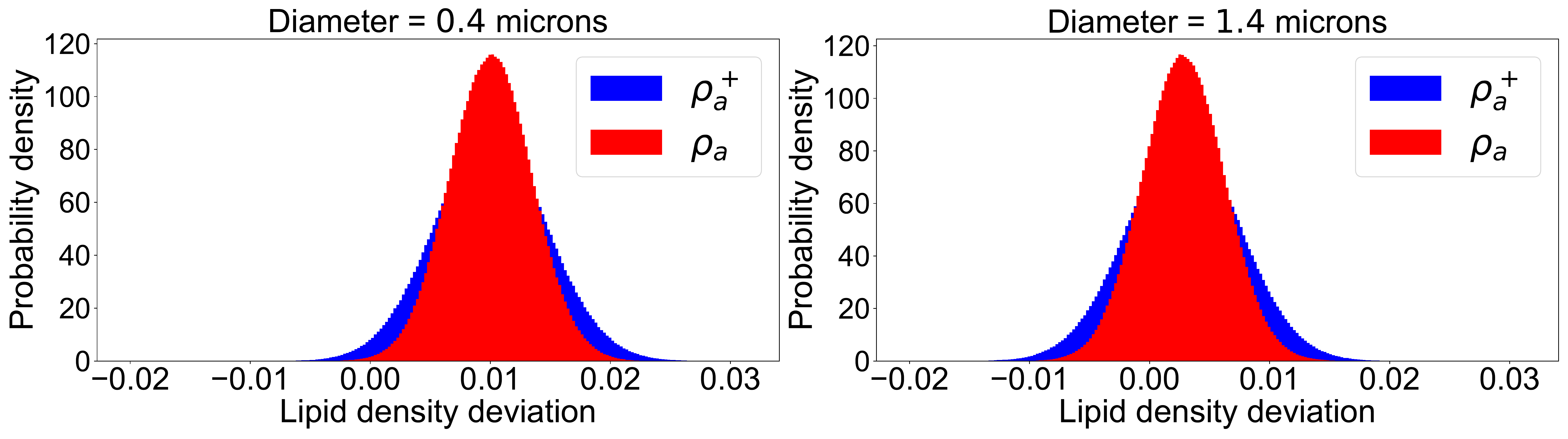}
    \caption{Histograms (normalized as probability densities) from simulations of beads of diameters $0.4$  $\mu$m (left) and $1.4$  $\mu$m (right), with $\gamma = 10^{13}$ J/m$^4$. In each case, $\rho_a$ and $\rho_a^+$ have the same mean values at steady-state, but $\rho_a^+$ has more variance. Parameters: Table \ref{tab:ModelParameters}.}
    \label{rhorhoplusBeadDist}
\end{figure}

To simulate fluctuations in $\rho^+$, we use the relation $\rho_\mathbf{q}^+ = \rho_\mathbf{q} + \bar{\rho}_\mathbf{q}$ in conjunction with the algorithms in Eq. (\ref{DensityFluctuationAlgorithmRho}) and Eq. (\ref{DensityFluctuationAlgorithmRhoBar}). We plot these histograms in Fig. \ref{rhorhoplusBeadDist}, and see, as we expect, that the mean value of $\rho_a$ and $\rho^+_a$ agree, but the variance of $\rho^+$ is larger. For the parameters used in Table \ref{tab:ModelParameters}, the simulated variance in $\rho^+_a$ is approximately $2.13 \times 10^{-5}$ (for the beads in Fig. \ref{rhorhoplusBeadDist}). This is in good agreement with the flat membrane theory variance of Eq. (\ref{app:eq:rhoplusquadrature}), which is approximately $2.11 \times 10^{-5}$ as computed by numerical quadrature.

\section{\label{app:BilayerParameterDependence}Dependence of bilayer bending modulus on monolayer thickness and area compressibility modulus}

In addition to understanding how curvature sensing efficacy depends explicitly on each physical parameter in our model, it may also be of interest to consider instances when these parameters are coupled. Although outside the scope of our  paper,  phenomenological evidence based on a polymer brush model \cite{rawicz2000effect, boal2012mechanics} suggests that the membrane bilayer's bending modulus is coupled to the monolayer's thickness and  area compressibility modulus as
\begin{equation}
    \kappa_\text{bilayer} = \frac{K_A d_{\text{bilayer}}^2}{\alpha} =  \frac{kd^2}{3},
\end{equation}
where $\alpha = 24$ is obtained as a fit parameter from data corresponding to various lipid species, $K_A = 2k$ is the bilayer's area compressibility modulus, and $d_\text{bilayer} = 2d$.

The renormalized membrane bending modulus can then be expressed as 
\begin{equation}
    \Tilde{\kappa} = \kappa_\text{bilayer} + 2d^2k = \frac{7kd^2}{3}.
\end{equation}

\begin{figure}[ht]
\centering
    \includegraphics[width=0.4\textwidth]{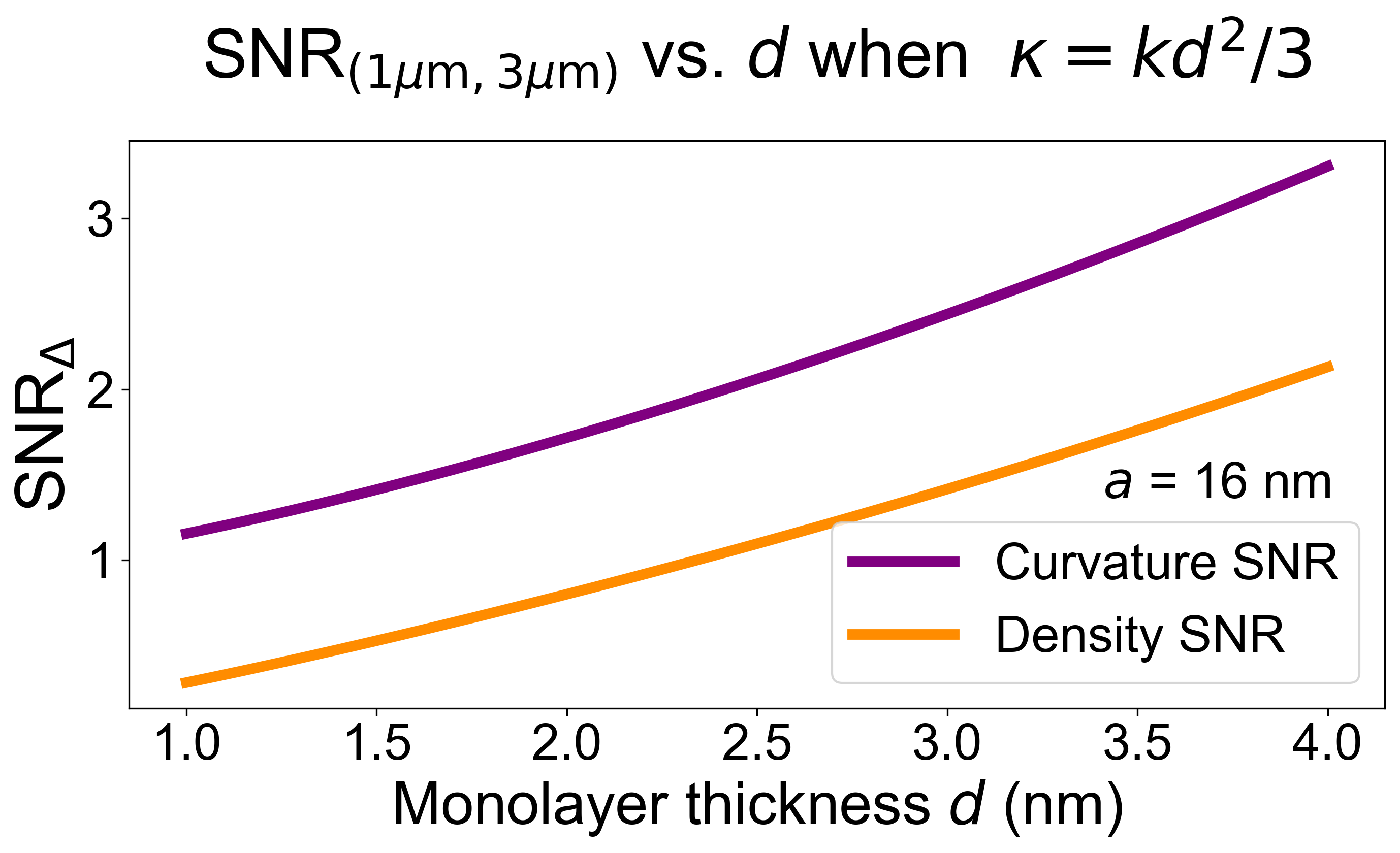}
    \caption{SNR$_C$ and SNR$_\rho$ for beads of diameter $\Delta$ =  (1 $\mu$m, 3 $\mu$m) as a function of the monolayer thickness $d$ when the membrane  bending modulus is coupled to the monolayer's thickness and area compressibility modulus as $\kappa = kd^2/3$.}
    \label{kappadependence}
\end{figure}

As shown in Fig. \ref{kappadependence}, this formulation allows us to compute the SNR  without explicitly choosing a bending modulus by substituting $\kappa = kd^2/3$ in Eq. (\ref{ContLimCurvVar_DimLess}) and Eq. (\ref{ContLimRhoVar_DimLess}). However, this $\alpha = 24$ is phenomenological and may not apply to all lipid species. Understanding, e.g. the role of lipid type on driving different association rates to beads may require systematic characterization of both $\kappa$ and $d$ for different lipid mixtures.

\section{\label{app:RelativeEfficacy}Relative efficacy of lipid density sensing and curvature sensing}

\begin{figure*}[ht]
     \centering
     \begin{subfigure}
         \centering
         \includegraphics[width=0.46\textwidth]{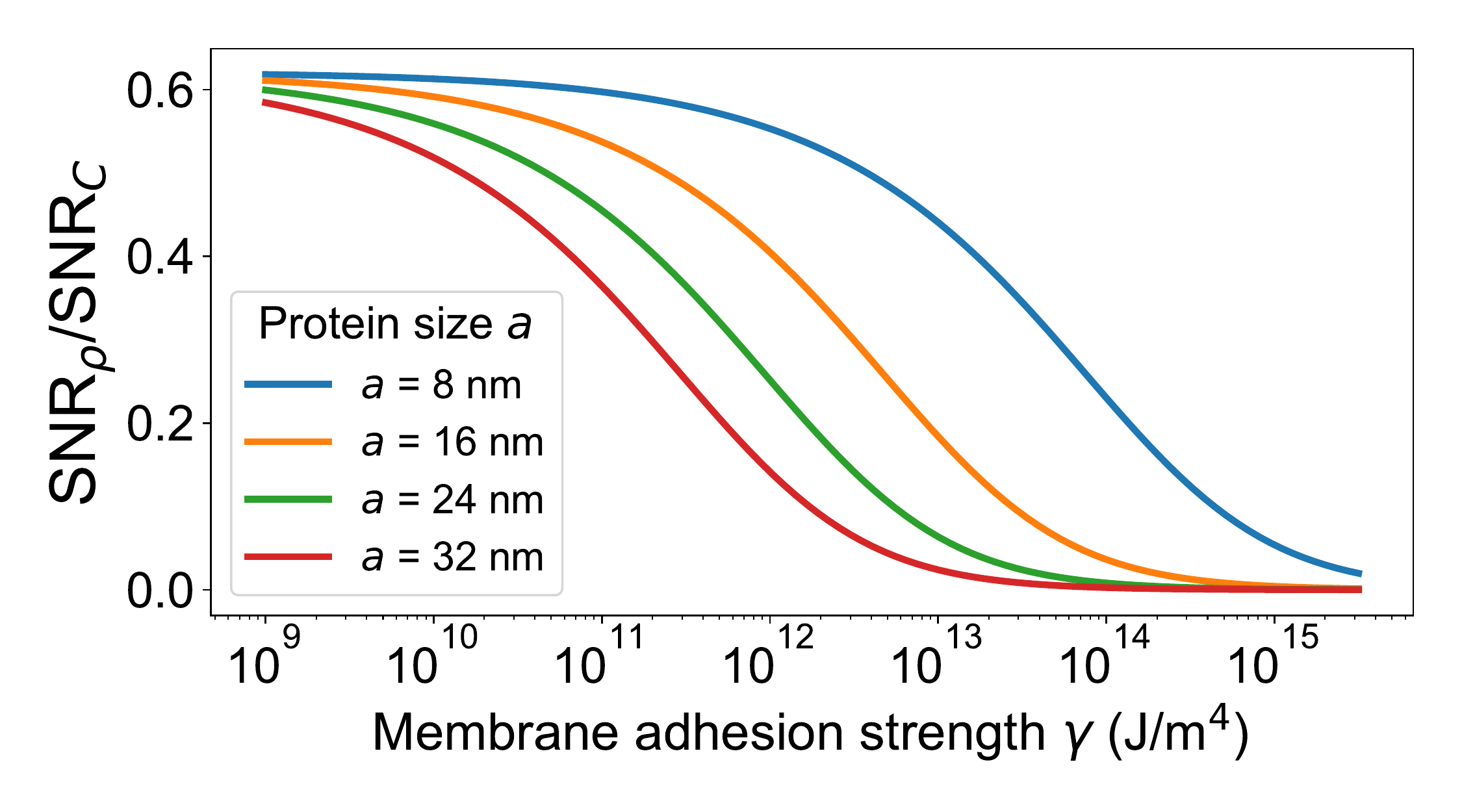}
     \end{subfigure}
     \begin{subfigure}
         \centering
         \includegraphics[width=0.46\textwidth]{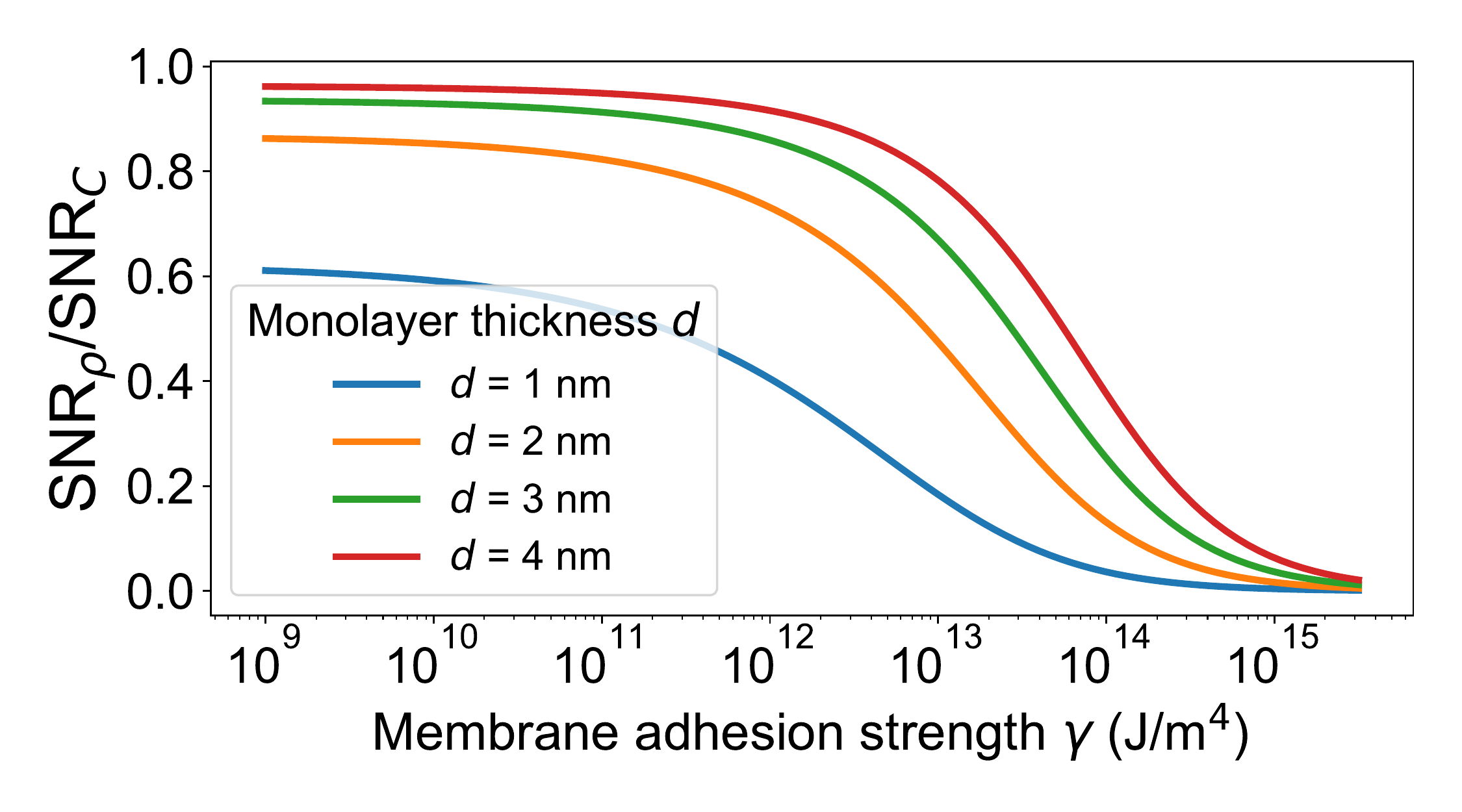}
     \end{subfigure}
     \begin{subfigure}
         \centering
         \includegraphics[width=0.46\textwidth]{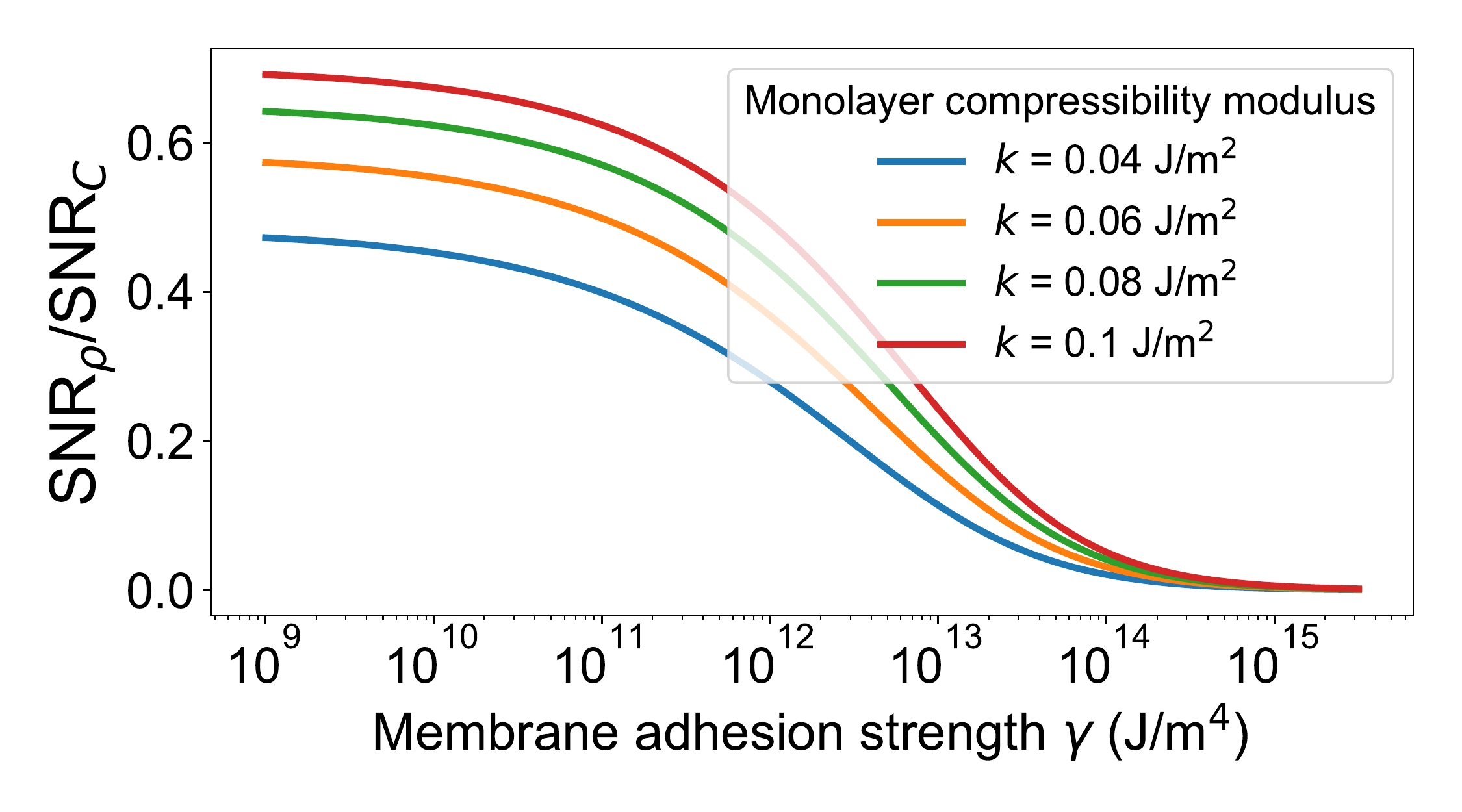}
     \end{subfigure}
     \begin{subfigure}
         \centering
         \includegraphics[width=0.46\textwidth]{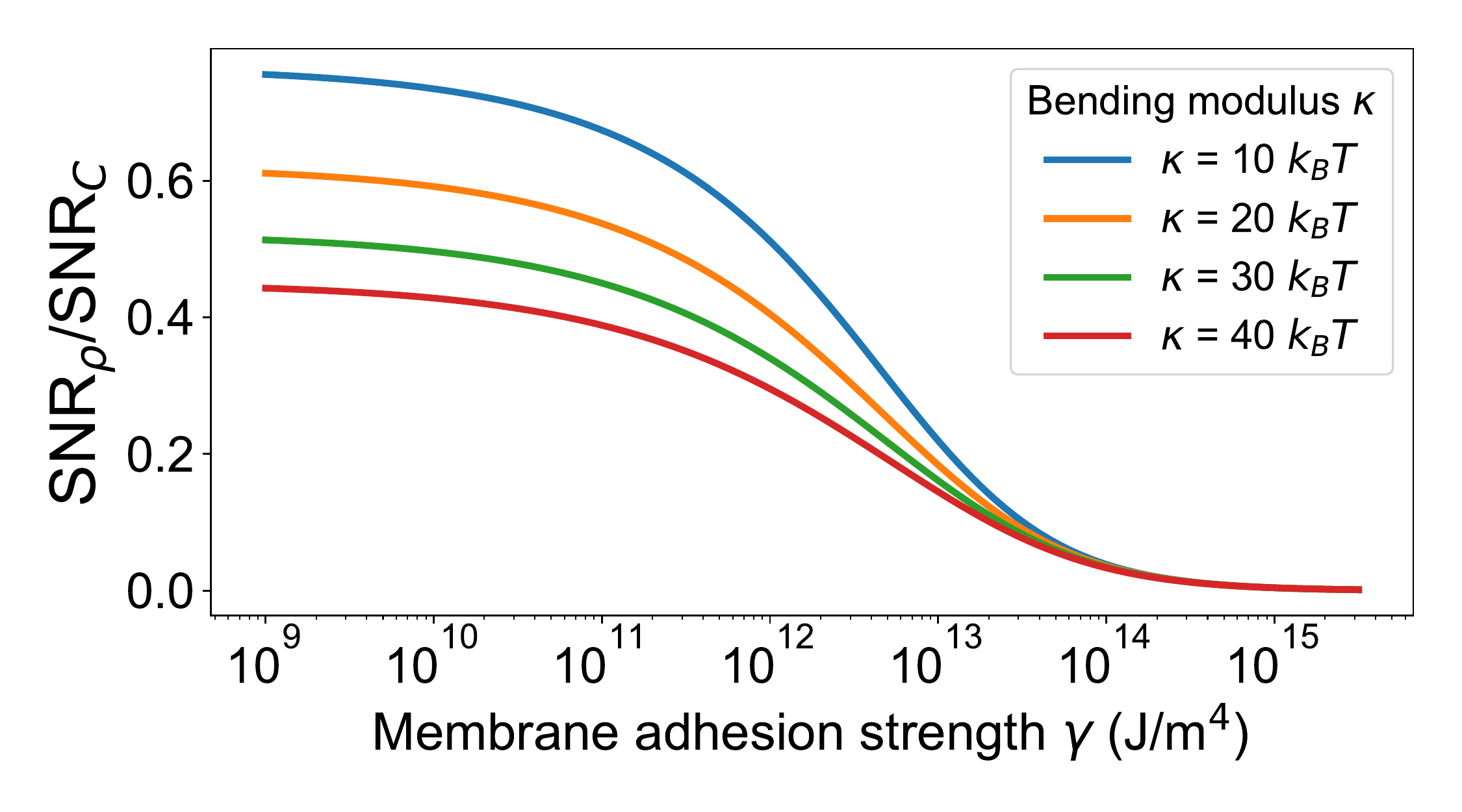}
     \end{subfigure}
     \caption{Ratios between the theoretically predicted $\rho$ SNR and curvature SNR for beads of diameter (1 $\mu$m, 3 $\mu$m) for various physical parameters as a function of increasing $\gamma$. For each plot, only the depicted parameters are varied, while the other parameters are the same as those referenced in Table \ref{tab:ModelParameters}.}
    \label{fig:relativesensingefficacy}
\end{figure*}

To better understand the relationship between various physical parameters and the relative sensing efficacy of lipid density sensing in comparison to local curvature sensing, we plot the ratio between SNR$_{\rho}$ and SNR$_{C}$ in Fig. \ref{fig:relativesensingefficacy} for beads of diameter 1 $\mu$m and 3 $\mu$m. At low $\gamma$, density sensing is a fairly effective sensing strategy compared to local curvature sensing, with the $\frac{SNR_\rho}{SNR_C}$ ratio approaching a value of 1 for thicker membranes with larger $d$. As $\gamma$ is increased, the variance in $\rho$ saturates and curvature sensing becomes significantly more effective as a sensing strategy. For smaller proteins, density sensing serves as a passable proxy for curvature sensing even at relatively high adhesion strengths, in contrast to larger proteins, for which SNR$_\rho/$SNR$_C$ decays more prominently as a function of $\gamma$.

\section{\label{app:FMCsimulations}Fourier Monte Carlo simulations}

We develop here an alternative method to simulate coupled fluctuations in the membrane's height and lipid densities based on a Fourier Monte Carlo (FMC) algorithm \cite{gouliaev1998simulations}, and use it to ensure our FSBD algorithm is correctly reproducing the thermal equilibrium.  For large membranes and membrane-adhered beads, FMC takes a much longer time than our FSBD algorithms to satisfactorily simulate, since a larger number of modes entails a substantial increase in the number of Monte Carlo steps (MCS) required for convergence. Therefore, we only use the FMC method to corroborate our FSBD simulations for small system-sizes. 

In Table \ref{tab:FSBD_FMC_Table}, we compare SNR$_C$ and SNR$_\rho$ for a pair of beads with small radii, as obtained from FSBD and FMC simulations. For beads as small as these, the simulated membrane has deviations from the simple theory result, as in Fig. \ref{fig:SNR_varyingbeads}, but nonetheless FSBD and FMC are in excellent agreement.

\begin{table}[ht]
\begin{tabular}{|c|c|c|}
\hline

Bead diameters: (0.1, 0.2) $\mu$m & FSBD            & FMC             \\ \hline
SNR$_C$                           & 394.2 $\pm$ 2.8 & 395.1 $\pm$ 2.3 \\ \hline
SNR$_\rho$                        & 82.6 $\pm$ 0.4  & 82.8 $\pm$ 0.6  \\ \hline
Bead diameters: (0.2, 0.3) $\mu$m & FSBD            & FMC             \\ \hline
SNR$_C$                           & 6.35 $\pm$ 0.07 & 6.35 $\pm$ 0.08 \\ \hline
SNR$_\rho$                        & 1.33 $\pm$ 0.03 & 1.29 $\pm$ 0.02 \\ \hline
\end{tabular}
\caption{\label{tab:FSBD_FMC_Table} Comparison of FSBD and FMC simulations for small system sizes. Parameters: $L = 400$ nm, $N$ = 21, $\gamma = 10^{13}$ J/m$^4$. FSBD: $t_\text{sim} = 0.016$ s, $\Delta t = 3.2$ ns. FMC: $5 \times 10^7$ attempts (Monte Carlo steps = attempts/$N^2$). The error bars denote standard errors, and were computed using the block averaging method (see Fig. \ref{fig:SNR_varyingbeads} caption). The data were separated into $N_\textrm{block} = 5$ blocks for both FSBD and FMC data, truncating the initial 40\% of FMC data to allow for equilibriation burn-in.}
\end{table}

In the FMC approach, we propose changes to only a single Fourier mode chosen at random for each attempt, compute the resultant change in the membrane energy, and use an acceptance criterion in accordance with the Metropolis rule to determine whether to accept or reject the proposed change. The size of the proposed change varies with each attempt; as we show subsequently, the proposed changes are scaled such that on average, 50\% of the proposals are accepted.

For a flat membrane subject to adhesion, the membrane energy is computed as 
\begin{equation}
    E_\text{tot} = \sum_{\mathbf{q}} \frac{1}{2L^2}\Big((\Tilde{\kappa} {q}^4 + \gamma)|h_{\qv}|^2 - 2kd{q}^2 \rho_{\qv}^* h_{\qv}\\
    - 2kd {q}^2 h_{\qv}^* \rho_{\qv} + 2k|\rho_{\qv}|^2 + 2k|\bar{\rho}_{\qv}|^2 \Big).
\end{equation}

For a membrane-adhered bead, the energy due to the harmonic potential must be explicitly accounted for, such that
\begin{equation}
    E_\text{tot} = \sum_{\mathbf{q}} \frac{1}{2L^2}\Big(\Tilde{\kappa} {q}^4 |h_{\qv}|^2 - 2kd{q}^2 \rho_{\qv}^* h_{\qv} - 2kd {q}^2 h_{\qv}^* \rho_{\qv} + 2k|\rho_{\qv}|^2 + 2k|\bar{\rho}_{\qv}|^2 \Big) + E_\text{adh}.
\end{equation}
where the adhesion energy is computed as a sum over the lattice in real-space as $E_\text{adh} =  \frac{(L/N)^2}{2} \sum_{\mathbf{r}} {\gamma(h(\mathbf{r}) - h_\text{bead}(\mathbf{r}))^2}$.

A single independent mode $\qv$ is selected at random (with the exception of the zeroth mode, which does not evolve), and changes to this mode (denoted by $\Delta$) are computed for the real and imaginary components of this mode $\qv$, for each Monte Carlo attempt, as
 \begin{align}
 \label{app:eq:FMCproposal_height}
     h_{\qv,\text{new}} = h_\qv + 2s_h\sqrt{\langle |h_\qv|^2 \rangle}(\text{rand} - 0.5) + 2 i s_h\sqrt{\langle |h_\qv|^2 \rangle} (\text{rand} - 0.5),\\
\label{app:eq:FMCproposal_rho}
     \rho_{\qv,\text{new}} = \rho_\qv + 2s_\rho \sqrt{\langle |\rho_\qv|^2 \rangle}(\text{rand} - 0.5) + 2is_\rho \sqrt{\langle |\rho_\qv|^2 \rangle}(\text{rand} - 0.5),\\
\label{app:eq:FMCproposal_rhobar}
     \bar{\rho}_{\qv,\text{new}} = \bar{\rho}_\qv + 2s_{\bar{\rho}}\sqrt{\langle |\bar{\rho}_\qv|^2 \rangle}(\text{rand} - 0.5) + 2is_{\bar{\rho}}\sqrt{\langle |\bar{\rho}_\qv|^2 \rangle}(\text{rand} - 0.5),
 \end{align}
  where rand indicates a random number between 0 and 1. Each use of rand here is a different random number, so the real and imaginary parts are updated with independent random values. $\langle |h_\qv|^2 \rangle$,$\langle |\rho_\qv|^2 \rangle$, and $\langle |\bar{\rho}_\qv|^2 \rangle$ are as derived in Eqs. (\ref{heightFlatFourier})--(\ref{rhobarFlatFourier}). $s_h$, $s_{\rho}$, and $s_{\bar{\rho}}$ are scaling factors that can be varied to influence how frequently the proposed changes are accepted. For the simulations in Table \ref{tab:FSBD_FMC_Table}, the scaling factors $s_h = 0.95$, $s_\rho = 1.3$, and  $s_{\bar{\rho}} = 1.3$ resulted in approximately 50\% acceptance.
 
To update the dependent modes, we conjugate the independent modes to ensure that the height and density variables in real-space are real-valued (see Appendix \ref{app:FourierConventions}). Therefore, $h_{-\qv, \text{new}} = h_{\qv, \text{new}}^*$, and similarly for the other fields. 
Subsequently, $h_{\mathbf{q}, \text{new}}, \rho_{\mathbf{q}, \text{new}}$, and  $\bar{\rho}_{\mathbf{q}, \text{new}}$ are inverse Fast Fourier transformed to obtain their corresponding real-space values.

We change one mode $\mathbf{q}$ at a time, and also only change one of the three fields $h_\mathbf{q}$, $\rho_\mathbf{q}$, and $\bar{\rho}_\mathbf{q}$ at a time. (We propose changes first for $h_\mathbf{q}$, then for $\rho_\mathbf{q}$, then $\bar{\rho}_\mathbf{q}$.) The usual Metropolis acceptance criterion is used, and applied after each change---i.e. we update the three fields separately, not simultaneously. This Metropolis criterion is: 
 \begin{equation}
     \text{rand} < \exp\left(\frac{-(E_\text{tot, new} - E_\text{tot})}{k_B T}\right).
 \end{equation}

 If this condition is fulfilled, then the change is accepted, and $E_\text{tot}$ and the corresponding height and density variables are updated to their new values and iterated for use in the next attempt.

\bibliographystyle{ieeetr}
\bibliography{bibliography.bib}

\begin{thebibliography}{10}

\bibitem{mcmahon2005membrane}
H.~T. McMahon and J.~L. Gallop, ``Membrane curvature and mechanisms of dynamic
  cell membrane remodelling,'' {\em Nature}, vol.~438, no.~7068, pp.~590--596,
  2005.

\bibitem{mcmahon2015membrane}
H.~T. McMahon and E.~Boucrot, ``Membrane curvature at a glance,'' {\em Journal
  of cell science}, vol.~128, no.~6, pp.~1065--1070, 2015.

\bibitem{stachowiak2013cost}
J.~C. Stachowiak, F.~M. Brodsky, and E.~A. Miller, ``A cost--benefit analysis
  of the physical mechanisms of membrane curvature,'' {\em Nature cell
  biology}, vol.~15, no.~9, pp.~1019--1027, 2013.

\bibitem{zimmerberg2006proteins}
J.~Zimmerberg and M.~M. Kozlov, ``How proteins produce cellular membrane
  curvature,'' {\em Nature reviews Molecular cell biology}, vol.~7, no.~1,
  pp.~9--19, 2006.

\bibitem{ramamurthi2010protein}
K.~S. Ramamurthi, ``Protein localization by recognition of membrane
  curvature,'' {\em Current opinion in microbiology}, vol.~13, no.~6,
  pp.~753--757, 2010.

\bibitem{singh2022sensing}
A.~R. Singh, T.~Leadbetter, and B.~A. Camley, ``Sensing the shape of a cell
  with reaction diffusion and energy minimization,'' {\em Proceedings of the
  National Academy of Sciences}, vol.~119, no.~31, p.~e2121302119, 2022.

\bibitem{bassereau20182018}
P.~Bassereau, R.~Jin, T.~Baumgart, M.~Deserno, R.~Dimova, V.~A. Frolov, P.~V.
  Bashkirov, H.~Grubm{\"u}ller, R.~Jahn, H.~J. Risselada, {\em et~al.}, ``The
  2018 biomembrane curvature and remodeling roadmap,'' {\em Journal of physics
  D: Applied physics}, vol.~51, no.~34, p.~343001, 2018.

\bibitem{lou2018role}
H.-Y. Lou, W.~Zhao, Y.~Zeng, and B.~Cui, ``The role of membrane curvature in
  nanoscale topography-induced intracellular signaling,'' {\em Accounts of
  chemical research}, vol.~51, no.~5, pp.~1046--1053, 2018.

\bibitem{bridges2016micron}
A.~A. Bridges, M.~S. Jentzsch, P.~W. Oakes, P.~Occhipinti, and A.~S.
  Gladfelter, ``Micron-scale plasma membrane curvature is recognized by the
  septin cytoskeleton,'' {\em Journal of Cell Biology}, vol.~213, no.~1,
  pp.~23--32, 2016.

\bibitem{beber2019membrane}
A.~Beber, C.~Taveneau, M.~Nania, F.-C. Tsai, A.~Di~Cicco, P.~Bassereau,
  D.~L{\'e}vy, J.~T. Cabral, H.~Isambert, S.~Mangenot, and A.~Bertin,
  ``Membrane reshaping by micrometric curvature sensitive septin filaments,''
  {\em Nature Communications}, vol.~10, no.~1, pp.~1--12, 2019.

\bibitem{cannon2017unsolved}
K.~S. Cannon, B.~L. Woods, and A.~S. Gladfelter, ``The unsolved problem of how
  cells sense micron-scale curvature,'' {\em Trends in biochemical sciences},
  vol.~42, no.~12, pp.~961--976, 2017.

\bibitem{cannon2019amphipathic}
K.~S. Cannon, B.~L. Woods, J.~M. Crutchley, and A.~S. Gladfelter, ``An
  amphipathic helix enables septins to sense micrometer-scale membrane
  curvature,'' {\em Journal of Cell Biology}, vol.~218, no.~4, pp.~1128--1137,
  2019.

\bibitem{shi2022kinetic}
W.~Shi, K.~S. Cannon, B.~N. Curtis, C.~Edelmaier, A.~S. Gladfelter, and
  E.~Nazockdast, ``A kinetic basis for curvature sensing by septins,'' {\em
  bioRxiv}, 2022.

\bibitem{safran2018statistical}
S.~Safran, {\em Statistical thermodynamics of surfaces, interfaces, and
  membranes}.
\newblock CRC Press, 2018.

\bibitem{drin2010amphipathic}
G.~Drin and B.~Antonny, ``Amphipathic helices and membrane curvature,'' {\em
  FEBS letters}, vol.~584, no.~9, pp.~1840--1847, 2010.

\bibitem{drin2007general}
G.~Drin, J.-F. Casella, R.~Gautier, T.~Boehmer, T.~U. Schwartz, and B.~Antonny,
  ``A general amphipathic $\alpha$-helical motif for sensing membrane
  curvature,'' {\em Nature structural \& molecular biology}, vol.~14, no.~2,
  pp.~138--146, 2007.

\bibitem{cui2011mechanism}
H.~Cui, E.~Lyman, and G.~A. Voth, ``Mechanism of membrane curvature sensing by
  amphipathic helix containing proteins,'' {\em Biophysical Journal}, vol.~100,
  no.~5, pp.~1271--1279, 2011.

\bibitem{antonny2011mechanisms}
B.~Antonny, ``Mechanisms of membrane curvature sensing,'' {\em Annual review of
  biochemistry}, vol.~80, pp.~101--123, 2011.

\bibitem{assoian2019cellular}
R.~K. Assoian, N.~D. Bade, C.~V. Cameron, and K.~J. Stebe, ``Cellular sensing
  of micron-scale curvature: a frontier in understanding the
  microenvironment,'' {\em Open biology}, vol.~9, no.~10, p.~190155, 2019.

\bibitem{baumgart2011thermodynamics}
T.~Baumgart, B.~R. Capraro, C.~Zhu, and S.~L. Das, ``Thermodynamics and
  mechanics of membrane curvature generation and sensing by proteins and
  lipids,'' {\em Annual review of physical chemistry}, vol.~62, pp.~483--506,
  2011.

\bibitem{fu2021continuum}
Y.~Fu, W.~F. Zeno, J.~C. Stachowiak, and M.~E. Johnson, ``A continuum membrane
  model can predict curvature sensing by helix insertion,'' {\em Soft Matter},
  vol.~17, no.~47, pp.~10649--10663, 2021.

\bibitem{campelo2014sensing}
F.~Campelo and M.~M. Kozlov, ``Sensing membrane stresses by protein
  insertions,'' {\em PLoS Computational Biology}, vol.~10, no.~4, p.~e1003556,
  2014.

\bibitem{berg1977physics}
H.~C. Berg and E.~M. Purcell, ``Physics of chemoreception,'' {\em Biophysical
  journal}, vol.~20, no.~2, pp.~193--219, 1977.

\bibitem{bialek2005physical}
W.~Bialek and S.~Setayeshgar, ``Physical limits to biochemical signaling,''
  {\em Proceedings of the National Academy of Sciences}, vol.~102, no.~29,
  pp.~10040--10045, 2005.

\bibitem{endres2009maximum}
R.~G. Endres and N.~S. Wingreen, ``Maximum likelihood and the single
  receptor,'' {\em Physical Review Letters}, vol.~103, no.~15, p.~158101, 2009.

\bibitem{kaizu2014berg}
K.~Kaizu, W.~De~Ronde, J.~Paijmans, K.~Takahashi, F.~Tostevin, and P.~R.
  Ten~Wolde, ``The berg-purcell limit revisited,'' {\em Biophysical Journal},
  vol.~106, no.~4, pp.~976--985, 2014.

\bibitem{ten2016fundamental}
P.~R. ten Wolde, N.~B. Becker, T.~E. Ouldridge, and A.~Mugler, ``Fundamental
  limits to cellular sensing,'' {\em Journal of Statistical Physics}, vol.~162,
  no.~5, pp.~1395--1424, 2016.

\bibitem{hu2010physical}
B.~Hu, W.~Chen, W.-J. Rappel, and H.~Levine, ``Physical limits on cellular
  sensing of spatial gradients,'' {\em Physical Review Letters}, vol.~105,
  no.~4, p.~048104, 2010.

\bibitem{camley2018collective}
B.~A. Camley, ``Collective gradient sensing and chemotaxis: modeling and recent
  developments,'' {\em Journal of Physics: Condensed Matter}, vol.~30, no.~22,
  p.~223001, 2018.

\bibitem{ipina2022collective}
E.~{Perez Ipi{\~n}a} and B.~A. Camley, ``Collective gradient sensing with
  limited positional information,'' {\em Phys. Rev. E}, vol.~105, p.~044410,
  2022.

\bibitem{mugler2016limits}
A.~Mugler, A.~Levchenko, and I.~Nemenman, ``Limits to the precision of gradient
  sensing with spatial communication and temporal integration,'' {\em
  Proceedings of the National Academy of Sciences}, vol.~113, no.~6,
  pp.~E689--E695, 2016.

\bibitem{nwogbaga2022physical}
I.~Nwogbaga, A.~H. Kim, and B.~A. Camley, ``Physical limits on galvanotaxis,''
  {\em arXiv preprint arXiv:2209.04742}, 2022.

\bibitem{fancher2020precision}
S.~Fancher, M.~Vennettilli, N.~Hilgert, and A.~Mugler, ``Precision of flow
  sensing by self-communicating cells,'' {\em Physical Review Letters},
  vol.~124, no.~16, p.~168101, 2020.

\bibitem{beroz2017physical}
F.~Beroz, L.~M. Jawerth, S.~M{\"u}nster, D.~A. Weitz, C.~P. Broedersz, and
  N.~S. Wingreen, ``Physical limits to biomechanical sensing in disordered
  fibre networks,'' {\em Nature Communications}, vol.~8, no.~1, pp.~1--11,
  2017.

\bibitem{beroz2020physical}
F.~Beroz, D.~Zhou, X.~Mao, and D.~K. Lubensky, ``Physical limits to sensing
  material properties,'' {\em Nature communications}, vol.~11, no.~1, pp.~1--9,
  2020.

\bibitem{brown2008elastic}
F.~L. Brown, ``Elastic modeling of biomembranes and lipid bilayers,'' {\em
  Annu. Rev. Phys. Chem.}, vol.~59, pp.~685--712, 2008.

\bibitem{schmidt2014signature}
D.~Schmidt, C.~Monzel, T.~Bihr, R.~Merkel, U.~Seifert, K.~Sengupta, and A.-S.
  Smith, ``Signature of a nonharmonic potential as revealed from a consistent
  shape and fluctuation analysis of an adherent membrane,'' {\em Physical
  Review X}, vol.~4, no.~2, p.~021023, 2014.

\bibitem{swain1999influence}
P.~S. Swain and D.~Andelman, ``The influence of substrate structure on membrane
  adhesion,'' {\em Langmuir}, vol.~15, no.~26, pp.~8902--8914, 1999.

\bibitem{seifert1993viscous}
U.~Seifert and S.~A. Langer, ``Viscous modes of fluid bilayer membranes,'' {\em
  EPL (Europhysics letters)}, vol.~23, no.~1, p.~71, 1993.

\bibitem{seifert1997configurations}
U.~Seifert, ``Configurations of fluid membranes and vesicles,'' {\em Advances
  in physics}, vol.~46, no.~1, pp.~13--137, 1997.

\bibitem{watson2011intermediate}
M.~C. Watson, Y.~Peng, Y.~Zheng, and F.~L. Brown, ``The intermediate scattering
  function for lipid bilayer membranes: From nanometers to microns,'' {\em The
  Journal of chemical physics}, vol.~135, no.~19, p.~194701, 2011.

\bibitem{helfrich1973elastic}
W.~Helfrich, ``Elastic properties of lipid bilayers: theory and possible
  experiments,'' {\em Zeitschrift f{\"u}r Naturforschung C}, vol.~28,
  no.~11-12, pp.~693--703, 1973.

\bibitem{hossein2020spontaneous}
A.~Hossein and M.~Deserno, ``Spontaneous curvature, differential stress, and
  bending modulus of asymmetric lipid membranes,'' {\em Biophysical Journal},
  vol.~118, no.~3, pp.~624--642, 2020.

\bibitem{camley2013diffusion}
B.~A. Camley and F.~L. Brown, ``Diffusion of complex objects embedded in free
  and supported lipid bilayer membranes: role of shape anisotropy and leaflet
  structure,'' {\em Soft Matter}, vol.~9, no.~19, pp.~4767--4779, 2013.

\bibitem{lin2004brownian}
L.~C.-L. Lin and F.~L. Brown, ``Brownian dynamics in fourier space: membrane
  simulations over long length and time scales,'' {\em Physical review
  letters}, vol.~93, no.~25, p.~256001, 2004.

\bibitem{lin2006simulating}
L.~C.-L. Lin and F.~L. Brown, ``Simulating membrane dynamics in nonhomogeneous
  hydrodynamic environments,'' {\em Journal of Chemical Theory and
  Computation}, vol.~2, no.~3, pp.~472--483, 2006.

\bibitem{gouliaev1998simulations}
N.~Gouliaev and J.~F. Nagle, ``Simulations of a single membrane between two
  walls using a monte carlo method,'' {\em Physical Review E}, vol.~58, no.~1,
  p.~881, 1998.

\bibitem{grossfield2018best}
A.~Grossfield, P.~N. Patrone, D.~R. Roe, A.~J. Schultz, D.~W. Siderius, and
  D.~M. Zuckerman, ``Best practices for quantification of uncertainty and
  sampling quality in molecular simulations [article v1. 0],'' {\em Living
  journal of computational molecular science}, vol.~1, no.~1, 2018.

\bibitem{deserno2003wrapping}
M.~Deserno and T.~Bickel, ``Wrapping of a spherical colloid by a fluid
  membrane,'' {\em EPL (Europhysics Letters)}, vol.~62, no.~5, p.~767, 2003.

\bibitem{kardar2007statistical}
M.~Kardar, {\em Statistical physics of particles}.
\newblock Cambridge University Press, 2007.

\bibitem{2020SciPy-NMeth}
P.~Virtanen, R.~Gommers, T.~E. Oliphant, M.~Haberland, T.~Reddy, D.~Cournapeau,
  E.~Burovski, P.~Peterson, W.~Weckesser, J.~Bright, S.~J. {van der Walt},
  M.~Brett, J.~Wilson, K.~J. Millman, N.~Mayorov, A.~R.~J. Nelson, E.~Jones,
  R.~Kern, E.~Larson, C.~J. Carey, {\.I}.~Polat, Y.~Feng, E.~W. Moore,
  J.~{VanderPlas}, D.~Laxalde, J.~Perktold, R.~Cimrman, I.~Henriksen, E.~A.
  Quintero, C.~R. Harris, A.~M. Archibald, A.~H. Ribeiro, F.~Pedregosa, P.~{van
  Mulbregt}, and {SciPy 1.0 Contributors}, ``{{SciPy} 1.0: Fundamental
  Algorithms for Scientific Computing in Python},'' {\em Nature Methods},
  vol.~17, pp.~261--272, 2020.

\bibitem{Rohatgi2022}
A.~Rohatgi, ``Webplotdigitizer: Version 4.6.
  (https://automeris.io/webplotdigitizer/),'' 2022.

\bibitem{newville2016lmfit}
M.~Newville, T.~Stensitzki, D.~B. Allen, M.~Rawlik, A.~Ingargiola, and
  A.~Nelson, ``Lmfit: Non-linear least-square minimization and curve-fitting
  for python,'' {\em Astrophysics Source Code Library}, pp.~ascl--1606, 2016.

\bibitem{biswas2017mapping}
A.~Biswas, A.~Alex, and B.~Sinha, ``Mapping cell membrane fluctuations reveals
  their active regulation and transient heterogeneities,'' {\em Biophysical
  journal}, vol.~113, no.~8, pp.~1768--1781, 2017.

\bibitem{oldham2008error}
K.~B. Oldham, J.~C. Myland, and J.~Spanier, ``The error function erf (x) and
  its complement erfc (x),'' in {\em An Atlas of Functions}, pp.~405--415,
  Springer, 2008.

\bibitem{jin2022curvature}
R.~Jin, R.~Cao, and T.~Baumgart, ``Curvature dependence of bar protein membrane
  association and dissociation kinetics,'' {\em Scientific Reports}, vol.~12,
  no.~1, pp.~1--9, 2022.

\bibitem{sengupta2018adhesion}
K.~Sengupta and A.-S. Smith, ``Adhesion of biological membranes,'' in {\em
  Physics of Biological Membranes}, pp.~499--535, Springer, 2018.

\bibitem{gordon2015membrane}
V.~D. Gordon, T.~O'Halloran, and O.~Shindell, ``Membrane adhesion and the
  formation of heterogeneities: biology, biophysics, and biotechnology,'' {\em
  Physical Chemistry Chemical Physics}, vol.~17, no.~24, pp.~15522--15533,
  2015.

\bibitem{hynes1999cell}
R.~O. Hynes, ``Cell adhesion: old and new questions,'' {\em Trends in
  Genetics}, vol.~15, no.~12, pp.~M33--M37, 1999.

\bibitem{sheetz2001cell}
M.~P. Sheetz, ``Cell control by membrane--cytoskeleton adhesion,'' {\em Nature
  Reviews Molecular Cell Biology}, vol.~2, no.~5, pp.~392--396, 2001.

\bibitem{ursell2011lipid}
T.~Ursell, A.~Agrawal, and R.~Phillips, ``Lipid bilayer mechanics in a pipette
  with glass-bilayer adhesion,'' {\em Biophysical journal}, vol.~101, no.~8,
  pp.~1913--1920, 2011.

\bibitem{anderson2009formation}
T.~H. Anderson, Y.~Min, K.~L. Weirich, H.~Zeng, D.~Fygenson, and J.~N.
  Israelachvili, ``Formation of supported bilayers on silica substrates,'' {\em
  Langmuir}, vol.~25, no.~12, pp.~6997--7005, 2009.

\bibitem{zwang2010quantification}
T.~J. Zwang, W.~R. Fletcher, T.~J. Lane, and M.~S. Johal, ``Quantification of
  the layer of hydration of a supported lipid bilayer,'' {\em Langmuir},
  vol.~26, no.~7, pp.~4598--4601, 2010.

\bibitem{israelachvili2011intermolecular}
J.~N. Israelachvili, {\em Intermolecular and surface forces}.
\newblock Academic press, 2011.

\bibitem{huang2010macromolecules}
K.~C. Huang and K.~S. Ramamurthi, ``Macromolecules that prefer their membranes
  curvy,'' {\em Molecular microbiology}, vol.~76, no.~4, pp.~822--832, 2010.

\bibitem{gill2015structural}
R.~L. Gill, J.-P. Castaing, J.~Hsin, I.~S. Tan, X.~Wang, K.~C. Huang, F.~Tian,
  and K.~S. Ramamurthi, ``Structural basis for the geometry-driven localization
  of a small protein,'' {\em Proceedings of the National Academy of Sciences},
  vol.~112, no.~15, pp.~E1908--E1915, 2015.

\bibitem{watson2011thermal}
M.~C. Watson, E.~S. Penev, P.~M. Welch, and F.~L. Brown, ``Thermal fluctuations
  in shape, thickness, and molecular orientation in lipid bilayers,'' {\em The
  Journal of Chemical Physics}, vol.~135, no.~24, p.~244701, 2011.

\bibitem{watson2012determining}
M.~C. Watson, E.~G. Brandt, P.~M. Welch, and F.~L. Brown, ``Determining
  biomembrane bending rigidities from simulations of modest size,'' {\em
  Physical Review Letters}, vol.~109, no.~2, p.~028102, 2012.

\bibitem{hopkins2020chemotaxis}
A.~Hopkins and B.~A. Camley, ``Chemotaxis in uncertain environments: Hedging
  bets with multiple receptor types,'' {\em Physical Review Research}, vol.~2,
  no.~4, p.~043146, 2020.

\bibitem{mehta2012energetic}
P.~Mehta and D.~J. Schwab, ``Energetic costs of cellular computation,'' {\em
  Proceedings of the National Academy of Sciences}, vol.~109, no.~44,
  pp.~17978--17982, 2012.

\bibitem{govern2014energy}
C.~C. Govern and P.~R. ten Wolde, ``Energy dissipation and noise correlations
  in biochemical sensing,'' {\em Physical review letters}, vol.~113, no.~25,
  p.~258102, 2014.

\bibitem{lang2014thermodynamics}
A.~H. Lang, C.~K. Fisher, T.~Mora, and P.~Mehta, ``Thermodynamics of
  statistical inference by cells,'' {\em Physical Review Letters}, vol.~113,
  no.~14, p.~148103, 2014.

\bibitem{alvarez2010diffusion}
N.~J. Alvarez, L.~M. Walker, and S.~L. Anna, ``Diffusion-limited adsorption to
  a spherical geometry: The impact of curvature and competitive time scales,''
  {\em Physical Review E}, vol.~82, no.~1, p.~011604, 2010.

\bibitem{pinot2018feedback}
M.~Pinot, S.~Vanni, E.~Ambroggio, D.~Guet, B.~Goud, and J.-B. Manneville,
  ``Feedback between membrane tension, lipid shape and curvature in the
  formation of packing defects,'' {\em bioRxiv}, p.~389627, 2018.

\bibitem{colom2018fluorescent}
A.~Colom, E.~Derivery, S.~Soleimanpour, C.~Tomba, M.~D. Molin, N.~Sakai,
  M.~Gonz{\'a}lez-Gait{\'a}n, S.~Matile, and A.~Roux, ``A fluorescent membrane
  tension probe,'' {\em Nature Chemistry}, vol.~10, no.~11, pp.~1118--1125,
  2018.

\bibitem{wasnik2015modeling}
V.~Wasnik, N.~S. Wingreen, and R.~Mukhopadhyay, ``Modeling curvature-dependent
  subcellular localization of the small sporulation protein spovm in bacillus
  subtilis,'' {\em PLoS One}, vol.~10, no.~1, p.~e0111971, 2015.

\bibitem{scomparin2009diffusion}
C.~Scomparin, S.~Lecuyer, M.~Ferreira, T.~Charitat, and B.~Tinland, ``Diffusion
  in supported lipid bilayers: Influence of substrate and preparation technique
  on the internal dynamics,'' {\em The European Physical Journal E}, vol.~28,
  no.~2, pp.~211--220, 2009.

\bibitem{gunderson2018liquid}
R.~S. Gunderson and A.~R. Honerkamp-Smith, ``Liquid-liquid phase transition
  temperatures increase when lipid bilayers are supported on glass,'' {\em
  Biochimica et Biophysica Acta (BBA)-Biomembranes}, vol.~1860, no.~10,
  pp.~1965--1971, 2018.

\bibitem{amjad2017membrane}
O.~A. Amjad, B.~M. Mognetti, P.~Cicuta, and L.~Di~Michele, ``Membrane adhesion
  through bridging by multimeric ligands,'' {\em Langmuir}, vol.~33, no.~5,
  pp.~1139--1146, 2017.

\bibitem{sigurdsson2013hybrid}
J.~K. Sigurdsson, F.~L. Brown, and P.~J. Atzberger, ``Hybrid continuum-particle
  method for fluctuating lipid bilayer membranes with diffusing protein
  inclusions,'' {\em Journal of Computational Physics}, vol.~252, pp.~65--85,
  2013.

\bibitem{camley2014fluctuating}
B.~A. Camley and F.~L. Brown, ``Fluctuating hydrodynamics of multicomponent
  membranes with embedded proteins,'' {\em The Journal of chemical physics},
  vol.~141, no.~7, p.~08B615\_1, 2014.

\bibitem{lin2004dynamics}
L.~C.-L. Lin and F.~L. Brown, ``Dynamics of pinned membranes with application
  to protein diffusion on the surface of red blood cells,'' {\em Biophysical
  journal}, vol.~86, no.~2, pp.~764--780, 2004.

\bibitem{harris2020array}
C.~R. Harris, K.~J. Millman, S.~J. van~der Walt, R.~Gommers, P.~Virtanen,
  D.~Cournapeau, E.~Wieser, J.~Taylor, S.~Berg, N.~J. Smith, R.~Kern, M.~Picus,
  S.~Hoyer, M.~H. van Kerkwijk, M.~Brett, A.~Haldane, J.~F. del R{\'{i}}o,
  M.~Wiebe, P.~Peterson, P.~G{\'{e}}rard-Marchant, K.~Sheppard, T.~Reddy,
  W.~Weckesser, H.~Abbasi, C.~Gohlke, and T.~E. Oliphant, ``Array programming
  with {NumPy},'' {\em Nature}, vol.~585, pp.~357--362, Sept. 2020.

\bibitem{boal2012mechanics}
D.~Boal and D.~H. Boal, {\em Mechanics of the Cell}.
\newblock Cambridge University Press, 2012.

\bibitem{faizi2021viscosity}
H.~A. Faizi, R.~Dimova, and P.~M. Vlahovska, ``Viscosity of fluid membranes
  measured from vesicle deformation,'' {\em arXiv preprint arXiv:2103.02106},
  2021.

\bibitem{doi1988theory}
M.~Doi, S.~F. Edwards, and S.~F. Edwards, {\em The theory of polymer dynamics},
  vol.~73.
\newblock oxford university press, 1988.

\bibitem{gardiner1985handbook}
C.~W. Gardiner {\em et~al.}, {\em Handbook of stochastic methods}, vol.~3.
\newblock springer Berlin, 1985.

\bibitem{kloeden2012numerical}
P.~E. Kloeden, E.~Platen, and H.~Schurz, {\em Numerical solution of SDE through
  computer experiments}.
\newblock Springer Science \& Business Media, 2012.

\bibitem{camley2010dynamic}
B.~A. Camley and F.~L. Brown, ``Dynamic simulations of multicomponent lipid
  membranes over long length and time scales,'' {\em Physical Review Letters},
  vol.~105, no.~14, p.~148102, 2010.

\bibitem{israelachvili1994strength}
J.~Israelachvili, ``Strength of van der waals attraction between lipid
  bilayers,'' {\em Langmuir}, vol.~10, no.~9, pp.~3369--3370, 1994.

\bibitem{rawicz2000effect}
W.~Rawicz, K.~C. Olbrich, T.~McIntosh, D.~Needham, and E.~Evans, ``Effect of
  chain length and unsaturation on elasticity of lipid bilayers,'' {\em
  Biophysical journal}, vol.~79, no.~1, pp.~328--339, 2000.

\end{thebibliography}

\end{document}